\documentclass [a4paper,12pt] {article}
\usepackage[psamsfonts]{amssymb}
\usepackage{latexsym}
\usepackage[dvips]{graphicx}
\usepackage{color}
\usepackage{bm}
\usepackage{amsmath}
\usepackage{comment}
\usepackage{caption,soul}

\usepackage{lscape} 
\usepackage{hyperref}

\usepackage[authoryear]{natbib}     

\usepackage{multirow} 
\usepackage{setspace}    

\usepackage[T1]{fontenc}
\usepackage[scaled=0.9]{beramono}
\usepackage{tikz}
\usetikzlibrary{decorations.pathreplacing}

\usepackage[margin=1in]{geometry}

\newcommand{\ml}{\mathcal}

\newcommand{\wh}{\widehat}
\newcommand{\wt}{\widetilde}

\newcommand\ds{\displaystyle}

\newcommand{\Bx}{\mathbf{x}}

\newcommand{\BG}{\bm{G}}

\newcommand{\BM}{\bm{M}}
\newcommand{\BO}{\bm{O}}

\newcommand{\BS}{\bm{S}}

\newcommand{\BU}{\bm{U}}
\newcommand{\BV}{\bm{V}}
\newcommand{\BW}{\bm{W}}
\newcommand{\BX}{\bm{X}}

\newcommand{\BZ}{\bm{Z}}

\newcommand{\TX}{\textbf{\texttt{X}}}
\newcommand{\Tx}{\textbf{\texttt{x}}}
\newcommand{\TW}{\textbf{\texttt{W}}}

\newcommand{\mN}{\ml{N}}

\newcommand{\mX}{\ml{X}}
\newcommand{\mZ}{\ml{Z}}

\newcommand{\hBG}{\wh{\bm{G}}}
\newcommand{\hBM}{\wh{\bm{M}}}

\newcommand{\hBS}{\wh\BS}

\newcommand{\heta}{\wh\eta}

\newcommand{\tTX}{\wt{\textbf{\texttt{X}}}}
\newcommand{\ttTX}{\wt{\wt{\textbf{\texttt{X}}}}}

\newcommand{\tBS}{\wt{\BS}}
\newcommand{\ttBS}{\wt{\wt{\BS}}}
\newcommand{\tF}{\wt{F}}

\newcommand{\ttF}{\wt{\wt{F}}}

\newcommand{\tBU}{\wt{\BU}}
\newcommand{\ttBU}{\wt{\wt{\BU}}}

\newcommand{\htBS}{\wh{\wt{\BS}}}

\newcommand{\bo}{\bm{o}}

\newcommand{\bb}{\bm\beta}
\newcommand{\ba}{\bm\alpha}
\newcommand{\bga}{\bm\gamma}

\newcommand{\bLambda}{\bm\Lambda}
\newcommand{\bPsi}{\bm\Psi}

\newcommand{\bDel}{\bm\Delta}

\newcommand{\bPhi}{\bm\Phi}
\newcommand{\bpi}{\bm\pi}
\newcommand{\bSigma}{\bm\Sigma}
\newcommand{\hpi}{\widehat{\pi}}
\newcommand{\hbpi}{\widehat{\bm\pi}}
\newcommand{\bPi}{\bm\Pi}

\newcommand{\hbPi}{\widehat{\bm\Pi}}
\newcommand{\bv}{\bm{v}}

\newcommand{\tbLambda}{\wt\bLambda}
\newcommand{\ttbLambda}{\wt{\wt\bLambda}}

\newcommand{\hbb}{\wh{\bm\beta}}

\newcommand{\hbDel}{\wh{\bm\Delta}}

\newcommand{\hbPhi}{\wh{\bm{\Phi}}}
\newcommand{\hbPsi}{\wh{\bm{\Psi}}}

\DeclareMathOperator{\E}{\mathrm{E}}
\DeclareMathOperator{\var}{\mathrm{Var}}

\def\tT{\texttt{T}}

\def\b0{\bm0}

\def\dovr{\buildrel{d}\over\longrightarrow}
\def\povr{\buildrel{p}\over\longrightarrow}

\def\2ovr{\buildrel{2}\over\longrightarrow}

\definecolor{brown1}{rgb}{1.00,0.25,0.25}
\definecolor{brown2}{rgb}{0.93,0.23,0.23}
\definecolor{brown3}{rgb}{0.80,0.20,0.20}
\definecolor{brown4}{rgb}{0.55,0.14,0.14}
\definecolor{brown}{rgb}{0.65,0.16,0.16}
\definecolor{green1}{rgb}{0.00,1.00,0.00}
\definecolor{green2}{rgb}{0.00,0.93,0.00}
\definecolor{green3}{rgb}{0.00,0.80,0.00}
\definecolor{green4}{rgb}{0.00,0.70,0.00}
\definecolor{steelblue}{rgb}{0.27,0.51,0.71}
\definecolor{DodgerBlue}{rgb}{0.12,0.56,1.00}
\definecolor{RoyalBlue}{rgb}{0.25,0.41,0.88}
\definecolor{lightgrey}{rgb}{0.5,0.5,0.5}
\definecolor{islamicgreen}{rgb}{0,0.56,0}
\definecolor{bubblegum}{rgb}{0.99, 0.76, 0.8}
\definecolor{lavenderpink}{rgb}{0.98, 0.68, 0.82}
\definecolor{lightsalmonpink}{rgb}{1.0, 0.6, 0.6}

\definecolor{babyblue}{rgb}{0.54, 0.81, 0.94}
\definecolor{babyblueeyes}{rgb}{0.63, 0.79, 0.95}
\definecolor{babypink}{rgb}{0.96, 0.76, 0.76}
\definecolor{ballblue}{rgb}{0.13, 0.67, 0.8}
\definecolor{bananamania}{rgb}{0.98, 0.91, 0.71}
\definecolor{beaublue}{rgb}{0.74, 0.83, 0.9}
\definecolor{brightgreen}{rgb}{0.4, 1.0, 0.0}

\definecolor{pastelpink}{rgb}{1.0, 0.82, 0.86}
\definecolor{pink}{rgb}{1.0, 0.75, 0.8}
\definecolor{bittersweet}{rgb}{1.0, 0.44, 0.37}	
\definecolor{ao(english)}{rgb}{0.0, 0.5, 0.0}
\definecolor{amaranth}{rgb}{0.9, 0.17, 0.31}
\definecolor{amber}{rgb}{1.0, 0.75, 0.0}
\definecolor{amber(sae/ece)}{rgb}{1.0, 0.49, 0.0}
\definecolor{black}{rgb}{0.0, 0.0, 0.0}

\newtheorem{theorem}{Theorem}
\newtheorem{lemma}{Lemma}

\begin{document}


\begin{center}
{\Large 
\textbf{Large-sample properties of multiple imputation estimators for parameters of logistic regression with covariates missing at random separately or simultaneously} }
\vskip 2mm
\centerline{Phuoc-Loc Tran$^1$, Shen-Ming Lee$^2$, Truong-Nhat Le$^3$, and Chin-Shang Li$^4$}
\vskip 0mm
$^1$Department of Mathematics, College of Natural Science, Can Tho University, Vietnam
\vskip 0mm
Email: tploc@ctu.edu.vn 
\vskip 0mm
$^2$Department of Statistics, Feng Chia University,	Taiwan, ROC.
\vskip 0mm 
Email: smlee@mail.fcu.edu.tw
\vskip 0mm
$^3$Faculty of Mathematics and Statistics, Ton Duc Thang University, Vietnam
\vskip 0mm
Email: letruongnhat@tdtu.edu.vn
\vskip 0mm
$^4$Division of Supportive Care in Cancer, Department of Surgery, University of Rochester Medical Center,
Rochester, NY, USA. 
\vskip 0mm
Email: csli2003@gmail.com
\end{center}

\date{\today}

\begin{abstract}
	We consider logistic regression including two sets of discrete or categorical covariates that are missing at random (MAR) separately or simultaneously. We examine 
	the asymptotic properties of two multiple imputation (MI) estimators, given in the study of \cite{lee2023estimation}, for the parameters of the logistic regression model  
	with both sets of discrete or categorical covariates that are MAR separately or simultaneously. The proposed estimated asymptotic variances of the two MI estimators address a limitation observed with Rubin's type estimated variances, which lead to underestimate the variances of the two MI estimators \citep{rubin1987statistical}.
	Simulation results demonstrate that our two proposed  MI methods outperform the complete-case, semiparametric inverse probability weighting, 
	random forest MI using chained equations, and stochastic approximation of expectation-maximization methods. 
	To illustrate the methodology's practical application, we provide 
	a real data example from a survey conducted in the Feng Chia night market  in Taichung City, Taiwan.	
\end{abstract}


\section{Introduction} 
\label{Introduction}
Missing data processing is one of the rapidly developing research areas of Applied Statistics and Data Science,  drawing the attention of researchers worldwide. 
Three primary types of missing mechanisms exist:  missing completely at random (MCAR), missing at random (MAR), and missing not at random (MNAR). 
MAR, which implies that missingness is solely related to variables with complete data and unrelated to variables with missing data, has been extensively studied due to its complexity when compared to MCAR; see, e.g., \cite{rubin1976inference,rubin1987statistical} for in-depth discussions.
Researchers in various fields, such as economics, agriculture, medicine, biology, social sciences, and data mining, often encounter challenges related to data processing, which can affect the accuracy of estimates and predictions. Consequently, one of the most critical considerations for improving the rationality and reliability of statistical models is the management of missing values.
 
Numerous studies have explored regression models involving missing data variables, with a particular emphasis on logistic regression, widely applied in various domains.
\cite{wang1997weighted} utilized the inverse probability weighting (IPW), in which the selection probabilities were estimated through kernel smoothers, to investigate the asymptotic properties of estimators of parameters of a regression model with covariates MAR.
 \cite{lukusa2016semiparametric} employed the semiparametric IPW (SIPW) method to estimate parameters in a zero-inflated Poisson (ZIP) regression model, analyzing  motorcycle traffic regulation data in Taiwan. While the SIPW method outperformed the complete-case (CC) method, 
 its primary limitation lay in its reliance on validation or CC data, leading to potential information loss from the data. 
 
 \cite{wang2002joint} delved into semiparametric estimation in logistic regression with covariates MAR, in which the joint conditional likelihood (JCL) approach, which combines
 the conditional likelihoods of validation and non-validation data sets, was used to 
 estimate model parameters and applied to analyze bladder cancer data. 
 Several researchers further refined and extended this technique, yielding promising results. 
 For example, \cite{lee2012semiparametric} employed the JCL method to estimate parameters in logistic regression where both dependent and independent variables were MAR. Recognizing that binary outcomes and covariates in logistic regression were MAR separately or simultaneously, 
  \cite{hsieh2013logistic} extended the JCL method to estimate model parameters and established their asymptotic properties. 
  \cite{tran2023} also applied the JCL method to estimate logistic regression parameters when covariates were MAR separately or simultaneously. 
  Although the results indicated that the JCL method outperformed other methods, it was noted for its complex calculation formulas and time-consuming computations \citep{lee2023estimation}. Additionally, \cite{jiang2020logistic} proposed a stochastic approximation of expectation-maximization (SAEM) 
  method for parameter estimation in logistic regression with missing covariates.

 Furthermore, in an effort to retain valuable information rather than discarding rows with missing data,
 \cite{rubin1976inference} introduced imputation, a method in  which missing data are replaced with plausible values to create an imputed (``completed'') data set for analysis. 
 This concept was subsequently expanded into the multiple imputation (MI) method, 
  which involves iterating the imputation process $M$ times to generate $M$ imputed (``completed'') data sets.
 This approach enhances the quality of the analysis by averaging results across these data sets. 
 However, determining the optimal imputation method remains challenging, as each technique can yield different analytical outcomes based on the method used for value replacement. 
 
 To address this, statisticians have continuously striven to develop and refine imputation methods.
 \cite{buuren2011mice} consolidated several popular MI methods into the well-known \texttt{mice} \textsf{R} package, implementing the multivariate imputation using chained equations (MICE) algorithm. \cite{austin2022effect}  utilized the predictive mean matching method from the \texttt{mice} package to investigate the performance of MI  in handling high prevalence of missing data in logistic regression. 
  \cite{jiang2020logistic} compared their SAEM method, implemented in the \texttt{misaem} package in \textsf{R}, with the random forest multiple imputation (RFMI) method from the \texttt{mice} package in \textsf{R} to evaluate their relative effectiveness. 
 
 In an alternative approach, \cite{lee2016estimation,lee2020estimation} amalgamated ideas from \cite{rubin1976inference}, \cite{fay1996alternative}, and \cite{wang2009empirical} 
 to {extend MI methods for a closed capture–recapture model and a ZIP regression model, both involving covariates MAR. \cite{lee2023estimation} employed  these approaches to propose two distinct MI methods, MI1 and MI2, for logistic regression with covariates MAR separately or simultaneously. 
 These two MI methods outperformed others and exhibited shorter computing times.
 However, it is noted that the estimated variances of the parameter estimators, obtained by using Rubin's type variance estimation  method, tended to be underestimated. See, 
 e.g.,  \cite{rubin1986multiple}, \cite{rubin1987statistical}, \cite{righi2014methods}, and \cite{lee2022goodness}. 
 Moreover, the large-sample properties of these two MI estimators have not been provided. 
 Consequently, given the aforementioned challenges and the widespread utility of MI methods in data analysis, our motivation lies in establishing the large-sample properties of these two MI estimators and presenting explicit variance estimation formulas to address the underestimation observed in \cite{lee2023estimation}.
 Additionally, we aim to investigate the computational efficiency and practical application of the proposed methods through numerical simulations and a real data set, comparing our proposed variance estimation methods for the variances of these two MI estimators with benchmark methods such as CC, SIPW, MI1,  MI2, RFMI, and SAEM.

In Section~\ref{Model and Methods}, we introduce the logistic regression model, along with its notations and underlying assumptions. We also review the estimation methods used in this study, including CC, SIPW, MI1, and MI2. 
In Section~\ref{Asymptotic}, we present the asymptotic results for the proposed methods, denoted as MI1n and MI2n, derived from their counterparts MI1 and MI2. This section constitutes the core results of this work.
 We evaluate the finite-sample performance of these methods through a comprehensive simulation analysis and apply the suggested estimation techniques to an actual data set 
in Section~\ref{SimulationAndRealData}. Section~\ref{Conclusion} offers some discussions and conclusions. The proofs of the main results presented in Section~\ref{Asymptotic} can be found in the Appendix.

\section{Model and Methods} \label{Model and Methods}
\subsection{Logistic Regression Model with Covariates Missing Values}
This work considers the following logistic regression model: 
\begin{align}\label{def: logistic regression model}
 	P(Y=1|\BX,\BZ)
        =H(\beta_0+\bb_1^{\tT}\TX_1+\bb_2^{\tT}\TX_2+\bb_3^{\tT}\BZ)
     =H(\bb^{\tT}\bm\mX).	
\end{align}
Here the dependent variable $Y$ is a dichotomy with $Y=1$ indicating the occurrence of an event of interest and $Y=0$ otherwise. $\bb=(\beta_0,\bb_1^{\tT},\bb_2^{\tT},\bb_3^{\tT})^{\tT}$ is a vector of the regression model parameters.
Two discrete covariate vectors, $\TX_1=(X_1,X_2,\dots,X_{s})^{\tT}$ and $\TX_2=(X_{s+1},X_{s+2},\dots,X_p)^{\tT}$, are assumed missing at random separately or simultaneously. 
Another discrete covariates $\BZ=(Z_1,Z_2,\dots,Z_q)^{\tT}$ are always observed. 
$\bm\mX=(1,\BX^{\tT},\BZ^{\tT})^{\tT}$ for $\BX=(\TX_1^{\tT},\TX_2^{\tT})^{\tT}$.
$H(u)=\{1+\exp(-u)\}^{-1}$.	
Let $\{(Y_i,\bm\mX_i): i=1,2,\dots,n\}$ be a random sample, where $\bm\mX_i=(1,\BX_i^{\tT},\BZ_i^{\tT})^{\tT}$ and $\BX_i=(\TX_{1i}^{\tT},\TX_{2i}^{\tT})^{\tT}$.

To identify the missingness statuses of $\BX_i=(\TX_{1i}^{\tT},\TX_{2i}^{\tT})^{\tT}$, define $\delta_{ij}$, $j=1,2,3,4$, where $\delta_{i1}=1$ if both $\TX_{1i}$ and $\TX_{2i}$ are
observed; 0 otherwise; $\delta_{i2}=1$ if $\TX_{1i}$ is missing and $\TX_{2i}$ is observed; 0 otherwise; $\delta_{i3}=1$ if 
$\TX_{1i}$ is observed and $\TX_{2i}$ is missing; 0 otherwise; $\delta_{i4}=1$ if both $\TX_{1i}$ and $\TX_{2i}$ are missing; 0 otherwise. 
In some studies, the variables with missing data can be easier guessed or predicted from other observed variables, called surrogate variables, 
which are related to them and unrelated to the dependent variable. Therefore, we also consider that $\TW_1$ and $\TW_2$ are two discrete surrogate variable vectors for 
$\TX_1$ and $\TX_2$, such that $\TW_1$ and $\TW_2$ are 
dependent on $\TX_1$ and $\TX_2$, respectively, and independent of $Y$ given $\BX$ and $\BZ$. More specifically, the model in Equation~\eqref{def: logistic regression model} can be written as $P(Y=1|\TX_1,\TX_2,\BZ,\TW_1,\TW_2)=P(Y=1|\TX_1,\TX_2,\BZ)
=H(\beta_0+\bb_1^{\tT}\TX_1+\bb_2^{\tT}\TX_2+\bb_3^{\tT}\BZ)$.
For more details, refer to, e.g., \cite{wang1997weighted,wang2002joint}, \cite{hsieh2010logistic,hsieh2013logistic},
and \cite{lee2011semiparametric,lee2012semiparametric,lee2020estimation,lee2023estimation}. 

Let $\BW_i=(\TW_{1i}^{\tT},\TW_{2i}^{\tT})^{\tT}$ and $\BV_i=(\BZ_i^{\tT},\BW_i^{\tT})^{\tT}$. 
Based on the assumption of MAR mechanism of $\TX_1$ and $\TX_2$ \citep{rubin1976inference},
the selection probabilities are given by
\begin{align} \label{def: pi}
	P(\delta_{ij}=1|Y_i,\TX_{1i},\TX_{2i},\BZ_i,\BW_i)=\pi_j(Y_i,\BZ_i,\BW_i)
  =\pi_j(Y_i,\BV_i), \ j=1,2,3,4.
\end{align}
Here $\pi_j(Y_i,\BV_i)$, $j=1,2,3,4$, are the unknown nuisance parameters and need to be estimated, and $\sum_{j=1}^{4}\pi_j(Y_i,\BV_i)=1$.

\subsection{Review of SIPW Estimation Methods}
There are various estimation methods that have been employed to estimate the parameters of the model in Equation~\eqref{def: logistic regression model}.
The simplest of these methods is the CC approach, which utilizes only the CC data ($\delta_{i1}=1$) to construct the following estimating equations: 
\begin{align} \label{eq: score cc}
 \BU_C(\bb)
 =\frac{1}{\sqrt{n}}\sum_{i=1}^{n}\delta_{i1}\bm\mX_i\left[Y_i-H(\bb^{\tT}\bm\mX_i)\right]=\b0.
\end{align}
The CC estimator $\hbb_C$ of $\bb$ is the solution to the estimating equations in Equation~\eqref{eq: score cc}.
 When the data are MCAR and sample size is large, $\hbb_C$ remains an asymptotically unbiased estimator of $\bb$. 
 However, due to the non-MCAR nature of $\BX$, CC estimation may compromise validity and introduce a higher level of bias \citep{wang1997weighted}. 
 To address these limitations and enhance accuracy, certain studies have implemented appropriate data processing techniques.  
 In this section, we exclusively employ the SIPW and MI estimation methods for parameter estimation.   

 As demonstrated in \cite{lee2023estimation},  assuming that $\BV_i$s are discrete variable vectors,
 	the nonparametric estimators of $\pi_j(Y_i,\BV_i)$ are given as follows:
	\begin{align} \label{def: pihat}
		\hpi_j(Y_i,\BV_i)=\frac{\sum_{k=1}^n\delta_{kj}I(Y_k=Y_i,\BV_k=\BV_i)}{\sum_{s=1}^nI(Y_s=Y_i,\BV_s=\BV_i)},\  j=1,2,3,4,
	\end{align}
where $I(\cdot)$ represents an indicator function.
The SIPW method introduces a weighted inverse term that is a nonparametric estimator of the selection probability given in Equation~\eqref{def: pi}. This results in the following set of estimating equations:  
\begin{align} \label{eq: score sipw}
 \BU_W(\bb,\hbpi_1)
=\frac{1}{\sqrt{n}}\sum_{i=1}^{n}\frac{\delta_{i1}}{\hpi_1(Y_i,\BV_i)}\bm\mX_i\left[Y_i-H(\bb^{\tT}\bm\mX_i)\right]=\b0,
\end{align}
where $\hbpi_1=\left(\hpi_{11},\hpi_{12},\dots,\hpi_{1n}\right)$ for $\hpi_{1i}=\hpi_1(Y_i,\BV_i)$, given in Equation~\eqref{def: pihat}, serving as an estimator of $\pi_{1i}=\pi_1(Y_i,\BV_i)$ defined in  Equation~\eqref{def: pi}.
For a more comprehensive understanding of IPW and SIPW methods, refer to, e.g., \cite{horvitz1952generalization}, \cite{zhao1992designs}, \cite{wang1997weighted} and \cite{wang2001note}, 
\cite{hsieh2010logistic}, \cite{lee2012semiparametric,lee2023estimation}, and \cite{tran2023}.

\subsection{Review of MI Estimation Method}
The MI method, introduced by \cite{rubin1978multiple,rubin1987statistical, rubin1996multiple}, offers a widely-used approach for addressing missing data. The fundamental concept behind MI involves replacing
missing values in an origin missing data set $M$ times with reasonable values obtained from specified methods, resulting in $M$ imputed (``completed'') data sets.
These imputed (``completed'') data sets are then analyzed by using chosen statistical methods, 
and the results are subsequently combined to yield the final results. 
Various extensions of Rubin's approach exist, with variations in the methods employed for imputation. For example, missing values can be replaced by  mean, median, predicted values from regression models,
random forests, etc. Relevant \textsf{R} packages such as \texttt{mice} \citep{buuren2011mice} and \texttt{misssMDA}  \citep{josse2016missmda} provide tools for implementing these techniques.

In their work, \cite{lee2023estimation} drew inspiration from the ideas presented by \cite{wang2009empirical} and \cite{fay1996alternative} to develop two distinct MI methods for estimating the parameters of logistic regression with covariates missing separately or simultaneously. 
Their approaches comprise two steps: (1) Imputation, which utilizes the empirical cumulative distribution functions (CDFs) to impute missing data  based on individual or all missing covariates, and 
(2) Analysis, where they solve the estimating equations that include an average term of $M$ functions to yield parameter estimates. A brief summary of these methods is provided below.

Define the CDFs of $\TX_{1i}$  given $(\TX_{2i},Y_i,\BV_i)$, $\TX_{2i}$ given $(\TX_{1i},Y_i,\BV_i)$, and $\BX_i=(\TX_{1i}^{\tT},\TX_{2i}^{\tT})^{\tT}$ 
given $(Y_i,\BV_i)$ are $F_{\TX_{1i}}(\Tx_1|\TX_{2i},Y_i,\BV_i)$, $F_{\TX_{2i}}(\Tx_2|\TX_{1i},Y_i,\BV_i)$, and $F_{\BX_i}(\Bx|Y_i,\BV_i)$, respectively, where  $\Bx=(\Tx_1^{\tT},\Tx_2^{\tT})^{\tT}$. 
\cite{lee2023estimation} considered the corresponding empirical CDFs to provide the MI1 method
\begin{align}
 \label{eq:tF3}	
\left.
\begin{array}{l}
	\tF_{\TX_{1i}}(\Tx_1|\TX_{2i},Y_i,\BV_i)
	=\ds\sum_{k=1}^{n}
	\left(\dfrac{\delta_{k1}I(Y_k=Y_i,\TX_{2k}=\TX_{2i},\BV_k=\BV_i)}
	{\sum_{s=1}^{n}\delta_{s1}I(Y_s=Y_i,\TX_{2s}=\TX_{2i},\BV_s=\BV_i)}\right) I(\TX_{1k}\le\Tx_1), \\ \\[-4mm] 
	\tF_{\TX_{2i}}(\Tx_2|\TX_{1i},Y_i,\BV_i)
	=\ds\sum_{k=1}^{n}
    \left(\dfrac{\delta_{k1}I(Y_k=Y_i,\TX_{1k}=\TX_{1i},\BV_k=\BV_i)}
	{\sum_{s=1}^{n}\delta_{s1}I(Y_s=Y_i,\TX_{1s}=\TX_{1i},\BV_s=\BV_i)}\right)
	I(\TX_{2k}\le\Tx_2), \\ \\[-4mm]
	\tF_{\BX_i}(\Bx|Y_i,\BV_i)
	=\ds\sum_{k=1}^{n}
	\left(\dfrac{\delta_{k1}I(Y_k=Y_i,\BV_k=\BV_i)}{\sum_{s=1}^{n}\delta_{s1}I(Y_s=Y_i,\BV_s=\BV_i)}\right)
	I(\BX_k\le\Bx),
\end{array}
\right.
\end{align}
and the following empirical CDFs for the MI2 method
\begin{align}
\label{eq:ttF3}	
	\left.
\begin{array}{l}
	\ttF_{\TX_{1i}}(\Tx_1|Y_i,\BV_i)
	=\ds\sum_{k=1}^{n}\left(\dfrac{(\delta_{k1}+\delta_{k3})I(Y_k=Y_i,\BV_k=\BV_i)}
	{\sum_{s=1}^{n}(\delta_{s1}+\delta_{s3})I(Y_s=Y_i,\BV_s=\BV_i)}\right)I(\TX_{1k}\le\Tx_1), \\ \\ [-4mm]
	\ttF_{\TX_{2i}}(\Tx_2|Y_i,\BV_i)
	=\ds\sum_{k=1}^{n}\left(\dfrac{(\delta_{k1}+\delta_{k2})I(Y_k=Y_i,\BV_k=\BV_i)}
	{\sum_{s=1}^{n}(\delta_{s1}+\delta_{s2})I(Y_s=Y_i,\BV_s=\BV_i)}\right)I(\TX_{2k}\le\Tx_2), \\ \\ [-4mm]
	\ttF_{\BX_i}(\Bx|Y_i,\BV_i)
	=\ds\sum_{k=1}^{n}\left(\dfrac{\delta_{k1}I(Y_k=Y_i,\BV_k=\BV_i)}{\sum_{s=1}^{n}\delta_{s1}I(Y_s=Y_i,\BV_s=\BV_i)}\right)
	I(\BX_k\le\Bx). 
\end{array}
\right.
\end{align}

Let $M$ be the number of imputations. The algorithms of the MI1 and MI2 methods are outlined as follows: 
\begin{description}
	\item[{\bf Step 1.}] \textbf{\textit{Imputation:}} For $v=1,2,\dots,M$, replace the missing values of $\BX_i=(\TX_{1i}^{\tT},\TX_{2i}^{\tT})^{\tT}$, $i=1,2,\dots,n$ 
	to create $M$ imputed (``completed'') data sets.  This process is performed as follows:
	\begin{itemize}
		\item When $\delta_{i1}=1$, retain the values of $\TX_{1i}$ and $\TX_{2i}$ and define $\bm\mX_i=(1,\TX_{1i}^{\tT},\TX_{2i}^{\tT},\BZ_i^{\tT})^{\tT}$ for all $v$.
		\item When $\delta_{i2}=1$, retain the values of $\TX_{2i}$ and apply the following steps for both MI1 and MI2 methods:
		\begin{itemize}
			\item[$\circ$] MI1: Generate $\tTX_{1iv}$ from $\tF_{\TX_{1i}}(\Tx_1|\TX_{2i},Y_i,\BV_i)$ to replace the missing values of $\TX_{1i}$, 
			  and define $\wt{\bm\mX}_{2iv}=(1,\tTX^{\tT}_{1iv},\TX_{2i}^{\tT},\BZ^{\tT}_i)^{\tT}$,
			\item[$\circ$] MI2: Generate $\ttTX_{1iv}$ from $\ttF_{\TX_{1i}}(\Tx_1|Y_i,\BV_i)$ to replace the missing values of $\TX_{1i}$, 
			  and define $\wt{\wt{\bm\mX}}_{2iv}=(1,\ttTX^{\tT}_{1iv},\TX_{2i}^{\tT},\BZ^{\tT}_i)^{\tT}$.
		\end{itemize}		 
		\item When $\delta_{i3}=1$, retain the values of $\TX_{1i}$ and  apply the following steps for both MI1 and MI2 methods:
		\begin{itemize}
		   \item[$\circ$] MI1: Generate $\tTX_{2iv}$ from $\tF_{\TX_{2i}}(\Tx_2|\TX_{1i},Y_i,\BV_i)$ to replace the missing values of $\TX_{2i}$, 
		     and define $\wt{\bm\mX}_{3iv}=(1,\TX_{1i}^{\tT},\tTX^{\tT}_{2iv},\BZ^{\tT}_i)^{\tT}$,
		   \item[$\circ$] MI2: Generate $\ttTX_{2iv}$ from $\ttF_{\TX_{2i}}(\Tx_2|Y_i,\BV_i)$ to replace the missing values of $\TX_{2i}$, 
		    and define $\wt{\wt{\bm\mX}}_{3iv}=(1,\TX_{1i}^{\tT},\ttTX^{\tT}_{2iv},\BZ^{\tT}_i)^{\tT}$.
		\end{itemize}
		\item When $\delta_{i4}=1$,  apply the following steps for both MI1 and MI2 methods:
		\begin{itemize}
			\item[$\circ$] MI1: Generate $\tTX_{1iv}$ and $\tTX_{2iv}$
			from $\tF_{\BX_i}(\Bx|Y_i,\BV_i)$  to replace the missing values of $\TX_{1i}$ and $\TX_{2i}$, and define
			$\wt{\bm\mX}_{4iv}=(1,\tTX^{\tT}_{1iv},\tTX^{\tT}_{2iv},\BZ^{\tT}_i)^{\tT}$,
				\item[$\circ$] MI2: Generate $\ttTX_{1iv}$ and $\ttTX_{2iv}$
			from $\ttF_{\BX_i}(\Bx|Y_i,\BV_i)$  to replace the missing values of $\TX_{1i}$ and $\TX_{2i}$, and define
			$\wt{\wt{\bm\mX}}_{4iv}=(1,\ttTX^{\tT}_{1iv},\ttTX^{\tT}_{2iv},\BZ^{\tT}_i)^{\tT}$.
		\end{itemize}
	\end{itemize}
	\item[{\bf Step 2.}] \textbf{\textit{Analysis:}} Obtain the value of the MI1 estimator, $\hbb_{M1}$, and that of the MI2 estimator, $\hbb_{M2}$, of $\bb$ 
	 by solving the estimating equations in Equations~\eqref{eq: score mi1} and \eqref{eq: score mi2}, respectively, as follows:
	\begin{align} 
		\BU_{M1}(\bb)
		&=\frac{1}{\sqrt{n}}
		\sum_{i=1}^{n}\left(\delta_{i1}\BS_i(\bb)+ \sum_{k=2}^{4}\delta_{ik}\tBS_{ki}(\bb)\right)
		=\b0, \label{eq: score mi1} \\
		\BU_{M2}(\bb)
		&=\frac{1}{\sqrt{n}}\sum_{i=1}^{n}
		\left(\delta_{i1}\BS_i(\bb)+ \sum_{k=2}^{4}\delta_{ik}\ttBS_{ki}(\bb) \right)
		=\b0, \label{eq: score mi2}
	\end{align}
where
\begin{align}
	\label{eq:BSs}	
	\left.
	\begin{array}{l}
	\BS_i(\bb)=\bm\mX_i\left(Y_i-H(\bb^{\tT}\bm\mX_i)\right), \\ \\[-4mm]
	\tBS_{ki}(\bb)=\dfrac{1}{M}\sum_{v=1}^{M}\wt{\bm\mX}_{kiv}\left(Y_i-H(\bb^{\tT}\wt{\bm\mX}_{kiv})\right) 
	= \dfrac{1}{M}\sum_{v=1}^M\tBS_{kiv}(\bb),	\\ \\[-4mm]
	\ttBS_{ki}(\bb)=\dfrac{1}{M}\sum_{v=1}^{M}\wt{\wt{\bm\mX}}_{kiv}\Big(Y_i-H(\bb^{\tT}\wt{\wt{\bm\mX}}_{kiv})\Big)
	= \dfrac{1}{M}\sum_{v=1}^M\ttBS_{kiv}(\bb),
   \end{array}
    \right. 
\end{align}	
$k=2,3,4$,	
for $\tBS_{kiv}(\bb)=\wt{\bm\mX}_{kiv}(Y_i-H(\bb^{\tT}\wt{\bm\mX}_{kiv}))$ and 
$\ttBS_{kiv}(\bb)=\wt{\wt{\bm\mX}}_{kiv}(Y_i-H(\bb^{\tT}\wt{\wt{\bm\mX}}_{kiv}))$.	
	
 The Rubin's type estimated variances \citep{rubin1987statistical} of $\hbb_{M1}$ and $\hbb_{M2}$, denoted as  $\wh\var(\hbb_{M1})$ and 
 $\wh\var(\hbb_{M2})$, are computed by using the formulas in Equations~\eqref{MI1: asymptotic var} and 
 \eqref{MI2: asymptotic var}, respectively, as follows: 
	\begin{align}
 \frac{1}{n}\BG_{M1}^{-1}(\hbb_{M1})\left\{\frac{1}{M}\sum_{v=1}^{M}\sum_{i=1}^{n}\tBU_{vi}^{\otimes2}(\hbb_{M1})
  +\left(1+\dfrac{1}{M}\right)\frac{\sum_{v=1}^{M}\tBU_v^{\otimes2}(\hbb_{M1})}{M-1}\right\}[\BG_{M1}^{-1}(\hbb_{M1})]^{\tT},  \label{MI1: asymptotic var}\\
  \frac{1}{n}\BG_{M2}^{-1}(\hbb_{M2})\left\{\frac{1}{M}\sum_{v=1}^{M}\sum_{i=1}^{n}\ttBU_{vi}^{\otimes2}(\hbb_{M2})
 +\left(1+\frac{1}{M}\right)\dfrac{\sum_{v=1}^{M}\ttBU_v^{\otimes2}(\hbb_{M2})}{M-1}\right\}[\BG_{M2}^{-1}(\hbb_{M2})]^{\tT}, \label{MI2: asymptotic var}
	\end{align}
where $\bm{a}^{\otimes2}=\bm{a}\bm{a}^{\tT}$ for $\bm{a}$ being a column vector.
 $\BG_{M1}(\bb)$ and $\BG_{M2}(\bb)$ represent the gradients of $-n^{-1/2}M^{-1}\sum_{v=1}^{M}\tBU_v(\bb)=- n^{-1/2}\BU_{M1}(\bb)$ and $-n^{-1/2}M^{-1}\sum_{v=1}^{M}\ttBU_v(\bb)=-n^{-1/2}\BU_{M2}(\bb)$, respectively. The expressions for $\tBU_{vi}(\bb)$, $\tBU_{v}(\bb)$, $\ttBU_{vi}(\bb)$, and $\ttBU_{v}(\bb)$ are defined as follows:
    \begin{align*}
	\tBU_{vi}(\bb)
 	 &=\dfrac{1}{\sqrt{n}}\Big(\delta_{i1}\BS_i(\bb)+\sum_{k=2}^{4}\delta_{ik}\tBS_{kiv}(\bb)\Big),\quad 
	\tBU_{v}(\bb)
	=\sum_{i=1}^{n}\tBU_{vi}(\bb),  \\
	\ttBU_{vi}(\bb)
 	 &=\dfrac{1}{\sqrt{n}}\Big(\delta_{i1}\BS_i(\bb)+\sum_{k=2}^{4}\delta_{ik}\ttBS_{kiv}(\bb)\Big),\quad 
	\ttBU_{v}(\bb)
 	 =\sum_{i=1}^{n}\ttBU_{vi}(\bb).
  \end{align*}
\end{description}

\section{Asymptotic Results} \label{Asymptotic}
This section discusses the asymptotic properties of $\hbb_{M1}$ and $\hbb_{M2}$ under the assumptions that $\TX_1$ and $\TX_2$ 
are MAR, and $\TX_1$, $\TX_2$, $\BW$, and $\BZ$ are discrete. To begin, define the following quantities:
	\begin{align} 
	\BU_{M1}(\bb)
	&=\frac{1}{\sqrt{n}}
	\sum_{i=1}^{n}\Big(\delta_{i1}\BS_i(\bb)+\sum_{k=2}^{4}\delta_{ik}\tBS_{ki}(\bb)\Big)
	= \frac{1}{\sqrt{n}}
	\sum_{i=1}^{n}\tbLambda_i(\bb), \label{eq: score mi1 new} \\
	\BU_{M2}(\bb)
	&=\frac{1}{\sqrt{n}}\sum_{i=1}^{n}
	\Big(\delta_{i1}\BS_i(\bb)+\sum_{k=2}^{4}\delta_{ik}\ttBS_{ki}(\bb)\Big)
	=\frac{1}{\sqrt{n}}
	\sum_{i=1}^{n}\ttbLambda_i(\bb), \label{eq: score mi2 new}
\end{align}
where $\tbLambda_i(\bb)$ and $\ttbLambda_i(\bb)$ denote
\begin{align}
	\tbLambda_i(\bb)=\delta_{i1}\BS_i(\bb) + \sum_{k=2}^{4}\delta_{ik}\tBS_{ki}(\bb), 	\label{eq:tbLambda.mi1} \\
	\ttbLambda_i(\bb)=\delta_{i1}\BS_i(\bb)+\sum_{k=2}^{4}\delta_{ik}\ttBS_{ki}(\bb).	\label{eq:ttbLambda.mi2} 
\end{align}
And further introduce
 \begin{align} 
	\BU_1(\bb,\bPi)
	&= \dfrac{1}{\sqrt{n}}
	\sum_{i=1}^{n}\bPhi_i(\bb,\bpi_i), \label{eq: U1 MI1} \\
	\BU_2(\bb,\bPi)
	&=\dfrac{1}{\sqrt{n}}\sum_{i=1}^{n} \bPsi_i(\bb,\bpi_i), \label{eq: U2 MI2}
\end{align}
where $\bPhi_i(\bb,\bpi_i)$ and $\bPsi_i(\bb,\bpi_i)$ represent
\begin{align}
   \bPhi_i(\bb,\bpi_i)
  &=\frac{\delta_{i1}\BS_i(\bb)}{\pi_1(Y_i,\BV_i)}
  + \sum_{k=2}^{4}\BS^*_{ki}(\bb)\left(\delta_{ik}-\frac{\delta_{i1}\pi_k(Y_i,\BV_i)}{\pi_1(Y_i,\BV_i)}\right),	\label{eq:bPhi.mi1} \\
	\bPsi_i(\bb,\bpi_i)
 &=\delta_{i1}\BS_i(\bb)+\sum_{k=2}^{4}\delta_{ik}\BS^*_i(\bb)+\left(\BS_i(\bb)-\BS_i^*(\bb)\right)\eta(Y_i,\BV_i).	\label{eq:bPsi.mi2}
\end{align}
Here 
\begin{align}
	\label{eq:BS*s}	
	\left.
	\begin{array}{l}
	\BS_{2i}^*(\bb)=\E(\BS_1(\bb)|Y_i,\TX_{i2},\BV_i), \\	
	\BS_{3i}^*(\bb)=\E(\BS_1(\bb)|Y_i,\TX_{i1},\BV_i), \\
	\BS_{4i}^*(\bb)=\E(\BS_1(\bb)|Y_i,\BV_i)=\BS_i^*(\bb),
    \end{array}
\right.
\end{align}			
$\bpi=(\pi_1,\pi_2,\pi_3,\pi_4)$, $\bpi_i=(\pi_{1i},\pi_{2i},\pi_{3i},\pi_{4i})$,
$\pi_{ki}=\pi_k(Y_i,\BV_i)$, $k=1,2,3,4$, {$\bPi=\{\bpi_i:i=1,2,\dots,n\}$, and 
\begin{align*}
  \eta(Y_i,\BV_i)=\frac{(\delta_{i1}+\delta_{i3})\pi_2(Y_i,\BV_i)}{\pi_1(Y_i,\BV_i)+
\pi_3(Y_i,\BV_i)}+\frac{(\delta_{i1}+\delta_{i2})\pi_3(Y_i,\BV_i)}{\pi_1(Y_i,\BV_i)+\pi_2(Y_i,\BV_i)}+\frac{\delta_{i1}\pi_4(Y_i,\BV_i)}{\pi_{1}(Y_i,\BV_i)}.
\end{align*}
It can be shown that $\E[\BU_{Mr}(\bb)]\to\b0$, as $M,n\to\infty$, $r=1,2$, which implies that 
the estimating function $\BU_{Mr}(\bb)$ is  an asymptotically unbiased estimating function of $\bb$. 
See the proof in the Appendix. The simplified expression for the conditional expectation of $\BS_1(\bb)$ given $(Y_i,\BV_i)$, denoted as
$\BS_i^*(\bb)=\E(\BS_1(\bb)|Y_i,\BV_i)=\BS_{4i}^*(\bb)$, is particularly useful in the context of the MI2 method.

Secondly, the following regularity conditions are required:
\begin{description}
	\item [(C1)]
	Let $\text{supp}(\BV)$ denote the support of $\BV$. Assume that
	$\text{supp}(\BV)$ is independent of $\bb$. Furthermore, for
	any $y = 0,1$ and $\bv\in\text{supp}(\BV)$, ensure that the selection
	probabilities $\pi_j(y,\bv)>0$, $j=1,2,3,4$, and $\sum_{j=1}^{4}\pi_j(y,\bv)=1$.
	\item [(C2)] 
	$\E\left[\bPhi_1^{\otimes2}(\bb,\bpi)\right]$ and $\E\left[\bPsi_1^{\otimes2}(\bb,\bpi)\right]$
	are positive definite within a neighborhood of the true $\bb$.
	\item [(C3)]
	The first derivatives of $\wt\bLambda_1(\bb)$ and $\wt{\wt\bLambda}_1(\bb)$ in Equations~\eqref{eq:tbLambda.mi1} and \eqref{eq:ttbLambda.mi2}, respectively,
	with respect to $\bb$ exist
	almost surely within a neighborhood of the true $\bb$. Additionally, in
	this neighborhood, these derivatives are bounded above by a function of $(Y_1,\BX_1,\BV_1)$, and the expectation of this function exists.
\end{description}

\begin{lemma} \label{lemma}
 Under conditions (C1)-(C3), the following statements hold:
	\begin{description}
		\item [(i)] $\BU_{M1}(\bb)-\BU_{1}(\bb,\bPi)=\bo_p(1)+\BO_p(M^{-1/2})\  \text{as}\ M,n \to\infty$.
		\item [(ii)] $\BU_{M2}(\bb)-\BU_{2}(\bb,\bPi)=\bo_p(1)+\BO_p(M^{-1/2})\ \text{as}\ M,n \to\infty$.
	\end{description}
Here $\bo_p(a_n)$ and $\BO_p(a_n)$ denote a column vector with elements that are uniformly 
	$o_p(a_n)$ and $O_p(a_n)$, respectively.
\end{lemma}
\begin{theorem} \label{theorem 1}
  Under conditions (C1)-(C3), $\hbb_{M1}$ is a consistent estimator of $\bb$. Furthermore, $\sqrt{n}(\hbb_{M1}-\bb)$ asymptotically follows a normal distribution with a  mean of $\b0$ 
  and a covariance matrix $\bDel_{M1}$ as $ M,n\to\infty$. Here $\bDel_{M1}=\BG^{-1}(\bb)\BM_1(\bb,\bpi_1)[\BG^{-1}(\bb)]^{\tT}$ with $\BM_1(\bb,\bpi_1)=\E[\bPhi_1^{\otimes2}(\bb,\bpi_1)]$ 
  and $\BG(\bb)=\E\left[\bm\mX_1^{\otimes2}H^{(1)}(\bb^{\tT}\bm\mX_1)\right]$.	
\end{theorem}

\begin{theorem} \label{theorem 2}
	Under conditions (C1)-(C3), $\hbb_{M2}$ is a consistent estimator of $\bb$. Moreover, $\sqrt{n}(\hbb_{M2}-\bb)$ asymptotically follows a normal distribution with a mean of $\b0$ 
	and a covariance matrix $\bDel_{M2}$ as $ M,n\to\infty$. Here $\bDel_{M2}=\BG^{-1}(\bb)\BM_2(\bb,\bpi_1)[\BG^{-1}(\bb)]^{\tT}$ with $\BM_2(\bb,\bpi_1)=\E\left[\bPsi_1^{\otimes2}(\bb,\bpi_1)\right]$
   and $\BG(\bb)=\E\left[\bm\mX_1^{\otimes2}H^{(1)}(\bb^{\tT}\bm\mX_1)\right]$.	
\end{theorem}

In practice, it is required to have consistent estimators of $\bDel_{M1}$ and $\bDel_{M2}$. Let
\begin{align}
  \hbDel_{M1}
 &=\BG_{M1}^{-1}(\hbb_{M1})\BM_1(\hbb_{M1},\hbPi)[\BG_{M1}^{-1}(\hbb_{M1})]^{\tT},
 \label{MI1n: asymptotic var}  \\
  \hbDel_{M2}
&=\BG_{M2}^{-1}(\hbb_{M2})\BM_2(\hbb_{M2},\hbPi)[\BG_{M2}^{-1}(\hbb_{M2})]^{\tT} 
 \label{MI2n: asymptotic var}  
\end{align}	
be consistent estimators of $\bDel_{M1}$ and $\bDel_{M2}$, respectively. Here
$\hbPi=\{\hbpi_i:i=1,2,\dots,n\}$, $\BG_{Mr}(\hbb_{Mr})$, and $\BM_r(\hbb_{Mr},\hbPi)$, $r=1,2$, 
be estimators of $\BG(\bb)$ and $\BM_r(\bb,\bpi)$, respectively, as follows: 
\begin{align*}
\BG_{M1}(\hbb_{M1})
&=\frac{1}{n}\sum_{i=1}^{n}\left\{\delta_{i1}\bm\mX_i^{\otimes2}H^{(1)}(\hbb_{M1}^{\tT}\bm\mX_i)+\frac{1}{M}\sum_{k=2}^{4}\sum_{v=1}^{M}\delta_{ik}\wt{\bm\mX}_{kiv}^{\otimes2}H^{(1)}(\hbb_{M1}^{\tT}\wt{\bm\mX}_{kiv})\right\}, \\
\BG_{M2}(\hbb_{M2})
&=\frac{1}{n}\sum_{i=1}^{n}\left\{\delta_{i1}\bm\mX_i^{\otimes2}H^{(1)}(\hbb_{M2}^{\tT}\bm\mX_i)+\frac{1}{M}\sum_{k=2}^{4}\sum_{v=1}^{M}\delta_{ik}\wt{\wt{\bm\mX}}_{kiv}^{\otimes2}H^{(1)}(\hbb_{M2}^{\tT}\wt{\wt{\bm\mX}}_{kiv}) \right\}, \\
\hBM_1(\hbb_{M1},\hbPi)
&=\frac{1}{n}\sum_{i=1}^{n}\hbPhi_i^{\otimes2}(\hbb_{M1},\hbpi_i), \nonumber  \\
\hBM_2(\hbb_{M2},\hbPi)
&=\frac{1}{n}\sum_{i=1}^{n}\hbPsi_i^{\otimes2}(\hbb_{M2},\hbpi_i), \nonumber 
\end{align*}
where
\begin{align*}
\hbPhi_i(\hbb_{M1},\hbpi_i)
&=\frac{\delta_{i1}\BS_i(\hbb_{M1})}{\hpi_1(Y_i,\BV_i)}
+
\sum_{k=2}^{4}\hBS^*_{ki}(\hbb_{M1})\left(\delta_{ik}-\frac{\delta_{i1}\hpi_k(Y_i,\BV_i)}{\hpi_1(Y_i,\BV_i)}\right), \\
\hbPsi_i(\hbb_{M2},\hbpi_i)
&=\delta_{i1}\BS_i(\hbb_{M2})+ \sum_{k=2}^{4}\delta_{ik}\hBS^*_i(\hbb_{M2}) 
+ \left(\htBS_i(\hbb_{M2})-\hBS_i^*(\hbb_{M2})\right)\heta(Y_i,\BV_i),
\end{align*}
and
\begin{align*}
	\heta(Y_i,\BV_i)
	&=\frac{(\delta_{i1}+\delta_{i3})\hpi_2(Y_i,\BV_i)}{\hpi_1(Y_i,\BV_i)+\hpi_3(Y_i,\BV_i)}+\frac{(\delta_{i1}+\delta_{i2})\hpi_{3}(Y_i,\BV_i)}{\hpi_{1}(Y_i,\BV_i)+\hpi_{2}(Y_i,\BV_i)}
	 + \frac{\delta_{i1}\hpi_{4}(Y_i,\BV_i)}{\hpi_{1}(Y_i,\BV_i)}, \\
	\hBS^*_{2i}(\hbb_{M1})
	&=\frac{\sum_{k=1}^{n}\delta_{k1}\BS_k(\hbb_{M1})I(Y_k=Y_i,\TX_{2k}=\TX_{2i},\BV_k=\BV_i)}
	{\sum_{s=1}^{n}\delta_{s1}I(Y_s=Y_i,\TX_{2s}=\TX_{2i},\BV_s=\BV_i)}, \\
	\hBS^*_{3i}(\hbb_{M1})
	&=\frac{\sum_{k=1}^{n}\delta_{k1}\BS_k(\hbb_{M1})I(Y_k=Y_i,\TX_{1k}=\TX_{1i},\BV_k=\BV_i)}
	{\sum_{s=1}^{n}\delta_{s1}I(Y_s=Y_i,\TX_{1s}=\TX_{1i},\BV_s=\BV_i)}, \\
	\hBS^*_{4i}(\hbb_{M1})
	&=\dfrac{\sum_{k=1}^{n}\delta_{k1}\BS_k(\hbb_{M1})I(Y_k=Y_i,\BV_k=\BV_i)}
	{\sum_{s=1}^{n}\delta_{s1}I(Y_s=Y_i,\BV_s=\BV_i)}, \\
	\hBS^*_i(\hbb_{M2})
	&=\frac{\sum_{k=1}^{n}\delta_{k1}\BS_k(\hbb_{M2})I(Y_k=Y_i,\BV_k=\BV_i)}
	{\sum_{s=1}^{n}\delta_{s1}I(Y_s=Y_i,\BV_s=\BV_i)},\\
	\htBS_i(\hbb_{M2})
&=\frac{1}{M}\sum_{v=1}^{M}\sum_{k=1}^{4}\delta_{ik}\wt{\wt{\bm\mX}}_{ kiv}\Big(Y_i-H(\hbb_{M2}^{\tT}\wt{\wt{\bm\mX}}_{kiv})\Big),
\end{align*}
for $\BS_i(\hbb_{Mr})=\bm\mX_i\big(Y_i-H(\hbb_{Mr}^{\tT}\bm\mX_i)\big)$, $r=1,2$.

From Theorems~\ref{theorem 1} and \ref{theorem 2}, one can conclude that $\hbb_{Mr}\povr\bb$, $r=1,2$. Each component of $\BG_{Mr}(\cdot)$ 
represents the sum of independent and identically distributed random variables, leading to $\BG_{Mr}(\hbb_{Mr})-\BG_{Mr}(\bb)\povr\bm0$.
One can show that $\BG_{Mr}(\bb)\povr\BG(\bb)$ by utilizing the weak law of large numbers and, hence, by Slutsky's theorem, 
$\BG_{Mr}(\hbb_{Mr})\povr\BG(\bb)$. Similarly, it can also be shown that  
$\hBM_r(\hbb_{Mr},\hbPi)\povr\BM_r(\bb,\bpi_1)$. 
Therefore, we can obtain consistent estimators of $\bDel_{Mr}$, denoted as 
$\hbDel_{Mr}=\hBG_{Mr}^{-1}(\hbb_{Mr})\hBM_r(\hbb_{Mr},\hbPi)[\hBG_{Mr}^{-1}(\hbb_{Mr})]^{\tT}$, $r=1,2$.

\section{Simulation and real data studies}
\label{SimulationAndRealData}
\subsection{Simulation studies} \label{subsec: simulation}
This section presents several Monte Carlo simulations to assess the finite-sample performance of the following methods:
\begin{description}
	\item [(1)] $\hbb_F$: full data maximum likelihood (ML) estimator used as a benchmark for comparisons
	\item [(2)] $\hbb_C$: CC estimator
	\item [(3)] $\hbb_W$: SIPW estimator that is the solution of $\BU_W(\bb,\hbpi_1)=\b0$ in  Equation~\eqref{eq: score sipw}
	\item [(4)] $\hbb_{M1}$: MI1 estimator that is the solution of $\BU_{M1}(\bb)=\b0$ in  
	Equation~\eqref{eq: score mi1} 
	with Rubin's type estimated variance in Equation~\eqref{MI1: asymptotic var} 
	\item [(5)] $\hbb_{M2}$: MI2 estimator that is the solution of $\BU_{M2}(\bb)=\b0$ in 
	Equation~\eqref{eq: score mi2} 
	with Rubin's type estimated variance in Equation~\eqref{MI2: asymptotic var} 
	\item [(6)] $\hbb_{M1n}$: MI1 estimator that is the solution of $\BU_{M1}(\bb)=\b0$ in
	 Equation~\eqref{eq: score mi1} 
	with proposed estimated variance given in Equation~\eqref{MI1n: asymptotic var}, called an MI1n estimation method 
	\item [(7)] $\hbb_{M2n}$: MI2 estimator that is the solution of $\BU_{M2}(\bb)=\b0$ in
	 Equation~\eqref{eq: score mi2} 
	with proposed estimated variance given in Equation~\eqref{MI2n: asymptotic var}, called an MI2n estimation method 
	\item [(8)] $\hbb_{RF}$:  RFMI estimator from \texttt{mice} package in \textsf{R}, used as a benchmark for comparisons
	\item [(9)] $\hbb_{EM}$: SAEM estimator from \texttt{misaem} package in \textsf{R}, used as a benchmark for comparisons.
\end{description}

Four experiments were considered to assess the performance of the estimation methods.
\begin{description}
 \item[(1) Study 1:] We examined the impact of the missing rate while keeping the sample size constant.
 \item[(2) Study 2:] The effect of sample size was the focus, with different sample sizes considered under a fixed missing rate.
 \item[(3) Study 3:] This study explored the performance of MI methods with varying numbers of imputations.
 \item[(4) Study 4:] In this case, we analyzed scenarios where all covariates are binary variables with three sets of selection probabilities.	
\end{description}
For each experiment,  $1,000$ replications were performed. The sample size was set to $n=1,000$ in most cases, except in Study 2 where sample sizes 
	$n=500$, $1,000$, and $1,500$ were used. 15 ($M=15$) imputations were used in general, except in Study 3 where $M=10$, $20$ and $30$ imputations were considered. 
	We calculated the bias, standard deviation (SD), asymptotic standard error (ASE),  mean square error (MSE), and coverage probability (CP) of 
	a 95\% confidence interval for each estimator. The MSE of an estimator was defined as the sum of the square of bias and the square of SD, i.e., $\text{MSE}=\text{bias}^2+\text{SD}^2$. 	
To assess the relative efficiencies (REs) of the estimators, we calculated the ratio of the ASEs of each estimator (excluding the full data ML estimator) to that of both the MI1 and MI2 estimators.

\textbf{Study 1:} 
This study aimed to assess the impact of varying missing rates while maintaining a constant sample size ($n=1,000$) and a fixed number of imputations ($M=15$). 
	We followed a data generation process similar to that described in \cite{lee2023estimation}.
\begin{description}
 \item[$\bullet$]  Data of   $\texttt{X}_1$ were generated from a discrete distribution that has four values $(-0.3,-0.1,0.4,1)$ with probabilities  $(0.2,0.3,0.3,0.2)$, respectively.
 \item[$\bullet$] Data of $\texttt{X}_2$ were generated from a discrete distribution with values  $(-1,-0.4,0.2,0.6)$ and probabilities $(0.1,0.3,0.3,0.3)$.
  \item[$\bullet$] Data for $Z$ were generated from a Bernoulli distribution with a probability of 0.4.
  \item[$\bullet$] Binary surrogate variables $W_1$ and $W_2$ for $\texttt{X}_1$ and $\texttt{X}_2$ were defined as $W_k=1$ if $\texttt{X}_k>0$ and $W_k=0$ if $\texttt{X}_k\leq0$ for $k=1,2$.  
\end{description}
Data of the binary outcome $Y$ were generated from the Bernoulli distribution with success probability 
\begin{align*}
P(Y=1|\texttt{X}_1,\texttt{X}_2,Z)=H(\beta_0+\beta_1\texttt{X}_1+\beta_2\texttt{X}_2+\beta_3Z), 
\end{align*}
where $\bb=(\beta_0,\beta_1,\beta_2,\beta_3)^{\tT}=(-1,1,0.7,-1)^{\tT}$. 

To introduce the missingness in $\texttt{X}_1$ and $\texttt{X}_2$ that adheres the MAR mechanism, we defined missing indicators $\delta_s$, $s=1,2,3,4$. 
	These were generated from the following multinomial logistic regression model:
\begin{align}
	\label{eq:lnsimulation}
	\ln\left(\frac{P(\delta_{ij}=1|Y_i,W_{1i},W_{2i},Z_i)}{P(\delta_{i4}=1|Y_i,W_{1i},W_{2i},Z_i)}\right)
	=\alpha_j+\gamma_1Y_i+\gamma_2W_{1i}+\gamma_3W_{2i}+\gamma_4Z_i,
\end{align}
where 
$i=1,2,\dots,n$, $j=1,2,3$, $\bga=(\gamma_1,\gamma_2,\gamma_3,\gamma_4)^{\tT}=(0.7,-0.2,0.1,-1.2)^{\tT}$, 
$\ba=(\alpha_1,\alpha_2,\alpha_3)^{\tT}=(2.6,0.6,0.6)^{\tT}$, $(1.6,0.6,0.6)^{\tT}$, and $(0.8,0.6,0.6)^{\tT}$.
Three sets of observed selection probabilities, $(0.72,0.10,0.10,0.08)$, $(0.48,0.18,0.18,0.16)$, and $(0.30,0.24,0.24,0.22)$, were considered. 
	For instance, the selection probabilities $(0.72,0.10,0.10,0.08)$ 
indicate that the percentages of complete cases, cases missing only $\texttt{X}_1$, cases missing only $\texttt{X}_2$, and cases missing both $\texttt{X}_1$ and $\texttt{X}_2$ were 
72\%, 10\%, 10\%, and l8\%, respectively.

The simulation results of Study 1 are presented in Tables~\ref{tab: simulation study 1} and \ref{tab: RE_01}. 
Overall, an increase in the missing rate led to a decrease in the performance of all estimation 
methods. Specifically, as seen from Table~\ref{tab: simulation study 1} when the CC percentage was decreased from 72\% to 30\%, the biases, SDs, and ASEs 
of most estimators tended to increase. Notably, the full data maximum likelihood estimator ($\hbb_F$), used as a benchmark, consistently outperformed the other estimators.
The CC estimators showed the worst performance. Both proposed MI methods demonstrated superior performance in terms of MSE compared to 
CC, SIPW, RFMI, and SAEM estimators. In most cases, the RFMI estimator exhibited larger bias values compared to SIPW, MI1, MI2, MI1n, MI2n, and SAEM estimators.
The empirical CPs based on all estimation methods were generally close to the nominal probability 95\% except for CC (for  $\beta_3$ when CC percentage = 0.30), 
SIPW (for $\beta_3$ when CC percentage = 0.30), MI1 (for $\beta_1$ when CC percentage = 0.30), and RFMI methods (for $\beta_1$ when CC percentage = 0.48 and $\beta_0$, $\beta_1$, 
and $\beta_2$ when CC percentage = 0.48 and 0.30).       

Table~\ref{tab: RE_01} displays the relative efficiencies (REs) of the CC, SIPW, MI1, MI2, RTMI, and SAEM estimators to the MI1n and MI2n estimators, 
 denoted as C$r$, W$r$, MI1$r$, MI2$r$, RF$r$, and EM$r$, $r=1,2$. The RE values of C$r$, W$r$, RF$r$, and EM$r$, $r=1,2$, were greater than one and 
 increased as the missing rate rose. This indicates  that the MI$r$n estimators were more efficient than the CC, SIPW, RFMI, and SAEM estimators 
 due to their smaller ASEs, especially as the missing rate was increased. Conversely, the RE values of M$rr$, $r=1,2$, were close to 1, 
 suggesting that the ASEs of the MI$r$n estimator were somewhat larger than those of the corresponding MI$r$ estimator. 
This is reasonable because the MI$r$ estimation method using Rubin's type variance estimator was tended to underestimate the variance, as discussed in Section~\ref{Introduction}. Therefore, the  MI1n and MI2n estimation methods, which employ the proposed ASEs, could overcome this limitation. Finally, in most cases, the ASEs of the MI2n estimator were larger than those of the MI1n estimator.

\textbf{\textit{Study 2.}} This study aimed to evaluate the influence of sample size, $n=500$, $1,000$ and $1,500$, on the performance of all 
methods under the same missing rate and a fixed number of  imputations $M=15$. The data of $\texttt{X}_1$ were generated from a discrete distribution 
with four values $(-0.3,-0.08,0.5,0.8)$ and corresponding probabilities $(0.1,0.3,0.3,0.3)$. 
Similarly, the data for $\texttt{X}_2$ were generated from a discrete distribution with values $(-0.8,-0.6,0.1,0.9)$ and  corresponding probabilities 
$(0.3,0.3,0.3,0.1)$. The data for $Z$ were generated from a Bernoulli distribution with $p=0.5$. The surrogate variables $W_k$ for $\texttt{X}_k$, $k=1,2$, 
were defined in a similar manner to Study 1. We set $\bb=(1.2,1,1,1)^{\tT}$ to determine the probability $H(\beta_0+\beta_1\texttt{X}_1+\beta_2\texttt{X}_2+\beta_3Z)$ 
for generating the data of $Y$. We used the same $\bga$ as in Study 1 and $\ba=(1.4,0.6,0.6)^{\tT}$ in 
Equation~\eqref{eq:lnsimulation}
to generate the data of $\delta_s$, $s=1,2,3,4$. The observed selection probabilities were $(0.45,0.20,0.20,0.15)$.

Tables~\ref{tab: simulation study 2} and \ref{tab: RE_02} present the simulation results of Study 2, which are quite similar to 
those of Study 1. The performance of all estimators was improved as indicated by the reduction in their bias, SD, and ASE when the sample size was increased 
from $500$ to $1,500$. The MSEs of the proposed  MI1n and MI2n estimators were consistently the smallest compared to those of the
CC, SIPW, RFMI, and SAEM estimators in nearly all cases. Overall, the empirical CPs  for all the estimation methods were close 
to the nominal probability 95\% except for the RFMI (for $\beta_2$ when $n=500$; $\beta_1$ and $\beta_2$ when $n=1,000$ and $1,500$), SIPW, MI$r$, 
and MI$r$n, $r=1,2$ (for $\beta_2$ when $n=500$ and $1,000$) methods. Table~\ref{tab: RE_02} also demonstrates that the ASEs of the proposed MIn$r$ 
estimators, $r=1,2$, were consistently smaller than those of the CC, SIPW, MI1, and MI2 estimators in most cases. 
The only exception was the ASEs of the MI2n estimator in comparison with those of the MI1, MI2, and MI1n estimators.

\textbf{\textit{Study 3.}} The purpose of this study was to examine the performance of the MI approaches with varying amounts of imputations under the same 
missing rate and sample size of $n=1,000$. We kept all settings as in Study 2, except for the number of imputations, which was set to $M=10,20,30$. 
The observed selection probabilities remained the same at $(0.45,0.20,0.20,0.15)$. The simulation outcomes  presented in 
Tables~\ref{tab: simulation study 3} and \ref{tab: RE_03} were virtually identical to those in Tables~\ref{tab: simulation study 2} and \ref{tab: RE_02} 
from Study 2 when $n=1,000$. The bias, SD, ASE, MSE, and CP of the proposed MI$r$n estimators, $r=1,2$, remained largely unchanged across the
three levels of $M$. This suggests that the proposed estimated variances of MI estimators were not significantly affected by the 
number of imputations when $M\ge10$. 

\textbf{\textit{Study 4.}} The fourth study aimed to examine a scenario in which all covariates were binary variables with three sets of selection probabilities, 
while keeping the sample size fixed at $n=1,500$ and using $M=15$ imputations. For this purpose, the data of all three covariates $\texttt{X}_1$, $\texttt{X}_2$, and $Z$ were generated from 
a Bernoulli distribution with a probability of $p=0.5$. To generate the surrogate variables, $W_k$ for $\texttt{X}_k$, $k=1,2$, we also used 
Bernoulli distributions. Given each value of $\texttt{X}_1=1$ and 0, the data of $W_1$ were obtained by Bernoulli distributions with success probabilities of
$0.6$ and $0.5$, respectively. Similarly, for each value of $\texttt{X}_2=1$ and 0, Bernoulli distributions were used to generate the data of $W_2$ with success 
probabilities of $0.55$ and $0.6$, respectively. The values of $\bb$ and $\bga$ were fixed similarly to those in Study 2. The value of $\ba$ was set as
$(2.4,0.6, 0.6)^{\tT}$, $(1.6,0.6,0.6)^{\tT}$, and $(1,0.6,0.6)^{\tT}$ to obtain three sets  
of observed selection probabilities, which were $(0.70,0.12,0.12,0.06)$, $(0.50,0.19,0.19,0.12)$, and $(0.36,0.24,0.24,0.16)$, respectively.      

Tables~\ref{tab: simulation study 4} and \ref{tab: RE_04} present the simulation results of Study 4.
The performance observed in this study was quite similar to that in the previous studies, with some noteworthy findings. The RFMI estimator displayed the highest bias, followed by the CC estimator. 
The MSEs from the MI$r$n estimators were smaller than those of the RFMI estimator but larger than those of the SAEM estimator in certain cases. 
The empirical CPs for almost all estimation methods were consistently close to the nominal probability 95\%, except for the RFMI method 
(for $\beta_0$ and $\beta_2$ under the three sets of selection probabilities, and for $\beta_3$ for the selection probabilities: $(0.36,0.24,0.24,0.16)$). 
Notably, the SIPW, MI$r$, and MI$r$n estimation methods, $r=1,2$, exhibited deviations in some cases 
for $\beta_k$, $k=0,1,2$, under the two sets of selection probabilities: $(0.50,0.19,0.19,0.12)$ and  $(0.36,0.24,0.24,0.16)$.

\subsection{Example}
\label{Example}
The proposed methodologies were applied to showcase their practical utility by using data collected from a survey of $1,634$ respondents who visited 
the Feng Chia night market (FCNM) in Taichung City, Taiwan, during the period from November 17 to 21, 2022. The survey questionnaire comprised 37 questions divided into three sections 
designed to gather essential information and evaluations from customers shopping at the FCNM.
These sections covered  general opinions, consumption habits, and culinary preferences, all aimed at enhancing and delivering improved consumer experiences. 

To construct the logistic regression model, we designated the question ``Did you stay in a nearby hotel or daily rental suite during your visit to the FCNM?'' 
as a binary outcome variable $Y$ (with $1=$ denoting Yes and $0=$ indicating No). The question, ``What is the number of times you went shopping at the FCNM in the last half year (including this time)?'', 
was selected as the first covariate $\texttt{X}_1$ ($1=1,2$, or 3 times, $0=$ more than three times). The  respondent's current city of residence was used as the second covariate 
$\texttt{X}_2$ (categorized as $1=$ Neighborhood of Taichung or Others City and $0=$ Taichung City). The question, ``How much did you spend each time you visit the FCNM?'', was employed as the covariate 
$Z$ ($Z=0.1,0.35,0.75,1.25,1.75,3$, which was obtained by taking the median and being divided by $1,000$, representing six levels of expenditure: $<200$ New Taiwan dollar (TWD), $200-500$ TWD, 
$500-1,000$ TWD, $1,000-1,500$ TWD, $1,500-2,000$ TWD, and $>2,000$ TWD). Unfortunately, there 
were instances where respondents did not provide an answer to the question, ``What is the number 
of times you went shopping at the FCNM in the last half year (including this time)?'', resulting in $\texttt{X}_1$ having 33.4\% missing values. This also indicates 
that the size of the CC data is 1,089 (66.6\%). The goal is to fit the following logistic regression model to the FCNM data set:
\begin{align} \label{example: logistic model 1}
	P(Y=1|\texttt{X}_1,\texttt{X}_2,Z) = H(\beta_0+\beta_1\texttt{X}_1+\beta_2\texttt{X}_2+\beta_3Z).
\end{align}

Let $W$ represent the random variable of response to the question ``How much time did you spend traveling to the FCNM?'', where $W$ is defined as follows: $W=1$ for travel times within 15 minutes, 
$W=2$ for travel times between 15 and 60 minutes, and $W=3$ for travel times exceeding 60 minutes. $W$ serves as a surrogate variable for $X_1$ because, in logistic 
regression with response $Y$, $W$ lacks significance. Moreover, $W$ exhibits a correlation with $X_1$ (Spearman's rank correlation coefficient of 0.17 with $p$-value $<0.001$) based on the CC data. 

Let $\wt\delta$ be a variable defined such that $\wt\delta=1$ if both $\texttt{X}_1$ and $\texttt{X}_2$ are observed; otherwise $\wt\delta=0$. It is essential to note that only $X_1$ is missing in this particular case. To explore the influence of $Y$, $Z$, and $W$ on the missingness mechanism of $X_1$, we utilize the logistic regression model 
$\text{logit}[P(\wt\delta=1|Y,Z,W)]= \nu_0+\nu_1Y+\nu_2Z+\nu_3W$.
This model helps us identify the effects of $Y$, $Z$, and $W$ on the missingness mechanism of $\texttt{X}_1$.
The $p$-values from the Wald chi-squared tests 
for the effects of $Y$ (i.e., testing $H_0:\nu_1=0$), $Z$ (i.e., testing $H_0:\nu_2=0$), and $W$ (i.e., testing $H_0:\nu_3=0$) are $0.82$, $0.006$, and $0.147$, respectively. 
These results indicate that $Z$ is statistically significantly associated with the missingness mechanism of $\texttt{X}_1$, allowing us to assume that $\texttt{X}_1$ is MAR.

Table~\ref{tab: example 1a} presents the analysis results from the logistic regression model in Equation~\eqref{example: logistic model 1} with the original missing data, i.e., the selection probabilities $(0.666,0.334,0,0)$.  All $p$-values for testing $\beta_k=0$ are $<0.05$, signifying statistical significance. The only exception is testing $\beta_3=0$, as determined by using the CC estimation method. The positive estimates of $\beta_1$, $\beta_2$, and $\beta_3$ suggest
that respondents were more likely to stay near a hotel or daily rental suite when they had more recent time to go to the FCNM, when their current city of residence 
was not Taichung city, and when they spent more money on shopping each time. The parameter estimates obtained via the SIPW, MI1, MI2, MI1n, MI2n, RFMI,
 and SAEM approaches exhibit both similarities and differences compared
 to those obtained through the CC method. Notably, the ASEs of the MI1n and MI2n estimators are smaller than those of the CC and SIPW estimators. 
 The ASEs of the MI2n estimators are larger than those of the MI1, MI2, RFMI, and SAEM estimators, except for $\beta_1$. The ASEs of the MI1n estimator are smaller than those of the MI1 (for $\beta_2$ and $\beta_3$), 
 MI2, MI2n (except for $\beta_1$), RFMI, and SAEM estimators (except for $\beta_3$). These results demonstrate the performance of the two proposed estimators for the variances of the two MI estimators and are consistent with the simulation results.

 It is evident that the proposed methods have been quite effective when applied to the FCNM data set. However, in this data set, only the variable $\texttt{X}_1$ contains missing values.
To provide a more comprehensive illustration of the proposed methods, we have generated a new data set, referred to as an FCNMn data set,  with $\texttt{X}_1$ and $\texttt{X}_2$ that are MAR either separately or simultaneously. 
Specifically, we generated binary data of $\tau$ from a Bernoulli distribution with a success probability of $H(-1+1.2Y+0.3Z-0.4W)$. The new data of $\texttt{X}_2$, denoted as $\texttt{X}_2^*$, were generated 
by preserving $\texttt{X}_2$ if $\tau=1$ and making it as $NA$ if $\tau=0$. 
Hence we can  obtain $\delta_{i1}=\wt\delta_i\tau_i$,
 $\delta_{i2}=(1-\wt\delta_i) \tau_i$,  $\delta_{i3}=\wt\delta_i(1-\tau_i)$, and $\delta_{i4}=(1-\wt\delta_i)(1-\tau_i)$, $i=1,2,\dots,n$.	
The FCNMn data set shares the same sample size as the original data set, FCNM data set, and selection
probabilities of $(0.485, 0.255, 0.181, 0.079)$. These selection probabilities correspond to the percentages of complete cases, cases with only $\texttt{X}_1$ missing, cases with only $\texttt{X}_2$ missing, and 
cases with both variables missing, which are 48.5\%, 25.5\%, 18.1\%, and 7.9\%, respectively. 

The results of analysis of the logistic regression model used to fit the FCNMn data set are presented in Table~\ref{tab: example 1b}. 
In general, these results exhibit slight variations compared to those from the original FCNM data set in Table~\ref{tab: example 1a}, primarily due to the increased missing rates.
Nonetheless, they effectively demonstrate the performance characteristics  of the suggested estimation methods. Specifically, the absolute values 
of the estimates of $\beta_k$, $k=0,1,2,3$, have  seen a slight increase when compared to the original FCNM data set. The results of testing $\beta_k=0$ are statistically 
significant except for testing $\beta_3=0$, as observed when using the CC and SIPW estimation methods. The ASEs of all estimators have increased, with the MI1n and MI2n methods still displaying smaller ASEs than the  CC and IPW methods, larger ASEs than the MI1 and MI2 method (except for $\beta_3$), and smaller ASEs than 
the RFMI and SAEM method (except for $\beta_0$). 
Therefore, it can be concluded that the proposed MI1n and MI2n methods 
maintain consistent performance with  the simulation results, even with higher missing rates.

\section{Conclusion}
\label{Conclusion}
We have established the asymptotic properties of the MI1 and MI2 estimators,  as discussed in the work of \cite{lee2023estimation}, for the parameters of logistic regression with two sets of discrete or categorical covariates that are MAR separately or simultaneously. The two proposed MI methods, MI1n and MI2n, utilize estimated variances derived from the asymptotic variances of the two MI estimators. Based on the results obtained in various scenarios, the two proposed MI methods performed exceptionally well, and they consistently outperformed the CC, SIPW, RFMI, and SAEM methods. 
Significantly, the two proposed variance estimation methods effectively resolve the issue of Rubin's type estimated variances \citep{rubin1987statistical} underestimating the variances of the two MI estimators described in the study by  \cite{lee2023estimation}.  To demonstrate the practical utility of these two proposed MI methods, real data from a survey conducted at the FCNM, Taiwan, have been considered.

Although our primary focus is on logistics regression with two sets of discrete or categorical covariates that are MAR separately or simultaneously, our proposed MI methods can also be 
applied to, e.g., proportional odds (PO) models and generalized linear models provided they are used in conjunction with an appropriate link function. Furthermore, these two proposed MI methods can be extended to the logistic regression models, PO  models, and generalized linear models with continuous covariates MAR separately or simultaneously.  Additionally, we are equipped to address the issue of goodness-of-fit for these regression models with covariates MAR separately or simultaneously.

\section*{Acknowledgments}
P.L. Tran was funded by the Postdoctoral Scholarship Programme of Vingroup Innovation Foundation (VINIF), code VINIF.2023.STS.47.
S.M. Lee's research was supported by a grant from the Ministry of Science and Technology (MOST) of Taiwan, ROC, under grant number MOST-109-2118-M-035-002-MY3.


\newpage
\clearpage

\section*{Appendix} \label{Appendix}
\subsection*{Proof of Lemma~\ref{lemma}}
\subsubsection*{Proof of Lemma~\ref{lemma} (i)}
The expression of $\BU_{M1}(\bb)$ in Equation~\eqref{eq: score mi1 new} can be rewritten as follows: 
\begin{align*}
\BU_{M1}(\bb)
  &=\frac{1}{\sqrt{n}}\sum_{i=1}^{n}
	\left[\delta_{i1}\BS_i(\bb)+\delta_{i2}\tBS_{2i}(\bb) +\delta_{i3}\tBS_{3i}(\bb) +\delta_{i4}\tBS_{4i}(\bb)\right]\\
  &=\frac{1}{\sqrt{n}}\sum_{i=1}^{n}\left[\delta_{i1}\BS_i(\bb)+\delta_{i2}\BS_{2i}^*(\bb)+\delta_{i3}\BS_{3i}^*(\bb)+\delta_{i4}\BS_{4i}^*(\bb)\right] \\
   &\hskip 2mm
	+\frac{1}{\sqrt{n}}\sum_{i=1}^{n}\delta_{i2}\left[\tBS_{2i}(\bb)-\E_{\tF_{\TX_{1i}}}\left(\BS_i(\bb)|Y_i,\TX_{2i},\BV_i\right)\right]\\
	&\hskip 2mm
	+\frac{1}{\sqrt{n}}\sum_{i=1}^{n}\delta_{i3}\left[\tBS_{3i}(\bb)-\E_{\tF_{\TX_{2i}}}\left(\BS_i(\bb)|Y_i,\TX_{1i},\BV_i\right)\right]\\
	&\hskip 2mm
	+\frac{1}{\sqrt{n}}\sum_{i=1}^{n}\delta_{i4}\left[\tBS_{4i}(\bb)-\E_{\tF_{\TX_{i}}}\left(\BS_i(\bb)|Y_i,\BX_i,\BV_i\right)\right] \\
	&\hskip 2mm
	 +\frac{1}{\sqrt{n}}\sum_{i=1}^{n}\delta_{i2}\left[\E_{\tF_{\TX_{1i}}}\left(\BS_i(\bb)|Y_i,\TX_{2i},\BV_i\right)-\BS_{2i}^*(\bb)\right]\\
	&\hskip 2mm
	 +\frac{1}{\sqrt{n}}\sum_{i=1}^{n}\delta_{i3}\left[\E_{\tF_{\TX_{2i}}}\left(\BS_i(\bb)|Y_i,\TX_{1i},\BV_i\right)-\BS_{3i}^*(\bb)\right] \\
	&\hskip 2mm
	+\frac{1}{\sqrt{n}}\sum_{i=1}^{n}\delta_{i4}\left[\E_{\tF_{\BX_i}}\left(\BS_i(\bb)|Y_i,\BX_i,\BV_i\right)-\BS_{4i}^*(\bb)\right].
\end{align*}
Here $\wt\BS_{2i}(\bb)$, $\wt\BS_{3i}(\bb)$, and $\wt\BS_{4i}(\bb)$ are given in Equation~\eqref{eq:BSs}.
	$\BS_{2i}^*(\bb)$, $\BS_{3i}^*(\bb)$, and $\BS_{4i}^*(\bb)$ are presented in Equation~\eqref{eq:BS*s}.
Firstly, $\tBS_{2i}(\bb)-\E_{\tF_{\TX_{1i}}}\left(\BS_i(\bb)|Y_i,\TX_{2i},\BV_i\right)$ can be expressed as
\begin{align}
\label{eq:fact1}
& \tBS_{2i}(\bb)-\E_{\tF_{\TX_{1i}}}\left(\BS_i(\bb)|Y_i,\TX_{2i},\BV_i\right) \nonumber \\
=&\
\frac{1}{M}\sum_{v=1}^{M} \left[\tBS_{2iv}(\bb)-\E_{\tF_{\TX_{1i}}}\left(\BS_i(\bb)|Y_i,\TX_{2i},\BV_i\right)\right] \nonumber\\
=&\
\frac{1}{\sqrt{M}}
\left\{\frac{1}{\sqrt{M}}\sum_{v=1}^{M}\left[\tBS_{2iv}(\bb)-\E_{\tF_{\TX_{1i}}}\left(\BS_i(\bb)|Y_i,\TX_{2i},\BV_i\right)\right]\right\}.
\end{align}
By employing the central limit theorem (CLT), it can yield
\begin{align*}
\frac{1}{\sqrt{M}}\sum_{v=1}^{M}\left[\tBS_{2iv}(\bb)-\E_{\tF_{\TX_{1i}}}\left(\BS_i(\bb)|Y_i,\TX_{2i},\BV_i\right)\right]
\dovr\mZ_i(\bb)\sim\mN_{p+1}(\bm{0},\bSigma_{\mZ_i}).
\end{align*}
Therefore, $\tBS_{2i}(\bb)-\E_{\tF_{\TX_{1i}}}\left(\BS_i(\bb)|Y_i,\TX_{2i},\BV_i\right)
=M^{-1/2}\mZ_i(\bb)+\bo_p(M^{-1/2})$. From Equation~\eqref{eq:fact1},
we now have 
\begin{align*}
& \frac{1}{\sqrt{n}}\sum_{i=1}^{n}\delta_{i2}\left[\tBS_{2i}(\bb)-\E_{\tF_{\TX_{1i}}}\left(\BS_i(\bb)|Y_i,\TX_{2i},\BV_i\right)\right]\\
=&\  
\frac{1}{\sqrt{M}}\frac{1}{\sqrt{n}}\sum_{i=1}^{n}\delta_{i2}\mZ_i(\bb)  \\
&\hskip 2mm
 + \frac{1}{M}\frac{1}{\sqrt{n}} \sum_{i=1}^{n}\sum_{v=1}^{M}\delta_{i2}\left\{\left[\tBS_{2iv}(\bb)-\E_{\tF_{\TX_{1i}}}\left(\BS_i(\bb)|Y_i,\TX_{2i},\BV_i\right)\right]-\mZ_i(\bb)\right\}\\
=&\ 
\frac{1}{\sqrt{M}} \frac{1}{\sqrt{n}}\sum_{i=1}^{n}\delta_{i2}\mZ_i(\bb) \\
 &\hskip 2mm
 + \frac{1}{M} \sum_{v=1}^{M}\left\{\frac{1}{\sqrt{n}}\sum_{i=1}^{n}\delta_{i2}\left\{\left[\tBS_{2iv}(\bb)-\E_{\tF_{\TX_{1i}}}\left(\BS_i(\bb)|Y_i,\TX_{2i},\BV_i\right)\right]-\mZ_i(\bb)\right\}\right\}.
\end{align*}
As $v$ is fixed, utilizing the CLT,  $n^{-1/2}\sum_{i=1}^{n}\delta_{i2}\{[\tBS_{2iv}(\bb)-\E_{\tF_{\TX_{1i}}}(\BS_i(\bb)|Y_i,\TX_{2i},\BV_i)]-\mZ_i(\bb)\}$ can be shown to converge to a normal distribution with a mean of zero.
Hence the second term in the above expression converges to $\b0$ in probability by the weak law of large numbers. 
Accordingly, we can obtain
\begin{align*}
\frac{1}{\sqrt{n}}\sum_{i=1}^{n}\delta_{i2}\left[\tBS_{2i}(\bb)-\E_{\tF_{\TX_{1i}}}\left(\BS_i(\bb)|Y_i,\TX_{2i},\BV_i\right)\right]
=\frac{1}{\sqrt{M}}\frac{1}{\sqrt{n}}\sum_{i=1}^{n}\delta_{i2}\mZ_i(\bb)+\bo_p(1).
\end{align*}
It can be shown via the CLT that
\begin{align*}
 \frac{1}{\sqrt{n}}\sum_{i=1}^{n}\delta_{i2}\mZ_i(\bb)\dovr\mZ,
\end{align*}
where $\mZ$ follows a multivariate normal distribution with mean $\b0$, so
\begin{align*}
 \frac{1}{\sqrt{n}}\sum_{i=1}^{n}\delta_{i2}\left[\tBS_{2i}(\bb)-\E_{\tF_{\TX_{1i}}}\left(\BS_i(\bb)|Y_i,\TX_{2i},\BV_i\right)\right]=\BO_p(M^{-1/2}).
\end{align*}
Similarly, one can show
\begin{align*}
   \frac{1}{\sqrt{n}}\sum_{i=1}^{n}\delta_{i3}\left[\tBS_{3i}(\bb)-\E_{\tF_{\TX_{2i}}}\left(\BS_i(\bb)|Y_i,\TX_{1i},\BV_i\right)\right]
 &=\BO_p(M^{-1/2}), \\
  \frac{1}{\sqrt{n}}\sum_{i=1}^{n}\delta_{i4}\left[\tBS_{4i}(\bb)-\E_{\tF_{\TX_i}}\left(\BS_i(\bb)|Y_i,\BX_i,\BV_i\right)\right]
 &=\BO_p(M^{-1/2}).
\end{align*}
In addition,
\begin{align}
\label{eq:fact2}
&\frac{1}{\sqrt{n}}\sum_{i=1}^{n}\delta_{i2}\left[\E_{\tF_{\TX_{1i}}}\left(\BS_i(\bb)|Y_i,\TX_{2i},\BV_i\right)-\BS_{2i}^*(\bb)\right]  \nonumber \\
=&\
\frac{1}{\sqrt{n}}\sum_{i=1}^{n}\delta_{i2}\left[\frac{\sum_{r=1}^{n}\delta_{r1}I(Y_r=Y_i,\TX_{2r}=\TX_{2i},\BV_r=\BV_i)\BS_r(\bb)}
	{\sum_{j=1}^{n}\delta_{j1}I(Y_j=Y_i,\TX_{2j}=\TX_{2i},\BV_j=\BV_i)}-\BS_{2i}^*(\bb)\right] \nonumber \\
=&\
\frac{1}{\sqrt{n}}\sum_{i=1}^{n}\delta_{i2}\left[\frac{\sum_{r=1}^{n}\delta_{r1}I(Y_r=Y_i,\TX_{2r}=\TX_{2i},\BV_r=\BV_i)\left[\BS_r(\bb)-\BS_{2i}^*(\bb)\right]}
	{\sum_{j=1}^{n}\delta_{j1}I(Y_j=Y_i,\TX_{2j}=\TX_{2i},\BV_j=\BV_i)}\right] \nonumber \\
=&\
\frac{1}{\sqrt{n}}\sum_{i=1}^{n}\delta_{i2}\left[\frac{\sum_{r=1}^{n}\delta_{r1}I(Y_r=Y_i,\TX_{2r}=\TX_{2i},\BV_r=\BV_i)\left[\BS_r(\bb)-\BS^*_{2r}(\bb)\right]}
	{\sum_{j=1}^{n}\delta_{j1}I(Y_j=Y_i,\TX_{2j}=\TX_{2i},\BV_j=\BV_i)}\right]  \nonumber \\
=&\
\frac{1}{\sqrt{n}}\sum_{i=1}^{n}\delta_{i2}\left[\sum_{r=1}^{n}\frac{\delta_{r1}I(Y_r=Y_i,\TX_{2r}=\TX_{2i},\BV_r=\BV_i)}
	{\sum_{j=1}^{n}\delta_{j1}I(Y_j=Y_i,\TX_{2j}=\TX_{2i},\BV_j=\BV_i)}\left[\BS_r(\bb)-\BS^*_{2r}(\bb)\right]\right] \nonumber \\
=&\
\frac{1}{\sqrt{n}}\sum_{i=1}^{n}\delta_{i2}\left[\sum_{r=1}^{n}\frac{\delta_{r1}I(Y_i=Y_r,\TX_{2i}=\TX_{2r},\BV_i=\BV_r)}
	{\sum_{j=1}^{n}\delta_{j1}I(Y_j=Y_r,\TX_{2j}=\TX_{2r},\BV_j=\BV_r)}\left[\BS_r(\bb)-\BS^*_{2r}(\bb)\right]\right] \nonumber \\
=&\
\frac{1}{\sqrt{n}}\sum_{r=1}^{n}\delta_{r1}\left[\BS_r(\bb)-\BS^*_{2r}(\bb)\right]
\left[\sum_{i=1}^{n}\frac{\delta_{i2}I(Y_i=Y_r,\TX_{2i}=\TX_{2r}, \BV_i=\BV_r)}{\sum_{j=1}^{n}\delta_{j1}I(Y_j=Y_r,\TX_{2j}=\TX_{2r},\BV_j=\BV_r)}\right].
\end{align}

Let $P_r=P(Y=Y_r,\TX_2=\TX_{2r},\BV=\BV_r)$ and
	\begin{align*}
		\wh\pi_{s}(Y_r,\BV_r)\wh{P}_r
		=\frac{1}{n}\sum_{i=1}^{n}\delta_{is}I(Y_i=Y_r,\TX_{2i}=\TX_{2r},\BV_i=\BV_r).
	\end{align*}
Note that $\delta_{is}I(Y_i=Y_r,\TX_{2i}=\TX_{2r},\BV_i=\BV_r)$ is a binary random variable with 
		mean $\E(\pi_s(Y_r,\BV_r)P_r)$, so $(\wh\pi_s(Y_r,\BV_r)\wh{P}_r-\E(\pi_s(Y_r,\BV_r)P_r))^2$ is $o_p(n^{-1/2})$,
		$s=1,2,3,4$.
Using a Taylor's expansion to $\wh\pi_2(Y_r,\BV_r)$ and $\wh\pi_1(Y_r,\BV_r)$ that are presented in Equation~\eqref{def: pihat},  the expression in Equation~\eqref{eq:fact2} can be re-written as follows:
\begin{align}
\label{eq:fact3}
&\frac{1}{\sqrt{n}}\sum_{i=1}^{n}\delta_{i2}\left[\E_{\tF_{\TX_{1i}}}\left(\BS_i(\bb)|Y_i,\TX_{2i},\BV_i\right)-\BS_{2i}^*(\bb)\right] \nonumber \\	
=&\ 
\frac{1}{\sqrt{n}}\sum_{r=1}^{n}\delta_{r1}\left[\BS_r(\bb)-\BS^*_{2r}(\bb)\right]
\left[\sum_{i=1}^{n}\frac{\delta_{i2}I(Y_i=Y_r,\TX_{2i}=\TX_{2r}, \BV_i=\BV_r)}{\sum_{j=1}^{n}\delta_{j1}I(Y_j=Y_r,\TX_{2j}=\TX_{2r},\BV_j=\BV_r)}\right] \nonumber \\
=&\
\frac{1}{\sqrt{n}}\sum_{r=1}^{n} \delta_{r1}\left[\BS_r(\bb)-\BS^*_{2r}(\bb)\right]\left[ 
		\dfrac{\wh\pi_2(Y_r,\BV_r)\wh{P}_r}{\wh\pi_1(Y_r,\BV_r)\wh{P}_r}\right]	\nonumber \\
=&\
\frac{1}{\sqrt{n}}\sum_{r=1}^{n} \delta_{r1}\left[\BS_r(\bb)-\BS^*_{2r}(\bb)\right]
\left\{\left[ 
		\frac{\pi_{2}(Y_r,\BV_r)}{\pi_{1}(Y_r,\BV_r)}\right]
		+\left[\frac{\wh\pi_2(Y_r,\BV_r)\wh{P}_r-\pi_2(Y_r,\BV_r)P_r}
		{\pi_{1}(Y_r,\BV_r)P_r}\right]\right.    \nonumber  \\
&\hskip 30mm +\left.
		\left[\frac{\pi_{2}(Y_r,\BV_r)P_r}
		{\pi_{1}^2(Y_r,\BV_r)P^2_r}\right]\left[
		\wh\pi_1(Y_r,\BV_r)\wh{P}_r	
		-\pi_1(Y_r,\BV_r)P_r\right]+o_p(n^{-1/2})\right\}.
\end{align}
Because $\sqrt{n}\big(\wh\pi_s(Y_r,\BV_r)\wh{P}_r-\pi_s(Y_r,\BV_r)P_r\big)\dovr
	{f_s(Y_r,\TX_{2r},\BV_r)}\sim\ml{N}\big(0,\sigma^2_{s_{Y_r,\TX_{2r},\BV_r}}\big)$, it can be shown that 
\begin{align}
\label{eq:fact4}	
&\frac{1}{n}\sum_{r=1}^{n}\delta_{r1}\left[\BS_r(\bb)-\BS^*_{2r}(\bb)\right]\left\{
		\frac{\sqrt{n}\left(\wh\pi_2(Y_r,\BV_r)\wh{P}_r-\pi_2(Y_r,\BV_r)P_r\right)}
		{\pi_{1}(Y_r,\BV_r)P_r}\right\}  \nonumber \\
 =&\
 \frac{1}{n}\sum_{r=1}^{n} \delta_{r1}\left[\BS_r(\bb)-\BS^*_{2r}(\bb)\right]
 \left\{\frac{f_2(Y_r,\TX_{2r},\BV_r)}{\pi_1(Y_r,\BV_r)P_r}+o_p(1)\right\}=\bo_p(1)
\end{align}
and
\begin{align}
\label{eq:fact5}	
 &\frac{1}{n}\sum_{r=1}^{n} \delta_{r1}\left[\BS_r(\bb)-\BS^*_{2r}(\bb)\right]
  \left\{\frac{\pi_2(Y_r,\BV_r)P_r\left[\sqrt{n}\left(\wh\pi_1(Y_r,\BV_r)\wh{P}_r-\pi_1(Y_r,\BV_r)P_r\right)\right]}
		{\pi^2_1(Y_r,\BV_r)P^2_r}\right\}  \nonumber \\
 =&\
 \frac{1}{n}\sum_{r=1}^{n}\delta_{r1}\left[\BS_r(\bb)-\BS^*_{2r}(\bb)\right]
 \left\{\frac{\pi_2(Y_r,\BV_r)P_rf_1(Y_r,\TX_{2r},\BV_r)}
		{\pi^2_{1}(Y_r,\BV_r)P^2_r}+o_p(1)\right\}=\bo_p(1).
\end{align}
Hence from Equations~\eqref{eq:fact3}-\eqref{eq:fact5}, it can yield 
\begin{align*}
&
\frac{1}{\sqrt{n}}\sum_{i=1}^{n}\delta_{i2}\left[\E_{\tF_{\TX_{1i}}}\left(\BS_i(\bb)|Y_i,\TX_{2i},\BV_i\right)-\BS_{2i}^*(\bb)\right] \\
= &\
\frac{1}{\sqrt{n}}\sum_{r=1}^{n} \delta_{r1}\left[\BS_r(\bb)-\BS^*_{2r}(\bb)\right]\left( 
		\frac{\pi_{2}(Y_r,\BV_r)}{\pi_{1}(Y_r,\BV_r)}\right)+\bo_p(1).
\end{align*}

Similarly, one can show that
\begin{align*}
&
\frac{1}{\sqrt{n}}\sum_{i=1}^{n}\delta_{i3}\left[\E_{\tF_{\TX_{2i}}}\left(\BS_i(\bb)|Y_i,\TX_{1i},\BV_i\right)-\BS^*_{3i}(\bb)\right]\\
=&\
\frac{1}{\sqrt{n}}\sum_{r=1}^{n}\delta_{r1}[\BS_{r}(\bb)-\BS^*_{3r}(\bb)]\left(\frac{\pi_3(Y_r,\BV_r)}{\pi_1(Y_r,\BV_r)}\right)+\bo_p(1), \\ 
&\
\frac{1}{\sqrt{n}}\sum_{i=1}^{n}\delta_{i4}\left[\E_{\tF_{\TX_{i}}}\left(\BS_i(\bb)|Y_i,\BX_i,\BV_i\right)-\BS_{4i}^*(\bb)\right]\\
=&\
\frac{1}{\sqrt{n}}\sum_{r=1}^{n}\delta_{r1}\left[\BS_r(\bb)-\BS^*_{4r}(\bb)\right]\left(\frac{\pi_4(Y_r,\BV_r)}{\pi_1(Y_r,\BV_r)}\right)+\bo_p(1).
\end{align*}
Therefore, $\BU_{M1}(\bb)$ can be re-expressed as follows:
\begin{align*}
&\BU_{M1}(\bb)   \nonumber \\
=&\
	\frac{1}{\sqrt{n}}\sum_{i=1}^{n}
	\left[\delta_{i1}\BS_i(\bb)+\delta_{i2}\tBS_{2i}(\bb) +\delta_{i3}\tBS_{3i}(\bb)+\delta_{i4}\tBS_{4i}(\bb)\right]  \nonumber \\
=&\
	\frac{1}{\sqrt{n}}\sum_{i=1}^{n}\left[\delta_{i1}\BS_i(\bb)+\delta_{i2}\BS^*_{2i}(\bb)+\delta_{i3}\BS^*_{3i}(\bb)+\delta_{i4}\BS^*_{4i}(\bb)\right]
	+\BO_p(M^{-1/2})  \nonumber \\
&\
	+\frac{1}{\sqrt{n}}\sum_{r=1}^{n}\delta_{r1}[\BS_r(\bb)-\BS^*_{2r}(\bb)]\left(\frac{\pi_2(Y_r,\BV_r)}{\pi_1(Y_r,\BV_r)}\right)  \nonumber \\
&\
	+\frac{1}{\sqrt{n}}\sum_{r=1}^{n}\delta_{r1}[\BS_r(\bb)-\BS^*_{3r}(\bb)]\left(\frac{\pi_3(Y_r,\BV_r)}{\pi_1(Y_r,\BV_r)}\right)  \nonumber \\
&\
	+\frac{1}{\sqrt{n}}\sum_{r=1}^{n}\delta_{r1}[\BS_r(\bb)-\BS^*_{4r}(\bb)]\left(\frac{\pi_4(Y_r,\BV_r)}{\pi_1(Y_r,\BV_r)}\right)+\bo_p(1)  \nonumber \\
=&\
	\frac{1}{\sqrt{n}}\sum_{r=1}^{n}\delta_{r1}\BS_r(\bb)\left[1+\frac{\pi_2(Y_r,\BV_r)+\pi_3(Y_r,\BV_r)+\pi_4(Y_r,\BV_r)}{\pi_1(Y_r,\BV_r)}\right]  \nonumber \\
&\
	+\frac{1}{\sqrt{n}}\sum_{r=1}^{n}\left[\delta_{r2}\BS^*_{2r}(\bb)+\delta_{r3}\BS^*_{3r}(\bb)+\delta_{r4}\BS^*_{4r}(\bb)\right]  \nonumber \\
&\
	-\frac{1}{\sqrt{n}}\sum_{r=1}^{n}\delta_{r1}\left[\frac{\BS^*_{2r}(\bb)\pi_2(Y_r,\BV_r)}{\pi_1(Y_r,\BV_r)}
	+\frac{\BS^*_{3r}(\bb)\pi_3(Y_r,\BV_r)}{\pi_1(Y_r, \BV_r)}+\frac{\BS^*_{4r}(\bb)\pi_4(Y_r,\BV_r)}{\pi_1(Y_r,\BV_r)}\right]  \nonumber \\
&\
	+\BO_p(M^{-1/2})+\bo_p(1)  \nonumber \\
=&\
	\frac{1}{\sqrt{n}}\sum_{r=1}^{n}\delta_{r1}\BS_r(\bb)\left(\frac{1}{\pi_1(Y_r,\BV_r)}\right)
	+\frac{1}{\sqrt{n}}\sum_{r=1}^{n}\BS^*_{2r}(\bb)\left[\delta_{r2}-\pi_2(Y_r,\BV_r)\left(\frac{\delta_{r1}}{\pi_1(Y_r,\BV_r)}\right)\right]  \nonumber \\
&\
	+\frac{1}{\sqrt{n}}\sum_{r=1}^{n}\BS^*_{3r}(\bb)\left[\delta_{r3}-\pi_3(Y_r,\BV_r)\left(\frac{\delta_{r1}}{\pi_1(Y_r,\BV_r)}\right)\right]  \nonumber \\
&\
	+\frac{1}{\sqrt{n}}\sum_{r=1}^{n}\BS^*_{4r}(\bb)\left[\delta_{r4}-\pi_4(Y_r,\BV_r)
	\left(\frac{\delta_{r1}}{\pi_1(Y_r,\BV_r)}\right)\right]
	+\BO_p(M^{-1/2})+\bo_p(1)  \nonumber \\
=&\
\frac{1}{\sqrt{n}}\sum_{r=1}^{n}\bPhi_r(\bb,\bpi_r)+\BO_p(M^{-1/2})+\bo_p(1)  \nonumber \\
=&\
\BU_1(\bb,\bPi)+\BO_p(M^{-1/2})+\bo_p(1),
\end{align*}
which implies $\BU_{M1}(\bb)-\BU_1(\bb,\bPi)=\BO_p(M^{-1/2})+\bo_p(1)$, where 
	the expression of $\bPhi_r(\bb,\bpi_r)$ is presented in Equation~\eqref{eq:bPhi.mi1} and 
\begin{align}	
\label{eq:BU1bbBpi}
  \BU_1(\bb,\bPi)=\frac{1}{\sqrt{n}} \sum_{r=1}^{n}\bPhi_r(\bb,\bpi_r).
\end{align}
 Additionally, it can be shown that $\E\left[\BU_{M1}(\bb)\right]\to\b0$, as $n, M\to\infty$, and, hence, $\BU_{M1}(\bb)$ is an asymptotically unbiased estimating function. We finish the proof 
of Lemma~\ref{lemma} (i).

\subsubsection*{Proof of Lemma~\ref{lemma} (ii)}}
Similarly, the expression of $\BU_{M2}(\bb)$ in Equation~\eqref{eq: score mi2 new} can be reformulated as follows: 
\begin{align*}
    \BU_{M2}(\bb)
  &=\frac{1}{\sqrt{n}}\sum_{i=1}^{n}
	\left(\delta_{i1}\BS_i(\bb)+\delta_{i2}\ttBS_{2i}(\bb)+\delta_{i3}\ttBS_{3i}(\bb) +\delta_{i4}\ttBS_{4i}(\bb)\right)\\
  &=\frac{1}{\sqrt{n}}\sum_{i=1}^{n}\left[\delta_{i1}\BS_i(\bb)+\delta_{i2}\BS_i^*(\bb)+\delta_{i3}\BS_i^*(\bb)+\delta_{i4}\BS_i^*(\bb)\right]\\
	&\hskip 2mm
	+\frac{1}{\sqrt{n}}\sum_{i=1}^{n}\delta_{i2}\left[\ttBS_{2i}(\bb)-\E_{\ttF_{\TX_{1i}}}\left(\BS_i(\bb)|Y_i,\BV_i\right)\right] \\
	& \hskip 2mm
	+\frac{1}{\sqrt{n}}\sum_{i=1}^{n}\delta_{i3}\left[\ttBS_{3i}(\bb)-\E_{\ttF_{\TX_{2i}}}\left(\BS_i(\bb)|Y_i,\BV_i\right)\right]\\
	&\hskip 2mm
	+\frac{1}{\sqrt{n}}\sum_{i=1}^{n}\delta_{i4}\left[\ttBS_{4i}(\bb)-\E_{\ttF_{\BX_i}}\left(\BS_i(\bb)|Y_i,\BV_i\right)\right] \\
	&\hskip 2mm
	 +\frac{1}{\sqrt{n}}\sum_{i=1}^{n}\delta_{i2}\left[\E_{\ttF_{\TX_{1i}}}\left(\BS_i(\bb)|Y_i,\BV_i\right)-\BS_i^*(\bb)\right]\\
	&\hskip 2mm +\frac{1}{\sqrt{n}}\sum_{i=1}^{n}\delta_{i3}\left[\E_{\ttF_{\TX_{2i}}}\left(\BS_i(\bb)|Y_i,\BV_i\right)-\BS_i^*(\bb)\right] \\
	&\hskip 2mm
	+\frac{1}{\sqrt{n}}\sum_{i=1}^{n}\delta_{i4}\left[\E_{\ttF_{\BX_i}}\left(\BS_i(\bb)|Y_i,\BV_i\right)-\BS_i^*(\bb)\right],\\
\end{align*} 
where recall that $\BS_i^*(\bb)=\E(\BS_1(\bb)|Y_i,\BV_i)=\BS_{4i}^*(\bb)$. Firstly, we have 
\begin{align*}
	\E_{\ttF_{\TX_{1i}}}\left(\BS_i(\bb)|Y_i,\BV_i\right)
  = \sum_{k=1}^{n}\frac{(\delta_{k1}+\delta_{k3})I(Y_k=Y_i,\BV_k=\BV_i)\BS_k(\bb)}
	{\sum_{r=1}^{n}(\delta_{r1}+\delta_{r3})I(Y_r=Y_i,\bm{V}_r=\BV_i)}. 
\end{align*}
Then, based on the above expression and the expression of $\hpi_j(Y_i,\BV_i)$ in Equation~\eqref{def: pihat}, we can have    
\begin{align*}
	&\sum_{i=1}^{n}\delta_{i2}\E_{\ttF_{\TX_{1i}}}\left(\BS_i(\bb)|Y_i,\BV_i\right)\\
   =&\
   \sum_{i=1}^{n}\delta_{i2}\left(\sum_{k=1}^{n}\frac{(\delta_{k1}+\delta_{k3})I(Y_k=Y_i,\BV_k=\BV_i)\BS_k(\bb)}
	{\sum_{r=1}^{n}(\delta_{r1}+\delta_{r3})I(Y_r=Y_i,\BV_r=\BV_i)}\right)\\
	=&\
	\sum_{k=1}^{n}(\delta_{k1}+\delta_{k3})\BS_k(\bb)\left(\sum_{i=1}^{n}\frac{\delta_{i2}I(Y_i=Y_k,\BV_i=\BV_k)}
	{\sum_{r=1}^{n}(\delta_{r1}+\delta_{r3})I(Y_r=Y_k,\BV_r=\BV_k)}\right)\\
	=&\
	\sum_{k=1}^{n}(\delta_{k1}+\delta_{k3})\BS_k(\bb)
	  \left(\sum_{i=1}^{n}\frac{\delta_{i2}I(Y_i=Y_k,\BV_i=\BV_k)/\sum_{s=1}^{n}I(Y_s=Y_k,\BV_s=\BV_k)}
	{\sum_{r=1}^{n}(\delta_{r1}+\delta_{r3})I(Y_r=Y_k,\BV_r=\BV_k)/\sum_{s=1}^{n}I(Y_s=Y_k,\BV_s=\BV_k)}\right)\\
	=&\
	\sum_{k=1}^{n}(\delta_{k1}+\delta_{k3})\BS_k(\bb)
	  \left(\frac{\hpi_2(Y_k,\BV_k)}{\hpi_1(Y_k,\BV_k)+\hpi_3(Y_k,\BV_k)}\right).
\end{align*}
Similarly, it can yield
\begin{align*}
	\sum_{i=1}^{n}\delta_{i3}\E_{\ttF_{\TX_{2i}}}\left(\BS_i(\bb)|Y_i,\BV_i\right)
   &=\sum_{k=1}^{n}(\delta_{k1}+\delta_{k2})\BS_k(\bb)
       \left(\frac{\hpi_3(Y_k,\BV_k)}{\hpi_1(Y_k,\BV_k)+\hpi_2(Y_k,\BV_k)}\right), \\
	\sum_{i=1}^{n}\delta_{i4}\E_{\ttF_{\BX_i}}\left(\BS_i(\bb)|Y_i,\BV_i\right)
   &=\sum_{k=1}^{n}\delta_{k1}\BS_k(\bb)\left(\frac{\hpi_4(Y_k,\BV_k)}{\hpi_1(Y_k,\BV_k)}\right).
\end{align*}

Secondly, $n^{-1/2}\sum_{i=1}^{n}\delta_{i2}\big[\E_{\ttF_{\TX_{1i}}}\left(\BS_i(\bb)|Y_i,\BV_i\right)-\BS_i^*(\bb)\big]$ can be expressed as
\begin{align*}
	&
	\frac{1}{\sqrt{n}}\sum_{i=1}^{n}\delta_{i2}\left[\E_{\ttF_{\TX_{1i}}}\left(\BS_i(\bb)|Y_i,\BV_i\right)-\BS_i^*(\bb)\right]\\
	=&\
	\frac{1}{\sqrt{n}}\sum_{i=1}^{n}\delta_{i2}\left[\frac{\sum_{r=1}^{n}\left (\delta_{r1}+\delta_{r3}\right)I(Y_r=Y_i,\BV_r=\BV_i)\BS_r(\bb)}{\sum_{j=1}^{n}
		\left(\delta_{j1}+\delta_{j3}\right)I(Y_j=Y_i,\BV_j=\BV_i)}-\BS_i^*(\bb)\right]\\
	=&\
	\frac{1}{\sqrt{n}}\sum_{i=1}^{n}\delta_{i2}\left[\frac{\sum_{r=1}^{n}
		\left(\delta_{r1}+\delta_{r3}\right)I(Y_r=Y_i,\BV_r=\BV_i)\left(\BS_r(\bb)-\BS_i^*(\bb)\right)}
		 {\sum_{j=1}^{n}\left(\delta_{j1}+\delta_{j3}\right)I(Y_j=Y_i,\BV_j=\BV_i)}\right]\\
	=&\
	\frac{1}{\sqrt{n}}\sum_{i=1}^{n}\delta_{i2}\left[\frac{\sum_{r=1}^{n}
		\left(\delta_{r1}+\delta_{r3}\right)I(Y_i=Y_r,\BV_i=\BV_r)\left(\BS_r(\bb)-\BS_r^*(\bb)\right)}
	{\sum_{j=1}^{n}\left(\delta_{j1}+\delta_{j3}\right)I(Y_j=Y_r,\BV_j=\BV_r)}\right]\\
	=&\
	\frac{1}{\sqrt{n}}\sum_{r=1}^{n}
	\left(\delta_{r1}+\delta_{r3}\right)\left[\BS_r(\bb)-\BS_r^*(\bb)\right]\left[\frac{\sum_{i=1}^{n}\delta_{i2}I(Y_i=Y_r,\BV_i=\BV_r)}{\sum_{j=1}^{n}
		\left(\delta_{j1}+\delta_{j3}\right)I(Y_j=Y_r,\BV_j=\BV_r)}\right]\\
	=&\
	\frac{1}{\sqrt{n}}\sum_{r=1}^{n}\left(\delta_{r1}+\delta_{r3}\right)\left[\BS_r(\bb)-\BS_r^*(\bb)\right]\\
	 & \times\left[\frac{\sum_{i=1}^{n}\delta_{i2}I(Y_i=Y_r,\BV_i=\BV_r)/(\sum_{v=1}^{n}I(Y_v=Y_r,\BV_v=\BV_r))}
{\sum_{j=1}^{n}\left(\delta_{j1}+\delta_{j3}\right)I(Y_j=Y_r,\BV_j=\BV_r)/(\sum_{v=1}^{n}I(Y_v=Y_r,\BV_v=\BV_r))}\right]\\
	=&\
	\frac{1}{\sqrt{n}}\sum_{r=1}^{n}
	\left(\delta_{r1}+\delta_{r3}\right)\left[\BS_r(\bb)-\BS_r^*(\bb)\right]\left[\frac{\hpi_2(Y_r,\BV_r)}{\hpi_1(Y_r,\BV_r)+
		\hpi_3(Y_r,\BV_r)}\right].
\end{align*}
By a Taylor's expansion,  one can have
\begin{align*}
  &
  \frac{\hpi_2(Y_r,\BV_r)}{\hpi_1(Y_r,\BV_r)+\hpi_3(Y_r,\BV_r)}-\frac{\pi_2(Y_r,\BV_r)}{\pi_1(Y_r,\BV_r)+\pi_3(Y_r,\BV_r)}\\
 =&\
	\left(\frac{1}{\pi_1(Y_r,\BV_r)+\pi_3(Y_r,\BV_r)}\right)\left[\hpi_2(Y_r,\BV_r)-\pi_2(Y_r,\BV_r)\right]\\
  &\
  -\left(\frac{\pi_2(Y_r,\BV_r)}{(\pi_1(Y_r,\BV_r)+\pi_3(Y_r,\BV_r))^2}\right)
   \left[\hpi_1(Y_r,\BV_r)-\pi_1(Y_r,\BV_r)\right]\\
 &\
  -\left(\frac{\pi_2(Y_r,\BV_r)}{(\pi_1(Y_r,\BV_r)+\pi_3(Y_r,\BV_r))^2}\right)
    \left[\hpi_3(Y_r,\BV_r)-\pi_3(Y_r,\BV_r)\right]\\
 &\
  +O_p\left[\left(\hpi_1(Y_r,\BV_r)-\pi_1(Y_r,\BV_r)\right)^2\right]
  +O_p\left[\left(\hpi_2(Y_r,\BV_r)-\pi_2(Y_r,\BV_r)\right)^2\right]\\
 &\
	+
	O_p\left[\left(\hpi_3(Y_r,\BV_r)-\pi_3(Y_r,\BV_r)\right)^2\right].
\end{align*}
Note that if $\eta_i$ follows a Bernoulli distribution with a probability of $\kappa$,
 $i=1,2,\dots,n$, and $\wh\kappa={n}^{-1}\sum_{i=1}^{n}\eta_i$, then $\E[(\wh\kappa-\kappa)^4]=O(n^{-3})$. This implies $O\left((\wh\kappa-\kappa)^2\right)=o_p(n^{-1/2})$. Using this property, it can be shown that
$\E[(\hpi_k(Y_r,\BV_r)-\pi_k(Y_r,\BV_r))^2]=O(n^{-1})$ and
$O_p[(\hpi_k(Y_r,\BV_r)-\pi_k(Y_r,\BV_r))^2]=o_p(n^{-1/2})$, $k=1,2,3$, and, hence,
\begin{align*}
	&
\frac{\hpi_2(Y_r,\BV_r)}{\hpi_1(Y_r,\BV_r)+\hpi_3(Y_r,\BV_r)}-\frac{\pi_2(Y_r,\BV_r)}{\pi_1(Y_r,\BV_r)+\pi_3(Y_r,\BV_r)}\\
	=&\
	\left(\frac{1}{\pi_1(Y_r,\BV_r)+\pi_3(Y_r,\BV_r)}\right)\left[\hpi_2(Y_r,\BV_r)-\pi_2(Y_r,\BV_r)\right]\\
	&\
	-\left(\frac{\pi_2(Y_r,\BV_r)}{\left(\pi_1(Y_r,\BV_r)+\pi_3(Y_r,\BV_r)\right)^2}\right)
	\left[\hpi_1(Y_r,\BV_r)-\pi_1(Y_r,\BV_r)\right]\\
	&\
	-\left(\frac{\pi_2(Y_r,\BV_r)}{\left(\pi_1(Y_r,\BV_r)+\pi_3(Y_r,\BV_r)\right)^2}\right)
	\left[\hpi_3(Y_r,\BV_r)-\pi_3(Y_r,\BV_r)\right]+o_p\left(n^{-1/2}\right).
\end{align*}
Using the expression of $\hpi_k(Y_r,\BV_r)$ in Equation~\eqref{def: pihat}, $k=1, 2,3$, it can yield
\begin{align*}
	\hpi_k(Y_r,\BV_r)-\pi_k(Y_r,\BV_r)
	&=
	\frac{\sum_{i=1}^{n}\left(\delta_{ik}-\pi_k(Y_r,\BV_r)\right)I(Y_i=Y_r,\BV_i=\BV_r)}{\sum_{j=1}^{n}I(Y_j=Y_r,\BV_j=\BV_r)}\\
	&=
	\frac{1}{\sqrt{n}}\frac{n^{-1/2}\sum_{i=1}^{n}\left(\delta_{ik}-\pi_k(Y_r,\BV_r)\right)I(Y_i=Y_r,\BV_i=\BV_r)}{\frac{1}{n}\sum_{j=1}^{n}I(Y_j=Y_r,\BV_j=\BV_r)}\\
	&=
	\frac{1}{\sqrt{n}}\frac{n^{-1/2}\sum_{i=1}^{n}\left(\delta_{ik}-\pi_k(Y_r,\BV_r)\right)I(Y_i=Y_r,\BV_i=\BV_r)}
	  {P(Y=Y_r,\BV=\BV_r)}+o_p(n^{-1/2}).
\end{align*}

According to the CLT, we can show that
\begin{align*}
\frac{1}{\sqrt{n}}\sum_{i=1}^{n}\left(\delta_{ik}-\pi_k(Y_r,\BV_r)\right)I(Y_i=Y_r,\BV_i=\BV_r)\dovr \mN\left(0,\pi_k(Y_r,\BV_r)(1-\pi_k(Y_r,\BV_r))\right).
\end{align*}
Therefore, for $k=1,2,3$,
\begin{align*}
	\hpi_k(Y_r,\BV_r)-\pi_k(Y_r,\BV_r)=\frac{\sum_{i=1}^{n}\left(\delta_{ik}-\pi_k(Y_r,\BV_r)\right)I(Y_i=Y_r,\BV_i=\BV_r)}{\sum_{j=1}^{n}I(Y_j=Y_r,\BV_j=\BV_r)}=O_p(n^{-1/2}).
\end{align*}
Then, it can be shown that
\begin{align*}
&
\frac{1}{\sqrt{n}}\sum_{i=1}^{n}\delta_{i2}\left[\E_{\ttF_{\TX_{1i}}}\left(\BS_i(\bb)|Y_i,\BV_i\right)-\BS_i^*(\bb)\right]\\
  =&\
	\frac{1}{\sqrt{n}}\sum_{r=1}^{n}
	\left(\delta_{r1}+\delta_{r3}\right)\left[\BS_r(\bb)-\BS_r^*(\bb)\right]\left[\frac{\hpi_2(Y_r,\BV_r)}{\hpi_1(Y_r,\BV_r)+\hpi_3(Y_r,\BV_r)}\right]\\
	=&
	\frac{1}{\sqrt{n}}\sum_{r=1}^{n}
	\left(\delta_{r1}+\delta_{r3}\right)\left[\BS_r(\bb)-\BS_r^*(\bb)\right]\left[\frac{\pi_2(Y_r,\BV_r)}{\pi_1(Y_r,\BV_r)+
		\pi_3(Y_r,\BV_r)}\right]\\
	&
	+\frac{1}{\sqrt{n}}\sum_{r=1}^{n}
	\left(\delta_{r1}+\delta_{r3}\right)\left[\BS_r(\bb)-\BS_r^*(\bb)\right]\left[\frac{1}{\pi_1(Y_r,\BV_r)+
		\pi_3(Y_r,\BV_r)}\right]\left[\hpi_2(Y_r,\BV_r)-\pi_2(Y_r,\BV_r)\right]\\
	&
	-\frac{1}{\sqrt{n}}\sum_{r=1}^{n}
	\left(\delta_{r1}+\delta_{r3}\right)\left[\BS_r(\bb)-\BS_r^*(\bb)\right]\left[\frac{\pi_2(Y_r,\BV_r)}{(\pi_1(Y_r,\BV_r) +
		\pi_3(Y_r,\BV_r))^2}\right]\left[\hpi_1(Y_r,\BV_r)-\pi_1(Y_r,\BV_r)\right]\\
	&
	-\frac{1}{\sqrt{n}}\sum_{r=1}^{n}
	\left(\delta_{r1}+\delta_{r3}\right)\left[\BS_r(\bb)-\BS_r^*(\bb)\right]\left[\frac{\pi_2(Y_r,\BV_r)}{(\pi_1(Y_r,\BV_r)+
		\pi_3(Y_r,\BV_r))^2}\right]\left[\hpi_3(Y_r,\BV_r)-\pi_3(Y_r,\BV_r)\right]\\
	&
	+\bo_p(1)\\
	=&\
	\frac{1}{\sqrt{n}}\sum_{r=1}^{n}
	\left(\delta_{r1}+\delta_{r3}\right)\left[\BS_r(\bb)-\BS_r^*(\bb)\right]\left[\frac{\pi_2(Y_r,\BV_r)}{\pi_1(Y_r,\BV_r)+
		\pi_3(Y_r,\BV_r)}\right]+\bo_p(1).
\end{align*}
Similarly, one can show
\begin{align*}
	&
	\frac{1}{\sqrt{n}}\sum_{i=1}^{n}\delta_{i3}\left[\E_{\ttF_{\TX_{2i}}}\left(\BS_i(\bb)|Y_i,\BV_i\right)-\BS_i^*(\bb)\right]\\
	=&\
	\frac{1}{\sqrt{n}}\sum_{i=1}^{n}(\delta_{i1}+\delta_{i2})\left[\BS_i(\bb)-\BS_i^*(\bb)\right]\left [\frac{\pi_3(Y_i,\BV_i)}{\pi_{1}(Y_i,\BV_i)+\pi_2(Y_i,\BV_i)}\right]+\bo_p(1),\\
	&\
	\frac{1}{\sqrt{n}}\sum_{i=1}^{n}{\delta_{i4}}\left[\E_{\ttF_{\BX_i}}\left(\BS_i(\bb)|Y_i,\BV_i\right)-\BS_i^*(\bb)\right]\\
	=&\
	\frac{1}{\sqrt{n}}\sum_{i=1}^{n}\delta_{i1}\left[\BS_i(\bb)-\BS_i^*(\bb)\right]\left [\frac{\pi_4(Y_i,\BV_i)}{\pi_1(Y_i,\BV_i)}\right]+\bo_p(1).
\end{align*}
Therefore, $\BU_{M2}(\bb)$ can be rewritten as follows:
\begin{align*}
 \BU_{M2}(\bb)
  &=\frac{1}{\sqrt{n}}\sum_{i=1}^{n}
	\left(\delta_{i1}\BS_i(\bb)+\delta_{i2}\ttBS_{2i}(\bb) +\delta_{i3}\ttBS_{3i}(\bb) +\delta_{i4}\ttBS_{4i}(\bb)\right) \nonumber \\
  &=\frac{1}{\sqrt{n}}\sum_{i=1}^{n}\left[\delta_{i1}\BS_i(\bb)+\delta_{i2}\BS_i^*(\bb)+\delta_{i3}\BS_i^*(\bb)+\delta_{i4}\BS_i^*(\bb)\right] \nonumber \\
  &\hskip 2mm
	+\frac{1}{\sqrt{n}}\sum_{i=1}^{n}\left (\delta_{i1}+\delta_{i3}\right)\left[\BS_i(\bb)-\BS_i^*(\bb)\right]\left[\frac{\pi_2(Y_i,\BV_i)}{\pi_1(Y_i,\BV_i)+
		\pi_3(Y_i,\BV_i)}\right] \nonumber \\
  &\hskip 2mm
	+\frac{1}{\sqrt{n}}\sum_{i=1}^{n}(\delta_{i1}+\delta_{i2})\left[\BS_i(\bb)-\BS_i^*(\bb)\right]
	\left[\frac{\pi_{3}(Y_i,\BV_i)}{\pi_{1}(Y_i,\BV_i)+\pi_{2}(Y_i,\BV_i)}\right] \nonumber \\
	&\hskip 2mm
	+\frac{1}{\sqrt{n}}\sum_{i=1}^{n}\delta_{i1}\left[\BS_i(\bb)-\BS_i^*(\bb)\right]
	\left[\frac{\pi_{4}(Y_i,\BV_i)}{\pi_{1}(Y_i,\BV_i)}\right]
	+\BO_p(M^{-1/2})+\bo_p(1)  \nonumber \\
 &=\frac{1}{\sqrt{n}}\sum_{i=1}^n\bPsi_i(\bb,\bpi_i)+\BO_p(M^{-1/2})+\bo_p(1) \nonumber \\
 &=\BU_2(\bb,\bPi)+\BO_p(M^{-1/2})+\bo_p(1), 
\end{align*}
which implies $\BU_{M2}(\bb)-\BU_{2}(\bb,\bPi)=\BO_p(M^{-1/2})+\bo_p(1)$, where
	the expression of $\bPsi_i(\bb,\bpi_i)$ is given in Equation~\eqref{eq:bPsi.mi2}
and 
\begin{align}
\label{eq:BU2bbbPi}
\BU_2(\bb,\bPi)=\dfrac{1}{\sqrt{n}}\sum_{i=1}^{n}\bPsi_i(\bb,\bpi_i).
\end{align}

Note that by utilizing the following relationship: 
\begin{align*}
&\delta_{i1}\BS_i(\bb)+\delta_{i2}\BS_i^*(\bb)+\delta_{i3}\BS_i^*(\bb)+\delta_{i4}\BS_i^*(\bb) \\
=&\
\delta_{i1}\BS_i(\bb)-\delta_{i1}\BS_i^*(\bb)+\left[\delta_{i1}\BS_i^*(\bb)+\delta_{i2}\BS_i^*(\bb)+\delta_{i3}\BS_i^*(\bb)+\delta_{i4}\BS_i^*(\bb)\right] \\
=&\ 
\delta_{i1}\left[\BS_i(\bb)-\BS_i^*(\bb)\right]+\BS_i^*(\bb),
\end{align*}
 $\bPsi_i(\bb,\bpi_i)$ can also be rewritten as
\begin{align*}
	\bPsi_i(\bb,\bpi_i)
	&=\BS_i^*(\bb)
	+\left(\delta_{i1}+\delta_{i3}\right)\left[\BS_i(\bb)-\BS_i^*(\bb)\right]
	    \left[\frac{\pi_2(Y_i,\BV_i)}{\pi_1(Y_i,\BV_i)+\pi_3(Y_r,\BV_r)}\right]\\
	&\hskip 2mm
	+(\delta_{i1}+\delta_{i2})\left[\BS_i(\bb)-\BS_i^*(\bb)\right]
	\left[\frac{\pi_3(Y_i,\BV_i)}{\pi_1(Y_i,\BV_i)+\pi_2(Y_i,\BV_i)}\right]\\
	&\hskip 2mm
	+\delta_{i1}\left[\BS_i(\bb)-\BS_i^*(\bb)\right]\left[\frac{\pi_4(Y_i,\BV_i)}{\pi_1(Y_i,\BV_i)}+1\right].
\end{align*}
If $\pi_2(Y_i,\BV_i)=\pi_3(Y_i,\BV_i)=0$ for all $i=1,2,\dots,n$, then $\delta_{i2}=\delta_{i3}=0$ and $\pi_4(Y_i,\BV_i)=1-\pi_1(Y_i,\BV_i)$, 
and, hence, $\BX_i=(\TX_{1i},\TX_{2i})$  degenerates 
to only missing simultaneously. In this case, $\bPsi_i(\bb,\bpi_i)$ can be simplified as
\begin{align*}
	\bPsi_i(\bb,\bpi_i)
	&=\BS_i^*(\bb)+\delta_{i1}\left[\BS_i(\bb)-\BS_i^*(\bb)\right]\left[\frac{\pi_4(Y_i,\BV_i)}{\pi_{1}(Y_i,\BV_i)}+1\right]\\
	&=\BS_i^*(\bb)+\delta_{i1}\left[\BS_i(\bb)-\BS_i^*(\bb)\right]\left[\frac{1-\pi_1(Y_i,\BV_i)}{\pi_{1}(Y_i,\BV_i)}+1\right]\\
	&=\BS_i^*(\bb)+\delta_{i1}\left[\BS_i(\bb)-\BS_i^*(\bb)\right]\left[\frac{1}{\pi_1(Y_i,\BV_i)}\right]\\
	&=\left[\frac{\delta_{i1}}{\pi_1(Y_i,\BV_i)}\right]\BS_i(\bb)
	 +\left[1-\frac{\delta_{i1}}{\pi_{1}(Y_i,\BV_i)}\right]\BS_i^*(\bb).
\end{align*}
This result corresponds to that of \cite{wang2001note}. 
The proof of Lemma~\ref{lemma} (ii) is complete.

\subsection*{Proof of Theorem~\ref{theorem 1}}
Let 
\begin{align*}
	\BG_{M1}(\bb)=-\frac{1}{\sqrt{n}}\left[\frac{\partial\BU_{M1}(\bb)}{\partial\bb}\right].
\end{align*}
It can then be shown via the result in Lemma~\ref{lemma} (i) that
\begin{align*}
	\BG_{M1}(\bb)
  &=-\frac{1}{\sqrt{n}}\left[\frac{\partial\BU_{1}(\bb,\bPi)}{\partial\bb}\right]
	+\BO_p(M^{-1/2}n^{-1/2})+\bo_p(n^{-1/2}) \\
  &\povr\E\left[\mX_1^{\otimes2}H^{(1)}(\bb^{\tT}\mX_1)\right]=\BG(\bb),\ \text{as}\ n, M \to\infty.
\end{align*}
Based on condition (C3), one can show that the convergence in probability of $\BG_{M1}(\bb)$ to $\BG(\bb)$ is uniform in a neighborhood of the true value of $\bb$. 
By the inverse function theorem of \cite{foutz1997unique}, along with condition (C2), there exists a unique consistent solution of $\BU_1(\bb,\bPi)=\b0$ for a fixed $\bPi$. 
Let $\hbb(\bPi)$ and $\hbb(\hbPi)$ be the solutions of $\BU_1(\bb,\bPi)=\b0$ and
$\BU_1(\bb,\hbPi)=\b0$, respectively. This implies $\hbb(\bPi)\povr\bb$. Because $\hbpi_i$ is a consistent estimator of $\bpi_i$, one can obtain
\begin{align*}
	\BU_1(\bb,\bPi)-\BU_1(\bb,\hbPi)\povr\b0, \\
	\hbb(\hbPi)-\hbb(\bPi)\povr\b0.
\end{align*}
Therefore, it can yield $\hbb(\hbPi)\povr\bb$. However, $\hbb_{M1}$ and $\hbb(\hbPi)$ are the solutions  of $\BU_{M1}(\bb)=\b0$ and $\BU_1(\bb,\hbPi)=\b0$, respectively, so 
\begin{align*}
\b0=\BU_{M1}(\hbb_{M1})
&=\BU_1(\hbb_{M1},\bPi)+\BO_p(M^{-1/2})+\bo_p(1)\\
&=\BU_1(\bb,\bPi)-\BG(\bb)\sqrt{n}\big(\hbb_{M1}-\bb\big)+\BO_p(M^{-1/2})+\bo_p(1)
\end{align*}
and 
\begin{align*}
 \b0=\BU_{1}(\hbb(\bPi),\bPi)=\BU_1(\bb,\bPi)-\BG(\bb)\sqrt{n}\big(\hbb(\bPi)-\bb\big)+\bo_p(1),
\end{align*}
which lead to
\begin{align*}
\b0=\BG(\bb)\sqrt{n}\big(\hbb(\bPi)-\hbb_{M1}\big)+\BO_p(M^{-1/2})+\bo_p(1).
\end{align*}
Hence, it can be shown that
\begin{align*}
	\sqrt{n}\big(\hbb(\bPi)-\hbb_{M1}\big)\povr\b0,\ \text{as}\ n, M\to\infty.
\end{align*}
This means that $\hbb(\bPi)$ and $\hbb_{M1}$ are asymptotically equivalent. Thus, $\hbb_{M1}$ is also a consistent estimator of $\bb$. 

In addition, because
\begin{align*}
	\b0=\BU_{M1}(\hbb_{M1})
	&=\BU_1(\hbb_{M1},\bm{\Pi})+\BO_p(M^{-1/2})+\bo_p(1)\\
	&=\BU_1(\bb,\bPi)-\BG(\bb)\sqrt{n}(\hbb_{M1}-\bb)+\BO_p(M^{-1/2})+\bo_p(1),
\end{align*}
by using $\BU_1(\bb,\bPi)=n^{-1/2}\sum_{i=1}^{n}\bPhi_i(\bb,\bpi_i)$ in Equation~\eqref{eq:BU1bbBpi},
	from the above expression, $\sqrt{n}(\hbb_{M1}-\bb)$ can be expressed as follows:
\begin{align*}
 \sqrt{n}(\hbb_{M1}-\bb)
 &=\left[\BG(\bb)\right]^{-1}\BU_1(\bb,\bPi)+\BO_p(M^{-1/2})+\bo_p(1) \\
 &=\left[\BG(\bb)\right]^{-1}\frac{1}{\sqrt{n}}\sum_{i=1}^{n}\bPhi_i(\bb,\bpi_i)+\BO_p(M^{-1/2})+\bo_p(1).
\end{align*}
According to the CLT, one can show
\begin{align*}
 \frac{1}{\sqrt{n}}\sum_{i=1}^{n}\bPhi_i(\bb,\bpi_i)\dovr\mN_{p+1}\left(\b0,\BM_1(\bb,\bpi_1)\right),
\end{align*}
where $\BM_1(\bb,\bpi_1)=\E\left[\bPhi_1^{\otimes 2}(\bb,\bpi_1)\right]$.
Finally, it can establish that
\begin{align*}
	\sqrt{n}\left(\hbb_{M1}-\bb\right)\dovr\mN_{p+1}\left(\b0,\BG^{-1}(\bb)\BM_1(\bb,\bpi_1)\left[\BG^{-1}(\bb)\right]^{\tT}\right),\
	\text{as}\ n, M\to\infty.
\end{align*}
Therefore, the proof of Theorem~\ref{theorem 1} is complete.

\subsection*{Proofs of Theorem~\ref{theorem 2}}
Let
\begin{align*}
	\BG_{M2}(\bb)=-\frac{1}{\sqrt{n}}\left[\frac{\partial \BU_{M2}(\bb)}{\partial\bb}\right].
\end{align*}
By employing the result in Lemma~\ref{lemma} (ii), we can show
\begin{align*}
	\BG_{M2}(\bb)
	&=-\frac{1}{\sqrt{n}}\left[\frac{\partial\BU_2(\bb,\bPi)}{\partial\bb}\right]
	+\BO_p(M^{-1/2}n^{-1/2})+\bo_p(n^{-1/2})\\
	&\povr\E\left[\mX_1^{\otimes2}H^{(1)}(\bb^{\tT}\mX_1)\right]=\BG(\bb),\ \text{as}\ n, M \to\infty.
\end{align*}
Given condition (C3), it can be shown that the convergence of $\BG_{M2}(\bb)$ to $\BG(\bb)$ in probability is uniform across a neighborhood of the true value of $\bb$. 
Based on  the inverse function theorem of \cite{foutz1997unique} and condition (C2), there exists a unique consistent solution of $\BU_2(\bb,\bPi)=\b0$ for a fixed $\bPi$. 
Assume that $\hbb(\bPi)$ and $\hbb(\hbPi)$ are the solutions of $\BU_2(\bb,\bPi)=\b0$ and   
$\BU_2(\bb,\hbPi)=\b0$, respectively. This implies $\hbb(\bPi)\povr\bb$.
$\hbPi$ is a consistent estimator of $\bPi$, so we can establish that
\begin{align*}
	\BU_2(\bb,\bPi)-\BU_2(\bb,\hbPi)\povr\b0, \\
	\hbb(\hbPi)-\hbb(\bPi)\povr\b0.
\end{align*}
Accordingly, one can obtain $\hbb(\hbPi)\povr\bb$. However, because $\hbb_{M2}$ and $\hbb(\bPi)$ are the solutions of $\BU_{M2}(\bb)=\b0$ and $\BU_2(\bb,\bPi)=\b0$, respectively, 
\begin{align*}
	\b0=\BU_{M2}(\hbb_{M2})
	&=\BU_2(\hbb_{M2},\bPi)+\BO_p(M^{-1/2})+\bo_p(1)\\
	&=\BU_2(\bb,\bPi)-\BG(\bb)\sqrt{n}(\hbb_{M2}-\bb)+\BO_p(M^{-1/2})+\bo_p(1)
\end{align*}
and 
\begin{align*}
  \b0=\BU_2\big(\hbb(\bPi),\bPi\big)
  =\BU_2(\bb,\bPi)-\BG(\bb)\sqrt{n}\big(\hbb(\bPi)-\bb\big)+\bo_p(1).
\end{align*}
Then, from the above two expressions, we can have
\begin{eqnarray*}
	\b0=\bm{G}(\bb)\sqrt{n}\left (\hbb(\bPi)-\hbb_{M2}\right)+\BO_p(M^{-1/2})+\bo_p(1).
\end{eqnarray*}
Thus, it can be established that $\hbb(\bPi)$ and $\hbb_{M2}$ are asymptotically equivalent, expressed as follows:
\begin{eqnarray*}
	\sqrt{n}\left (\hbb(\bPi)-\hbb_{M2}\right)\povr\b0,\ \text{as}\ n, M\to\infty,
\end{eqnarray*}
which implies that $\hbb_{M2}$ is also a consistent estimator of $\bb$. 

Furthermore,
\begin{align*}
   \b0=\BU_{M2}(\hbb_{M2})
	&=\BU_2(\hbb_{M2},\bPi)+\BO_p(M^{-1/2})+\bo_p(1)\\
	&=\BU_2\bb,\bPi)-\BG(\bb)\sqrt{n}(\hbb_{M2}-\bb)+\BO_p(M^{-1/2})+\bo_p(1),
\end{align*}
so from the  above expression and $\BU_2(\bb,\bPi)=n^{-1/2}\sum_{i=1}^{n}\bPsi_i(\bb,\bpi_i)$ in Equation~\eqref{eq:BU2bbbPi}, we can write $\sqrt{n}(\hbb_{M2}-\bb)$ as follows:
\begin{align*}
  \sqrt{n}(\hbb_{M2}-\bb)
  &=\left[\BG(\bb)\right]^{-1}\BU_2(\bb,\bPi)+\BO_p(M^{-1/2})+\bo_p(1)\\
  &=\left[\BG(\bb)\right]^{-1}\frac{1}{\sqrt{n}}\sum_{i=1}^{n}\bPsi_i(\bb,\bpi_i)+\BO_p(M^{-1/2})+\bo_p(1).
\end{align*}
It can be shown via the CLT that
\begin{align*}
  \frac{1}{\sqrt{n}}\sum_{i=1}^{n}\bPsi_i(\bb,\bpi_i)\dovr\mN_{p+1}\left(\b0,\BM_2(\bb,\bpi_1)\right),
\end{align*}
where
$\BM_2(\bb,\bpi_1)=\E\left[\bPsi_1^{\otimes2}(\bb,\bpi_1)\right]$.
Finally, we can show
\begin{align*}
 \sqrt{n}(\hbb_{M2}-\bb)\dovr\mN_{p+1}\left(\b0,\BG^{-1}(\bb)\BM_2(\bb,\bpi_1)\left[\BG^{-1}(\bb)\right]^{\tT}\right),\
	\text{as}\ n, M\to\infty
\end{align*}
to complete the proof of Theorem~\ref{theorem 2}.

\newpage
\clearpage
\vspace{-0.8cm}
\begin{table}[h!]
	{\tiny
	\caption{Simulation results of Study 1 ($M=15$; $n=1,000$);
			$\bb=(-1,1,0.7,-1)^{\tT}$; $\bga=(0.7,-0.2,0.1,-1.2)^{\tT}$; three sets of $\ba$ are $(2.6,0.6,0.6)^{\tT}$, $(1.6,0.6,0.6)^{\tT}$, and $(0.8,0.6,0.6)^{\tT}$ 
			to create observed selection probabilities: 
			$(0.72,0.10,0.10,0.08)$, $(0.48,0.18,0.18,0.16)$, and $(0.30,0.24,0.24,0.22)$, respectively}
		\vspace{-0.8cm}
		\setlength\tabcolsep{10.0 pt} 
		\begin{center}
			\begin{tabular}{llrrrrrrrrr}
				\hline  \\ [-0.7em]
		&&$\hbb_F$&$\hbb_C$&$\hbb_{W}$&$\hbb_{M1}$&$\hbb_{M2}$&$\hbb_{M1n}$&$\hbb_{M2n}$&$\hbb_{RF}$&$\hbb_{EM}$\\
				\hline  \\ [-0.7em]
				\multicolumn{11}{c}{Observed selection probabilities: $(0.72, 0.10, 0.10, 0.08)$; $\ba=(2.6,0.6,0.6)^{\tT}$}\\
				$\beta_0$				
				&Bias& -0.0067 & 0.0176 & -0.0073 & -0.0070 & -0.0071 & -0.0070 & -0.0071 & 0.0190 & -0.0092 \\ 
				&SD  & 0.1022 & 0.1229 & 0.1037 & 0.1028 & 0.1030 & 0.1028 & 0.1030 & 0.1019 & 0.1063 \\ 
				&ASE & 0.1029 & 0.1187 & 0.1042 & 0.1035 & 0.1035 & 0.1035 & 0.1039 & 0.1045 & 0.1056 \\ 
				&MSE & 0.0105 & 0.0154 & 0.0108 & 0.0106 & 0.0107 & 0.0106 & 0.0107 & 0.0107 & 0.0114 \\ 
				&CP  & 0.9520 & 0.9430 & 0.9510 & 0.9490 & 0.9520 & 0.9490 & 0.9520 & 0.9490 & 0.9390 \\ 
				
				\cline{2-11}  \\ [-0.7em]
				$\beta_1$		
				&Bias& 0.0088 & 0.0214 & 0.0110 & 0.0095 & 0.0095 & 0.0095 & 0.0095 & -0.0610 & 0.0205 \\ 
				&SD  & 0.1599 & 0.1886 & 0.1684 & 0.1649 & 0.1649 & 0.1649 & 0.1649 & 0.1635 & 0.1779 \\ 
				&ASE & 0.1608 & 0.1875 & 0.1685 & 0.1645 & 0.1646 & 0.1651 & 0.1671 & 0.1754 & 0.1759 \\ 
				&MSE & 0.0256 & 0.0360 & 0.0285 & 0.0273 & 0.0273 & 0.0273 & 0.0273 & 0.0305 & 0.0321 \\ 
				&CP  & 0.9620 & 0.9500 & 0.9500 & 0.9510 & 0.9480 & 0.9500 & 0.9530 & 0.9490 & 0.9570 \\  
				
				\cline{2-11}   \\ [-0.7em]
				$\beta_2$		
				&Bias& 0.0111 & 0.0086 & 0.0105 & 0.0110 & 0.0107 & 0.0110 & 0.0107 & -0.0544 & 0.0089 \\ 
				&SD  & 0.1577 & 0.1863 & 0.1668 & 0.1646 & 0.1646 & 0.1646 & 0.1646 & 0.1589 & 0.1752 \\ 
				&ASE & 0.1557 & 0.1814 & 0.1616 & 0.1584 & 0.1586 & 0.1587 & 0.1606 & 0.1693 & 0.1703 \\ 
				&MSE & 0.0250 & 0.0348 & 0.0279 & 0.0272 & 0.0272 & 0.0272 & 0.0272 & 0.0282 & 0.0308 \\ 
				&CP  & 0.9560 & 0.9450 & 0.9540 & 0.9500 & 0.9500 & 0.9510 & 0.9540 & 0.9490 & 0.9530 \\ 
				
				\cline{2-11}  \\ [-0.7em]
				$\beta_3$				
				&Bias& -0.0148 & 0.0468 & -0.0155 & -0.0149 & -0.0149 & -0.0149 & -0.0149 & 0.0011 & -0.0119 \\ 
				&SD  & 0.1701 & 0.2035 & 0.1709 & 0.1709 & 0.1710 & 0.1709 & 0.1710 & 0.1695 & 0.1724 \\ 
				&ASE & 0.1676 & 0.1983 & 0.1683 & 0.1680 & 0.1680 & 0.1678 & 0.1682 & 0.1679 & 0.1688 \\ 
				&MSE & 0.0292 & 0.0436 & 0.0294 & 0.0294 & 0.0295 & 0.0294 & 0.0295 & 0.0287 & 0.0299 \\ 
				&CP  & 0.9430 & 0.9420 & 0.9420 & 0.9420 & 0.9410 & 0.9410 & 0.9430 & 0.9460 & 0.9420 \\   
				
				\hline  \\ [-0.7em]
				\multicolumn{11}{c}{Observed selection probabilities: $(0.48, 0.18, 0.18, 0.16)$; $\ba=(1.6,0.6,0.6)^{\tT}$}\\
				$\beta_0$		
				&Bias& -0.0067 & 0.0401 & -0.0081 & -0.0067 & -0.0064 & -0.0067 & -0.0064 & 0.0422 & -0.0084 \\ 
				&SD  & 0.1022 & 0.1454 & 0.1075 & 0.1045 & 0.1043 & 0.1045 & 0.1043 & 0.1008 & 0.1083 \\ 
				&ASE & 0.1029 & 0.1418 & 0.1064 & 0.1040 & 0.1040 & 0.1044 & 0.1049 & 0.1061 & 0.1086 \\ 
				&MSE & 0.0105 & 0.0228 & 0.0116 & 0.0110 & 0.0109 & 0.0110 & 0.0109 & 0.0120 & 0.0118 \\ 
				&CP  & 0.9520 & 0.9360 & 0.9480 & 0.9500 & 0.9500 & 0.9510 & 0.9540 & 0.9420 & 0.9470 \\ 
				
				\cline{2-11}   \\ [-0.7em]
				$\beta_1$		
				&Bias& 0.0088 & 0.0148 & 0.0092 & 0.0070 & 0.0066 & 0.0070 & 0.0066 & -0.1276 & 0.0195 \\ 
				&SD  & 0.1599 & 0.2303 & 0.1847 & 0.1737 & 0.1723 & 0.1737 & 0.1723 & 0.1672 & 0.1958 \\ 
				&ASE & 0.1608 & 0.2259 & 0.1807 & 0.1672 & 0.1674 & 0.1707 & 0.1736 & 0.1892 & 0.1932 \\ 
				&MSE & 0.0256 & 0.0533 & 0.0342 & 0.0302 & 0.0297 & 0.0302 & 0.0297 & 0.0442 & 0.0387 \\ 
				&CP  & 0.9620 & 0.9440 & 0.9470 & 0.9480 & 0.9480 & 0.9510 & 0.9550 & 0.9110 & 0.9410 \\ 
				
				\cline{2-11}   \\ [-0.7em]
				$\beta_2$				
				&Bias& 0.0111 & 0.0088 & 0.0178 & 0.0134 & 0.0116 & 0.0134 & 0.0116 & -0.1074 & 0.0069 \\ 
				&SD  & 0.1577 & 0.2169 & 0.1735 & 0.1666 & 0.1659 & 0.1666 & 0.1659 & 0.1574 & 0.1894 \\ 
				&ASE & 0.1557 & 0.2186 & 0.1706 & 0.1602 & 0.1606 & 0.1620 & 0.1648 & 0.1809 & 0.1869 \\ 
				&MSE & 0.0250 & 0.0471 & 0.0304 & 0.0279 & 0.0277 & 0.0279 & 0.0277 & 0.0363 & 0.0359 \\ 
				&CP  & 0.9560 & 0.9530 & 0.9490 & 0.9470 & 0.9500 & 0.9510 & 0.9550 & 0.9320 & 0.9500 \\ 
				
				\cline{2-11}  \\ [-0.7em]
				$\beta_3$			
				&Bias& -0.0148 & 0.0878 & -0.0188 & -0.0147 & -0.0145 & -0.0147 & -0.0145 & 0.0109 & -0.0122 \\ 
				&SD  & 0.1701 & 0.2460 & 0.1754 & 0.1718 & 0.1717 & 0.1718 & 0.1717 & 0.1685 & 0.1750 \\ 
				&ASE & 0.1676 & 0.2425 & 0.1696 & 0.1682 & 0.1682 & 0.1675 & 0.1687 & 0.1679 & 0.1699 \\ 
				&MSE & 0.0292 & 0.0682 & 0.0311 & 0.0297 & 0.0297 & 0.0297 & 0.0297 & 0.0285 & 0.0308 \\ 
				&CP  & 0.9430 & 0.9300 & 0.9410 & 0.9410 & 0.9400 & 0.9390 & 0.9410 & 0.9440 & 0.9390 \\ 
				
				\hline \\ [-0.7em]
				\multicolumn{11}{c}{Observed selection probabilities: $(0.30, 0.24, 0.24, 0.22)$; $\ba=(0.8,0.6,0.6)^{\tT}$}\\
				$\beta_0$	
				&Bias& -0.0067 & 0.0560 & -0.0139 & -0.0097 & -0.0079 & -0.0097 & -0.0079 & 0.0583 & -0.0114 \\ 
				&SD  & 0.1022 & 0.1801 & 0.1118 & 0.1064 & 0.1052 & 0.1064 & 0.1052 & 0.1009 & 0.1120 \\ 
				&ASE & 0.1029 & 0.1793 & 0.1107 & 0.1044 & 0.1044 & 0.1056 & 0.1059 & 0.1075 & 0.1123 \\ 
				&MSE & 0.0105 & 0.0356 & 0.0127 & 0.0114 & 0.0111 & 0.0114 & 0.0111 & 0.0136 & 0.0127 \\ 
				&CP  & 0.9520 & 0.9370 & 0.9530 & 0.9530 & 0.9530 & 0.9540 & 0.9550 & 0.9250 & 0.9540 \\ 
				
				\cline{2-11}   \\ [-0.7em]
				$\beta_1$			
				&Bias& 0.0088 & 0.0383 & 0.0262 & 0.0162 & 0.0122 & 0.0162 & 0.0122 & -0.1708 & 0.0373 \\ 
				&SD  & 0.1599 & 0.2807 & 0.2040 & 0.1840 & 0.1761 & 0.1840 & 0.1761 & 0.1684 & 0.2095 \\ 
				&ASE & 0.1608 & 0.2884 & 0.2032 & 0.1694 & 0.1697 & 0.1796 & 0.1797 & 0.2020 & 0.2127 \\ 
				&MSE & 0.0256 & 0.0803 & 0.0423 & 0.0341 & 0.0312 & 0.0341 & 0.0312 & 0.0576 & 0.0453 \\ 
				&CP  & 0.9620 & 0.9610 & 0.9490 & 0.9240 & 0.9420 & 0.9430 & 0.9600 & 0.8970 & 0.9490 \\ 
				
				\cline{2-11}   \\ [-0.7em]
				$\beta_2$			
				&Bias& 0.0111 & 0.0317 & 0.0298 & 0.0209 & 0.0159 & 0.0209 & 0.0159 & -0.1441 & 0.0165 \\ 
				&SD  & 0.1577 & 0.2903 & 0.1988 & 0.1786 & 0.1719 & 0.1786 & 0.1719 & 0.1627 & 0.2130 \\ 
				&ASE & 0.1557 & 0.2793 & 0.1865 & 0.1614 & 0.1622 & 0.1665 & 0.1682 & 0.1906 & 0.2059 \\ 
				&MSE & 0.0250 & 0.0853 & 0.0404 & 0.0323 & 0.0298 & 0.0323 & 0.0298 & 0.0472 & 0.0456 \\ 
				&CP  & 0.9560 & 0.9470 & 0.9350 & 0.9300 & 0.9420 & 0.9360 & 0.9480 & 0.9040 & 0.9430 \\  
				
				\cline{2-11}  \\ [-0.7em]
				$\beta_3$
				&Bias& -0.0148 & 0.1000 & -0.0398 & -0.0164 & -0.0155 & -0.0164 & -0.0155 & 0.0183 & -0.0116 \\ 
				&SD  & 0.1701 & 0.3142 & 0.1889 & 0.1719 & 0.1715 & 0.1719 & 0.1715 & 0.1668 & 0.1747 \\ 
				&ASE & 0.1676 & 0.3134 & 0.1727 & 0.1684 & 0.1684 & 0.1661 & 0.1693 & 0.1679 & 0.1710 \\ 
				&MSE & 0.0292 & 0.1087 & 0.0373 & 0.0298 & 0.0297 & 0.0298 & 0.0297 & 0.0282 & 0.0307 \\ 
				&CP  & 0.9430 & 0.9270 & 0.9150 & 0.9400 & 0.9390 & 0.9380 & 0.9400 & 0.9490 & 0.9430 \\ 
				\hline  \\ [-0.7em]
				\multicolumn{11}{l}{$\hbb_{M1}$: MI1 estimator whose value is the solution of $\BU_{M1}(\bb)=\b0$ in Equation~\eqref{eq: score mi1} with Rubin's type }\\
				\multicolumn{11}{l}{estimated variance in Equation~\eqref{MI1: asymptotic var}} \\
				\multicolumn{11}{l}{$\hbb_{M2}$: MI2 estimator whose value is the solution of $\BU_{M2}(\bb)=\b0$ in Equation~\eqref{eq: score mi2} with Rubin's type}\\
				\multicolumn{11}{l}{estimated variance in Equation~\eqref{MI2: asymptotic var}} \\
				\multicolumn{11}{l}{$\hbb_{M1n}$: MI1 estimator whose value is the solution of $\BU_{M1}(\bb)=\b0$ in Equation~\eqref{eq: score mi1} with proposed}\\
				\multicolumn{11}{l}{estimated variance in Equation~\eqref{MI1n: asymptotic var}} \\
				\multicolumn{11}{l}{$\hbb_{M2n}$: MI2 estimator whose value is the solution of $\BU_{M2}(\bb)=\b0$ in Equation~\eqref{eq: score mi2} with proposed} \\
				\multicolumn{11}{l}{estimated variance in Equation~\eqref{MI2n: asymptotic var}}
			\end{tabular}
		\end{center}
		\label{tab: simulation study 1}
	}
\end{table}

\newpage
\clearpage
\begin{table}[h!] 
	{\footnotesize
		\caption{Relative efficiencies in Study 1 ($M=15$; $n=1,000$); $\bb=(-1,1,0.7,-1)^{\tT}$; $\bga=(0.7,-0.2,0.1,-1.2)^{\tT}$}
		\vspace{-0.8cm}
		\setlength\tabcolsep{5.5 pt} 
		\begin{center}
			\begin{tabular}{cccccccccccccc}
				\hline  \\[-0.9em]
				\multicolumn{14}{c}{Relative efficiency}\\
				\hline \\[-0.9em]
				
				$\bb$&$C1$&$W1$&$M11$ & $M21$ & $RF1$&$EM1$& $C2$&$W2$&$M12$ & $M22$ & $RF2$&$EM2$&$M12n$\\
				\hline  \\[-0.9em]
				\multicolumn{14}{c}{Observed selection probabilities: $(0.72,0.10,0.10,0.08)$; $\ba=(2.6,0.6,0.6)^{\tT}$}\\				
				$\beta_0$ & 1.146 & 1.006 & 0.999 & 1.000 & 1.009 & 1.019 & 1.142 & 1.002 & 0.996 & 0.996 & 1.006 & 1.016 & 0.996 \\ 
				$\beta_1$ & 1.136 & 1.021 & 0.997 & 0.997 & 1.063 & 1.066 & 1.122 & 1.008 & 0.984 & 0.985 & 1.050 & 1.053 & 0.988 \\ 
				$\beta_2$ & 1.143 & 1.018 & 0.998 & 0.999 & 1.067 & 1.073 & 1.130 & 1.006 & 0.986 & 0.987 & 1.054 & 1.061 & 0.988 \\ 
				$\beta_3$ & 1.182 & 1.003 & 1.001 & 1.001 & 1.001 & 1.006 & 1.179 & 1.001 & 0.999 & 0.999 & 0.998 & 1.004 & 0.998 \\

				\hline \\[-0.9em]
				\multicolumn{14}{c}{Observed selection probabilities: $(0.48,0.18,0.18,0.16)$; $\ba=(1.6,0.6,0.6)^{\tT}$}\\					
				$\beta_0$ & 1.358 & 1.019 & 0.996 & 0.996 & 1.016 & 1.040 & 1.351 & 1.014 & 0.991 & 0.991 & 1.011 & 1.035 & 0.995 \\ 
				$\beta_1$ & 1.323 & 1.059 & 0.980 & 0.981 & 1.108 & 1.132 & 1.301 & 1.041 & 0.963 & 0.964 & 1.089 & 1.113 & 0.983 \\ 
				$\beta_2$ & 1.350 & 1.053 & 0.989 & 0.991 & 1.117 & 1.154 & 1.327 & 1.035 & 0.972 & 0.975 & 1.098 & 1.134 & 0.983 \\ 
				$\beta_3$ & 1.448 & 1.013 & 1.004 & 1.004 & 1.002 & 1.014 & 1.437 & 1.005 & 0.997 & 0.997 & 0.995 & 1.007 & 0.993 \\ 
				
				\hline \\[-0.9em]
				\multicolumn{14}{c}{Observed selection probabilities: $(0.30, 0.24, 0.24, 0.22)$; $\ba=(0.8,0.6,0.6)^{\tT}$}\\
				$\beta_0$ & 1.697 & 1.048 & 0.989 & 0.988 & 1.017 & 1.063 & 1.693 & 1.045 & 0.986 & 0.986 & 1.015 & 1.060 & 0.998 \\ 
				$\beta_1$ & 1.606 & 1.131 & 0.943 & 0.945 & 1.125 & 1.184 & 1.605 & 1.130 & 0.943 & 0.944 & 1.124 & 1.183 & 0.999 \\ 
				$\beta_2$ & 1.677 & 1.120 & 0.969 & 0.974 & 1.144 & 1.236 & 1.661 & 1.109 & 0.959 & 0.965 & 1.133 & 1.224 & 0.990 \\ 
				$\beta_3$ & 1.887 & 1.039 & 1.014 & 1.014 & 1.011 & 1.030 & 1.851 & 1.020 & 0.995 & 0.995 & 0.992 & 1.010 & 0.981 \\ 		
				\hline \\ [-0.9em]
				\multicolumn{14}{l}{$Ar$, $A\in\{C,W,M1,M2,RF,EM\}$, $r=1,2$, are the relative efficiencies of CC, SIPW, MI1, MI2,} \\
				\multicolumn{14}{l}{RFMI, and SAEM estimators to MI$r$n estimators for each $\beta_i$, $i=0,1,2,3$. }\\
				\multicolumn{14}{l}{$Ar$ is the ratio of ASE of $A$ estimator to ASE of MI$r$n estimator for each $\beta_i$.}\\
				\multicolumn{14}{l}{$Ar=ASE_A/ASE_{MIrn}$, $M12n=ASE_{MI1n}/ASE_{MI2n}$} \\
				\multicolumn{14}{l}{MI1: MI1 estimation method  with Rubin's type estimated variance in Equation~\eqref{MI1: asymptotic var}} \\
		    	\multicolumn{14}{l}{MI2: MI2 estimation method  with Rubin's type estimated variance in Equation~\eqref{MI2: asymptotic var}} \\
			   \multicolumn{14}{l}{MI1n: MI1 estimation method with proposed estimated variance in Equation~\eqref{MI1n: asymptotic var}} \\
		 	   \multicolumn{14}{l}{MI2n: MI2 estimation method with proposed estimated variance in Equation~\eqref{MI2n: asymptotic var} }	
			\end{tabular}
			\label{tab: RE_01}
		\end{center}
	}
\end{table}

\newpage
\clearpage
\begin{table}[h!]
	{\tiny 
	\caption{Simulation results of Study 2 ($M=15$; $n=500,1,000,1,500$);
		$\bb=(1.2,1,1,1)^{\tT}$; $\ba=(1.4,0.6,0.6)^{\tT}$; $\bga=(0.7,-0.2,0.1,-1.2)^{\tT}$; observed selection probabilities: $(0.45, 0.20,0.20,0.15)$}
	\vspace{-0.8cm}
	\setlength\tabcolsep{10.0 pt} 
	\begin{center}
		\begin{tabular}{llrrrrrrrrr}
			\hline \\ [-0.7em]
			&&$\hbb_F$&$\hbb_C$&$\hbb_{W}$&$\hbb_{M1}$&$\hbb_{M2}$&$\hbb_{M1n}$&$\hbb_{M2n}$&$\hbb_{RF}$&$\hbb_{EM}$\\
			\hline \\ [-0.7em]
			&\multicolumn{10}{l}{$n=500$}\\
			$\beta_0$			
			&Bias& 0.0197 & 0.0981 & 0.0612 & 0.0370 & 0.0333 & 0.0370 & 0.0333 & -0.0356 & 0.0323 \\ 
			&SD  & 0.2119 & 0.3316 & 0.2564 & 0.2275 & 0.2255 & 0.2275 & 0.2255 & 0.2006 & 0.2505 \\ 
			&ASE & 0.2114 & 0.3207 & 0.2243 & 0.2135 & 0.2138 & 0.2081 & 0.2153 & 0.2263 & 0.2417 \\ 
			&MSE & 0.0453 & 0.1196 & 0.0695 & 0.0531 & 0.0520 & 0.0531 & 0.0520 & 0.0415 & 0.0638 \\ 
			&CP  & 0.9470 & 0.9530 & 0.9310 & 0.9340 & 0.9350 & 0.9270 & 0.9390 & 0.9680 & 0.9560 \\ 
			
			\cline{2-11}  \\ [-0.7em]
			$\beta_1$		
			&Bias& 0.0091 & 0.0518 & 0.0379 & 0.0151 & 0.0163 & 0.0151 & 0.0163 & -0.2059 & 0.0416 \\ 
			&SD  & 0.3033 & 0.4678 & 0.3548 & 0.3307 & 0.3265 & 0.3307 & 0.3265 & 0.3070 & 0.4049 \\ 
			&ASE & 0.3048 & 0.4707 & 0.3253 & 0.3068 & 0.3080 & 0.3054 & 0.3128 & 0.3624 & 0.3909 \\ 
			&MSE & 0.0921 & 0.2215 & 0.1273 & 0.1096 & 0.1069 & 0.1096 & 0.1069 & 0.1367 & 0.1657 \\ 
			&CP  & 0.9530 & 0.9530 & 0.9320 & 0.9390 & 0.9440 & 0.9360 & 0.9440 & 0.9390 & 0.9440 \\ 
			
			\cline{2-11}  \\ [-0.7em]
			$\beta_2$		
			&Bias& 0.0317 & 0.0536 & 0.1222 & 0.0716 & 0.0646 & 0.0716 & 0.0646 & -0.2358 & 0.0543 \\ 
			&SD  & 0.2883 & 0.4386 & 0.3651 & 0.3169 & 0.3141 & 0.3169 & 0.3141 & 0.2543 & 0.3773 \\ 
			&ASE & 0.2808 & 0.4341 & 0.3039 & 0.2816 & 0.2831 & 0.2728 & 0.2862 & 0.3268 & 0.3583 \\ 
			&MSE & 0.0841 & 0.1952 & 0.1482 & 0.1056 & 0.1028 & 0.1056 & 0.1028 & 0.1203 & 0.1453 \\ 
			&CP  & 0.9470 & 0.9560 & 0.8990 & 0.9170 & 0.9170 & 0.9060 & 0.9220 & 0.9090 & 0.9400 \\ 
			
			\cline{2-11}  \\ [-0.7em]
			$\beta_3$			
			&Bias& 0.0172 & 0.1480 & 0.0863 & 0.0216 & 0.0211 & 0.0216 & 0.0211 & -0.0084 & 0.0339 \\ 
			&SD  & 0.2589 & 0.4220 & 0.2971 & 0.2611 & 0.2611 & 0.2611 & 0.2611 & 0.2566 & 0.2708 \\ 
			&ASE & 0.2591 & 0.4191 & 0.2650 & 0.2594 & 0.2594 & 0.2504 & 0.2563 & 0.2589 & 0.2664 \\ 
			&MSE & 0.0673 & 0.1999 & 0.0957 & 0.0686 & 0.0686 & 0.0686 & 0.0686 & 0.0659 & 0.0745 \\ 
			&CP  & 0.9570 & 0.9540 & 0.9220 & 0.9530 & 0.9550 & 0.9430 & 0.9530 & 0.9560 & 0.9500 \\ 
			
			\hline \\ [-0.7em]
			&\multicolumn{10}{l}{$n=1,000$}\\
			$\beta_0$	
			&Bias& 0.0109 & 0.0827 & 0.0189 & 0.0142 & 0.0128 & 0.0142 & 0.0128 & -0.0462 & 0.0134 \\ 
			&SD  & 0.1521 & 0.2257 & 0.1661 & 0.1585 & 0.1565 & 0.1585 & 0.1565 & 0.1388 & 0.1712 \\ 
			&ASE & 0.1479 & 0.2223 & 0.1544 & 0.1493 & 0.1495 & 0.1484 & 0.1511 & 0.1583 & 0.1673 \\ 
			&MSE & 0.0233 & 0.0578 & 0.0279 & 0.0253 & 0.0247 & 0.0253 & 0.0247 & 0.0214 & 0.0295 \\ 
			&CP  & 0.9490 & 0.9430 & 0.9350 & 0.9390 & 0.9390 & 0.9350 & 0.9400 & 0.9600 & 0.9530 \\ 
			
			\cline{2-11}  \\ [-0.7em]
			$\beta_1$	
			&Bias& 0.0016 & 0.0310 & 0.0112 & 0.0055 & 0.0057 & 0.0055 & 0.0057 & -0.2123 & 0.0263 \\ 
			&SD  & 0.2131 & 0.3394 & 0.2331 & 0.2231 & 0.2213 & 0.2231 & 0.2213 & 0.2125 & 0.2753 \\ 
			&ASE & 0.2143 & 0.3291 & 0.2280 & 0.2173 & 0.2177 & 0.2192 & 0.2220 & 0.2578 & 0.2728 \\ 
			&MSE & 0.0454 & 0.1161 & 0.0544 & 0.0498 & 0.0490 & 0.0498 & 0.0490 & 0.0902 & 0.0765 \\ 
			&CP  & 0.9530 & 0.9500 & 0.9510 & 0.9460 & 0.9470 & 0.9480 & 0.9510 & 0.9060 & 0.9480 \\ 
			\cline{2-11}  \\ [-0.7em]
			$\beta_2$				
			&Bias& 0.0090 & 0.0206 & 0.0281 & 0.0169 & 0.0144 & 0.0169 & 0.0144 & -0.2594 & 0.0170 \\ 
			&SD  & 0.2053 & 0.3019 & 0.2406 & 0.2236 & 0.2207 & 0.2236 & 0.2207 & 0.1781 & 0.2591 \\ 
			&ASE & 0.1964 & 0.3015 & 0.2125 & 0.1994 & 0.2002 & 0.1981 & 0.2047 & 0.2321 & 0.2486 \\ 
			&MSE & 0.0422 & 0.0916 & 0.0587 & 0.0503 & 0.0489 & 0.0503 & 0.0489 & 0.0990 & 0.0674 \\ 
			&CP  & 0.9460 & 0.9530 & 0.9200 & 0.9190 & 0.9250 & 0.9170 & 0.9300 & 0.8230 & 0.9480 \\ 
			
			\cline{2-11}  \\ [-0.7em]
			$\beta_3$			
			&Bias& 0.0112 & 0.1394 & 0.0287 & 0.0126 & 0.0125 & 0.0126 & 0.0125 & -0.0128 & 0.0229 \\ 
			&SD  & 0.1952 & 0.2982 & 0.2052 & 0.1959 & 0.1960 & 0.1959 & 0.1960 & 0.1923 & 0.1988 \\ 
			&ASE & 0.1824 & 0.2930 & 0.1837 & 0.1825 & 0.1825 & 0.1804 & 0.1822 & 0.1819 & 0.1862 \\ 
			&MSE & 0.0382 & 0.1083 & 0.0429 & 0.0385 & 0.0386 & 0.0385 & 0.0386 & 0.0371 & 0.0401 \\ 
			&CP  & 0.9310 & 0.9350 & 0.9170 & 0.9310 & 0.9310 & 0.9270 & 0.9310 & 0.9330 & 0.9390 \\ 
					
			\hline \\ [-0.7em]
			&\multicolumn{10}{l}{$n=1,500$}\\
			$\beta_0$		
			&Bias& 0.0031 & 0.0734 & 0.0102 & 0.0066 & 0.0053 & 0.0066 & 0.0053 & -0.0530 & 0.0014 \\ 
			&SD  & 0.1238 & 0.1834 & 0.1312 & 0.1285 & 0.1282 & 0.1285 & 0.1282 & 0.1145 & 0.1409 \\ 
			&ASE & 0.1203 & 0.1798 & 0.1256 & 0.1217 & 0.1218 & 0.1217 & 0.1231 & 0.1299 & 0.1354 \\ 
			&MSE & 0.0153 & 0.0390 & 0.0173 & 0.0166 & 0.0165 & 0.0166 & 0.0165 & 0.0159 & 0.0199 \\ 
			&CP  & 0.9510 & 0.9450 & 0.9430 & 0.9410 & 0.9430 & 0.9400 & 0.9460 & 0.9480 & 0.9460 \\ 
			
			\cline{2-11}  \\ [-0.7em]
			$\beta_1$	
			&Bias& 0.0078 & 0.0356 & 0.0143 & 0.0108 & 0.0105 & 0.0108 & 0.0105 & -0.2013 & 0.0390 \\ 
			&SD  & 0.1759 & 0.2794 & 0.1928 & 0.1858 & 0.1843 & 0.1858 & 0.1843 & 0.1772 & 0.2279 \\ 
			&ASE & 0.1745 & 0.2668 & 0.1859 & 0.1774 & 0.1776 & 0.1796 & 0.1813 & 0.2122 & 0.2216 \\ 
			&MSE & 0.0310 & 0.0793 & 0.0374 & 0.0346 & 0.0341 & 0.0346 & 0.0341 & 0.0719 & 0.0534 \\ 
			&CP  & 0.9470 & 0.9370 & 0.9400 & 0.9360 & 0.9350 & 0.9390 & 0.9410 & 0.8640 & 0.9450 \\ 
			
			\cline{2-11}  \\ [-0.7em]
			$\beta_2$	
			&Bias& 0.0080 & 0.0209 & 0.0261 & 0.0166 & 0.0138 & 0.0166 & 0.0138 & -0.2568 & 0.0095 \\ 
			&SD  & 0.1670 & 0.2544 & 0.1871 & 0.1783 & 0.1773 & 0.1783 & 0.1773 & 0.1482 & 0.2150 \\ 
			&ASE & 0.1598 & 0.2442 & 0.1737 & 0.1633 & 0.1638 & 0.1636 & 0.1677 & 0.1919 & 0.2013 \\ 
			&MSE & 0.0280 & 0.0652 & 0.0357 & 0.0321 & 0.0316 & 0.0321 & 0.0316 & 0.0879 & 0.0463 \\ 
			&CP  & 0.9450 & 0.9390 & 0.9370 & 0.9300 & 0.9320 & 0.9310 & 0.9370 & 0.7590 & 0.9440 \\ 
			
			\cline{2-11}  \\ [-0.7em]
			$\beta_3$	
			&Bias& 0.0096 & 0.1226 & 0.0177 & 0.0106 & 0.0105 & 0.0106 & 0.0105 & -0.0158 & 0.0196 \\ 
			&SD  & 0.1458 & 0.2339 & 0.1512 & 0.1464 & 0.1463 & 0.1464 & 0.1463 & 0.1445 & 0.1496 \\ 
			&ASE & 0.1484 & 0.2363 & 0.1493 & 0.1486 & 0.1486 & 0.1478 & 0.1487 & 0.1479 & 0.1513 \\ 
			&MSE & 0.0213 & 0.0698 & 0.0232 & 0.0215 & 0.0215 & 0.0215 & 0.0215 & 0.0211 & 0.0228 \\ 
			&CP  & 0.9530 & 0.9340 & 0.9490 & 0.9510 & 0.9520 & 0.9510 & 0.9520 & 0.9600 & 0.9560 \\ 			
			\hline \\ [-0.7em]
			\multicolumn{11}{l}{$\hbb_{M1}$: MI1 estimator whose value is the solution of $\BU_{M1}(\bb)=\b0$ in Equation~\eqref{eq: score mi1} with Rubin's type}\\
			\multicolumn{11}{l}{estimated variance in Equation~\eqref{MI1: asymptotic var}} \\
			\multicolumn{11}{l}{$\hbb_{M2}$: MI2 estimator whose value is the solution of $\BU_{M2}(\bb)=\b0$ in Equation~\eqref{eq: score mi2} with Rubin's type}\\
			\multicolumn{11}{l}{estimated variance in Equation~\eqref{MI2: asymptotic var}} \\
			\multicolumn{11}{l}{$\hbb_{M1n}$: MI1 estimator whose value is the solution of $\BU_{M1}(\bb)=\b0$ in Equation~\eqref{eq: score mi1} with proposed}\\
			\multicolumn{11}{l}{estimated variance in Equation~\eqref{MI1n: asymptotic var}} \\
			\multicolumn{11}{l}{$\hbb_{M2n}$: MI2 estimator whose value is the solution of $\BU_{M2}(\bb)=\b0$ in Equation~\eqref{eq: score mi2} with proposed} \\
			\multicolumn{11}{l}{estimated variance in Equation~\eqref{MI2n: asymptotic var}}
		\end{tabular}
	\end{center}
	\label{tab: simulation study 2}
}
\end{table}

\newpage
\clearpage
\begin{table}[h!] 
	{\footnotesize
		\caption{Relative efficiencies in Study 2 ($M=15$; $n=500,1,000,1,500$);
			$\bb=(1.2,1,1,1)^{\tT}$; $\ba=(1.4,0.6,0.6)^{\tT}$; $\bga=(0.7,-0.2,0.1,-1.2)^{\tT}$; observed selection probabilities: $(0.45, 0.20,0.20,0.15)$}
		\vspace{-0.8cm}
		\setlength\tabcolsep{5.5 pt} 
		\begin{center}
			\begin{tabular}{cccccccccccccc}
				\hline  \\[-0.9em]
				\multicolumn{14}{c}{Relative efficiency}\\
				\hline \\[-0.9em]
				
				$\bb$&$C1$&$W1$&$M11$ & $M21$ & $RF1$&$EM1$& $C2$&$W2$&$M12$ & $M22$ & $RF2$&$EM2$&$M12n$\\
				\hline  \\[-0.9em]
				\multicolumn{14}{l}{$n=500$}\\
	
				$\beta_0$ & 1.541 & 1.078 & 1.026 & 1.028 & 1.088 & 1.162 & 1.489 & 1.042 & 0.992 & 0.993 & 1.051 & 1.123 & 0.966 \\ 
				$\beta_1$ & 1.542 & 1.065 & 1.005 & 1.009 & 1.187 & 1.280 & 1.505 & 1.040 & 0.981 & 0.985 & 1.159 & 1.250 & 0.976 \\ 
				$\beta_2$ & 1.591 & 1.114 & 1.032 & 1.038 & 1.198 & 1.314 & 1.517 & 1.062 & 0.984 & 0.989 & 1.142 & 1.252 & 0.953 \\ 
				$\beta_3$ & 1.673 & 1.058 & 1.036 & 1.036 & 1.034 & 1.064 & 1.635 & 1.034 & 1.012 & 1.012 & 1.010 & 1.039 & 0.977 \\
				
				\hline \\[-0.9em]
				\multicolumn{14}{l}{$n=1,000$}\\					
		
				$\beta_0$ & 1.498 & 1.041 & 1.006 & 1.007 & 1.067 & 1.127 & 1.472 & 1.022 & 0.988 & 0.989 & 1.048 & 1.107 & 0.982 \\ 
				$\beta_1$ & 1.501 & 1.040 & 0.991 & 0.993 & 1.176 & 1.244 & 1.482 & 1.027 & 0.979 & 0.981 & 1.161 & 1.229 & 0.987 \\ 
				$\beta_2$ & 1.522 & 1.073 & 1.007 & 1.011 & 1.172 & 1.255 & 1.473 & 1.038 & 0.974 & 0.978 & 1.134 & 1.215 & 0.968 \\ 
				$\beta_3$ & 1.624 & 1.019 & 1.012 & 1.012 & 1.008 & 1.032 & 1.608 & 1.008 & 1.001 & 1.002 & 0.998 & 1.022 & 0.990 \\ 
				
				\hline \\[-0.9em]
				\multicolumn{14}{l}{$n=1,500$}\\

				$\beta_0$ & 1.477 & 1.032 & 1.000 & 1.001 & 1.067 & 1.113 & 1.460 & 1.020 & 0.988 & 0.989 & 1.055 & 1.100 & 0.988 \\ 
				$\beta_1$ & 1.486 & 1.035 & 0.988 & 0.989 & 1.182 & 1.234 & 1.472 & 1.026 & 0.978 & 0.980 & 1.170 & 1.223 & 0.991 \\ 
				$\beta_2$ & 1.492 & 1.062 & 0.998 & 1.001 & 1.173 & 1.230 & 1.456 & 1.036 & 0.974 & 0.977 & 1.145 & 1.201 & 0.976 \\ 
				$\beta_3$ & 1.599 & 1.010 & 1.005 & 1.005 & 1.001 & 1.024 & 1.589 & 1.004 & 0.999 & 0.999 & 0.995 & 1.018 & 0.994 \\
				
				\hline \\ [-2mm]
				\multicolumn{14}{l}{$Ar$, $A\in\{C,W,M1,M2,RF,EM\}$, $r=1,2$, are the relative efficiencies of CC, SIPW, MI1, MI2,} \\
				\multicolumn{14}{l}{RFMI, and SAEM estimators to MI$r$n estimators for each $\beta_i$, $i=0,1,2,3$.}\\
				\multicolumn{14}{l}{$Ar$ is the ratio of ASE of $A$ estimator to ASE of MI$r$n estimator for each $\beta_i$.}\\
				\multicolumn{14}{l}{$Ar=ASE_A/ASE_{MIrn}$, $M12n=ASE_{MI1n}/ASE_{MI2n}$}  \\
				\multicolumn{14}{l}{MI1: MI1 estimation method  with Rubin's type estimated variance in Equation~\eqref{MI1: asymptotic var}} \\
            	\multicolumn{14}{l}{MI2: MI2 estimation method  with Rubin's type estimated variance in Equation~\eqref{MI2: asymptotic var}} \\
	            \multicolumn{14}{l}{MI1n: MI1 estimation method with proposed estimated variance in Equation~\eqref{MI1n: asymptotic var}} \\
            	\multicolumn{14}{l}{MI2n: MI2 estimation method with proposed estimated variance in Equation~\eqref{MI2n: asymptotic var} }				
			\end{tabular}
			\label{tab: RE_02}
		\end{center}
	}
\end{table}

\newpage
\clearpage
\begin{landscape}
	\begin{table}[h!]
		{\footnotesize
		\caption{Simulation results of Study 3 $(n=1,000)$;
			$\bb=(1.2,1,1,1)^{\tT}$; $\ba=(1.4,0.6,0.6)^{\tT}$; $\bga=(0.7,-0.2,0.1,-1.2)^{\tT}$; observed selection probabilities: $(0.45, 0.20, 0.20, 0.15)$; $M=(10,20,30)$}
		\vspace{-0.7cm}
		\setlength\tabcolsep{3 pt} 
		\begin{center}
				\begin{tabular}{ccrrrrr c rrrrr c rrrrr c rrrrr}   
					\hline \\ [-0.8em]
					\multicolumn{6}{c}{}
					&\multicolumn{5}{c}{$M=10$}&
					&\multicolumn{5}{c}{$M=20$}&
					&\multicolumn{5}{c}{$M=30$}\\
					\cline{8-12} \cline{14-18} \cline{20-24} \\  [-0.5em]
			&&$\hbb_F$&$\hbb_C$&$\hbb_{W}$&$\hbb_{EM}$&&$\hbb_{M1}$&$\hbb_{M2}$&$\hbb_{M1n}$&$\hbb_{M2n}$&$\hbb_{RF}$&&$\hbb_{M1}$&$\hbb_{M2}$&$\hbb_{M1n}$&$\hbb_{M2n}$&$\hbb_{RF}$&&$\hbb_{M1}$&$\hbb_{M2}$&$\hbb_{M1n}$&$\hbb_{M2n}$&$\hbb_{RF}$\\
					\hline \\ [-0.5em]
					$\beta_0$				
					&Bias& 0.0109 & 0.0827 & 0.0189 & 0.0134 && 0.0145 & 0.0130 & 0.0145 & 0.0130 & -0.0466 && 0.0143 & 0.0126 & 0.0143 & 0.0126 & -0.0465 && 0.0142 & 0.0125 & 0.0142 & 0.0125 & -0.0466 \\ 
					&SD  & 0.1521 & 0.2257 & 0.1661 & 0.1712 && 0.1585 & 0.1567 & 0.1585 & 0.1567 & 0.1398 && 0.1583 & 0.1565 & 0.1583 & 0.1565 & 0.1394 && 0.1584 & 0.1565 & 0.1584 & 0.1565 & 0.1384 \\ 
					&ASE & 0.1479 & 0.2223 & 0.1544 & 0.1673 && 0.1494 & 0.1495 & 0.1484 & 0.1511 & 0.1586 && 0.1493 & 0.1494 & 0.1484 & 0.1511 & 0.1585 && 0.1492 & 0.1494 & 0.1483 & 0.1510 & 0.1581 \\ 
					&MSE & 0.0233 & 0.0578 & 0.0279 & 0.0295 && 0.0253 & 0.0247 & 0.0253 & 0.0247 & 0.0217 && 0.0253 & 0.0246 & 0.0253 & 0.0246 & 0.0216 && 0.0253 & 0.0246 & 0.0253 & 0.0246 & 0.0213 \\ 
					&CP  & 0.9490 & 0.9430 & 0.9350 & 0.9530 && 0.9400 & 0.9380 & 0.9360 & 0.9410 & 0.9580 && 0.9400 & 0.9370 & 0.9360 & 0.9400 & 0.9590 && 0.9390 & 0.9400 & 0.9350 & 0.9410 & 0.9590 \\ 
					\cline{2-24}  \\ [-1em]
					$\beta_1$				
					&Bias& 0.0016 & 0.0310 & 0.0112 & 0.0263 && 0.0056 & 0.0059 & 0.0056 & 0.0059 & -0.2149 && 0.0058 & 0.0058 & 0.0058 & 0.0058 & -0.2103 && 0.0057 & 0.0054 & 0.0057 & 0.0054 & -0.2118 \\ 
					&SD  & 0.2131 & 0.3394 & 0.2331 & 0.2753 && 0.2232 & 0.2211 & 0.2232 & 0.2211 & 0.2132 && 0.2232 & 0.2213 & 0.2232 & 0.2213 & 0.2122 && 0.2232 & 0.2212 & 0.2232 & 0.2212 & 0.2100 \\ 
					&ASE & 0.2143 & 0.3291 & 0.2280 & 0.2728 && 0.2174 & 0.2178 & 0.2192 & 0.2220 & 0.2588 && 0.2173 & 0.2177 & 0.2192 & 0.2220 & 0.2569 && 0.2172 & 0.2176 & 0.2192 & 0.2219 & 0.2566 \\ 
					&MSE & 0.0454 & 0.1161 & 0.0544 & 0.0765 && 0.0499 & 0.0489 & 0.0499 & 0.0489 & 0.0916 && 0.0498 & 0.0490 & 0.0498 & 0.0490 & 0.0893 && 0.0498 & 0.0490 & 0.0498 & 0.0490 & 0.0889 \\ 
					&CP  & 0.9530 & 0.9500 & 0.9510 & 0.9480 && 0.9460 & 0.9460 & 0.9470 & 0.9510 & 0.8960 && 0.9470 & 0.9490 & 0.9470 & 0.9530 & 0.8950 && 0.9440 & 0.9480 & 0.9470 & 0.9510 & 0.9030 \\ 
					\cline{2-24}  \\ [-1em]
					$\beta_2$			
					&Bias& 0.0090 & 0.0206 & 0.0281 & 0.0170 && 0.0176 & 0.0149 & 0.0176 & 0.0149 & -0.2615 && 0.0172 & 0.0141 & 0.0172 & 0.0141 & -0.2587 && 0.0170 & 0.0138 & 0.0170 & 0.0138 & -0.2594 \\ 
					&SD  & 0.2053 & 0.3019 & 0.2406 & 0.2591 && 0.2241 & 0.2211 & 0.2241 & 0.2211 & 0.1759 && 0.2236 & 0.2208 & 0.2236 & 0.2208 & 0.1782 && 0.2236 & 0.2209 & 0.2236 & 0.2209 & 0.1747 \\ 
					&ASE & 0.1964 & 0.3015 & 0.2125 & 0.2486 && 0.1996 & 0.2003 & 0.1982 & 0.2048 & 0.2316 && 0.1993 & 0.2001 & 0.1981 & 0.2047 & 0.2316 && 0.1993 & 0.2001 & 0.1981 & 0.2046 & 0.2307 \\ 
					&MSE & 0.0422 & 0.0916 & 0.0587 & 0.0674 && 0.0505 & 0.0491 & 0.0505 & 0.0491 & 0.0993 && 0.0503 & 0.0489 & 0.0503 & 0.0489 & 0.0987 && 0.0503 & 0.0490 & 0.0503 & 0.0490 & 0.0978 \\ 
					&CP  & 0.9460 & 0.9530 & 0.9200 & 0.9480 && 0.9200 & 0.9240 & 0.9170 & 0.9290 & 0.8180 && 0.9210 & 0.9260 & 0.9190 & 0.9310 & 0.8180 && 0.9200 & 0.9250 & 0.9210 & 0.9300 & 0.8280 \\ 
					\cline{2-24}  \\ [-1em]
					$\beta_3$
					&Bias& 0.0112 & 0.1394 & 0.0287 & 0.0229 && 0.0126 & 0.0125 & 0.0126 & 0.0125 & -0.0130 && 0.0127 & 0.0125 & 0.0127 & 0.0125 & -0.0127 && 0.0127 & 0.0125 & 0.0127 & 0.0125 & -0.0130 \\ 
					&SD  & 0.1952 & 0.2982 & 0.2052 & 0.1988 && 0.1959 & 0.1960 & 0.1959 & 0.1960 & 0.1923 && 0.1959 & 0.1960 & 0.1959 & 0.1960 & 0.1921 && 0.1960 & 0.1960 & 0.1960 & 0.1960 & 0.1923 \\ 
					&ASE & 0.1824 & 0.2930 & 0.1837 & 0.1862 && 0.1825 & 0.1825 & 0.1804 & 0.1822 & 0.1819 && 0.1825 & 0.1825 & 0.1804 & 0.1822 & 0.1819 && 0.1825 & 0.1825 & 0.1804 & 0.1822 & 0.1819 \\ 
					&MSE & 0.0382 & 0.1083 & 0.0429 & 0.0401 && 0.0386 & 0.0386 & 0.0386 & 0.0386 & 0.0371 && 0.0385 & 0.0386 & 0.0385 & 0.0386 & 0.0371 && 0.0386 & 0.0386 & 0.0386 & 0.0386 & 0.0372 \\ 
					&CP  & 0.9310 & 0.9350 & 0.9170 & 0.9390 && 0.9310 & 0.9310 & 0.9280 & 0.9310 & 0.9340 && 0.9310 & 0.9320 & 0.9300 & 0.9310 & 0.9310 && 0.9310 & 0.9320 & 0.9300 & 0.9300 & 0.9310 \\ 
					\hline  \\ [-0.7em]
					\multicolumn{24}{l}{$\hbb_{M1}$: MI1 estimator whose value is the solution of $\BU_{M1}(\bb)=\b0$ in Equation~\eqref{eq: score mi1} with Rubin's type estimated variance in Equation~\eqref{MI1: asymptotic var}} \\
					\multicolumn{24}{l}{$\hbb_{M2}$: MI2 estimator whose value is the solution of $\BU_{M2}(\bb)=\b0$ in Equation~\eqref{eq: score mi2} with Rubin's type estimated variance in Equation~\eqref{MI2: asymptotic var}} \\
					\multicolumn{24}{l}{$\hbb_{M1n}$: MI1 estimator whose value is the solution of $\BU_{M1}(\bb)=\b0$ in Equation~\eqref{eq: score mi1} with proposed estimated variance in Equation~\eqref{MI1n: asymptotic var}} \\
					\multicolumn{24}{l}{$\hbb_{M2n}$: MI2 estimator whose value is the solution of $\BU_{M2}(\bb)=\b0$ in Equation~\eqref{eq: score mi2} with proposed estimated variance in Equation~\eqref{MI2n: asymptotic var}}
			\end{tabular}
		\end{center}
		\label{tab: simulation study 3}
	}
	\end{table}
\end{landscape}

\newpage
\clearpage
\begin{table}[h!] 
	{\footnotesize
		\caption{Relative efficiencies in Study 3 $(n=1,000)$;
			$\bb=(1.2,1,1,1)^{\tT}$; $\ba=(1.4,0.6,0.6)^{\tT}$; $\bga=(0.7,-0.2,0.1,-1.2)^{\tT}$; observed selection probabilities: $(0.45, 0.20, 0.20, 0.15)$; $M=10,20,30$}
		\vspace{-0.8cm}
		\setlength\tabcolsep{5.5 pt} 
		\begin{center}
			\begin{tabular}{cccccccccccccc}
				\hline  \\[-0.9em]
				\multicolumn{14}{c}{Relative efficiency}\\
				\hline \\[-0.9em]
	
				$\bb$&$C1$&$W1$&$M11$ & $M21$ & $RF1$&$EM1$& $C2$&$W2$&$M12$ & $M22$ & $RF2$&$EM2$&$M12n$\\
				\hline  \\[-0.9em]
				\multicolumn{14}{l}{$M=10$}\\
				$\beta_0$ & 1.498 & 1.040 & 1.006 & 1.008 & 1.069 & 1.127 & 1.471 & 1.022 & 0.988 & 0.989 & 1.050 & 1.107 & 0.982 \\ 
				$\beta_1$ & 1.501 & 1.040 & 0.992 & 0.994 & 1.180 & 1.244 & 1.482 & 1.027 & 0.979 & 0.981 & 1.166 & 1.229 & 0.987 \\ 
				$\beta_2$ & 1.522 & 1.072 & 1.007 & 1.011 & 1.169 & 1.255 & 1.472 & 1.038 & 0.975 & 0.978 & 1.131 & 1.214 & 0.968 \\ 
				$\beta_3$ & 1.624 & 1.019 & 1.012 & 1.012 & 1.008 & 1.032 & 1.608 & 1.008 & 1.001 & 1.002 & 0.998 & 1.022 & 0.990 \\ 				
				
				\hline \\[-0.9em]
				\multicolumn{14}{l}{$M=20$}\\					
	
				$\beta_0$ & 1.498 & 1.041 & 1.006 & 1.007 & 1.068 & 1.127 & 1.472 & 1.022 & 0.988 & 0.989 & 1.049 & 1.107 & 0.982 \\ 
				$\beta_1$ & 1.501 & 1.040 & 0.991 & 0.993 & 1.172 & 1.244 & 1.482 & 1.027 & 0.979 & 0.981 & 1.157 & 1.229 & 0.987 \\ 
				$\beta_2$ & 1.522 & 1.073 & 1.006 & 1.010 & 1.169 & 1.255 & 1.473 & 1.038 & 0.974 & 0.978 & 1.132 & 1.215 & 0.968 \\ 
				$\beta_3$ & 1.624 & 1.019 & 1.012 & 1.012 & 1.008 & 1.032 & 1.608 & 1.008 & 1.002 & 1.002 & 0.998 & 1.022 & 0.990 \\ 
						
				\hline \\[-0.9em]
				\multicolumn{14}{l}{$M=30$}\\

				$\beta_0$ & 1.499 & 1.041 & 1.006 & 1.007 & 1.066 & 1.127 & 1.472 & 1.022 & 0.988 & 0.989 & 1.047 & 1.107 & 0.983 \\ 
				$\beta_1$ & 1.501 & 1.040 & 0.991 & 0.993 & 1.171 & 1.245 & 1.483 & 1.027 & 0.979 & 0.980 & 1.156 & 1.229 & 0.988 \\ 
				$\beta_2$ & 1.522 & 1.073 & 1.006 & 1.010 & 1.165 & 1.255 & 1.473 & 1.039 & 0.974 & 0.978 & 1.127 & 1.215 & 0.968 \\ 
				$\beta_3$ & 1.624 & 1.019 & 1.012 & 1.012 & 1.008 & 1.032 & 1.608 & 1.008 & 1.002 & 1.002 & 0.998 & 1.022 & 0.990 \\  
							
				\hline \\ [-2mm]
				\multicolumn{14}{l}{$Ar$, $A\in\{C,W,M1,M2,RF,EM\}$, $r=1,2$, are the relative efficiencies of CC, SIPW, MI1, MI2,} \\
			\multicolumn{14}{l}{ RFMI, and SAEM estimators to MI$r$n estimators for each $\beta_i$, $i=0,1,2,3$. }\\
				\multicolumn{12}{l}{$Ar$ is the ratio of ASE of $A$ estimator to ASE of MI$r$n estimator for each $\beta_i$.}\\
				\multicolumn{14}{l}{$Ar=ASE_A/ASE_{MIrn}$, $Mn12=ASE_{MI1n}/ASE_{MI2n}$} \\
				\multicolumn{14}{l}{MI1: MI1 estimation method  with Rubin's type estimated variance in Equation~\eqref{MI1: asymptotic var}} \\
				\multicolumn{14}{l}{MI2: MI2 estimation method  with Rubin's type estimated variance in Equation~\eqref{MI2: asymptotic var}} \\
				\multicolumn{14}{l}{MI1n: MI1 estimation method with proposed estimated variance in Equation~\eqref{MI1n: asymptotic var}} \\
				\multicolumn{14}{l}{MI2n: MI2 estimation method with proposed estimated variance in Equation~\eqref{MI2n: asymptotic var} }			
			\end{tabular}
			\label{tab: RE_03}
		\end{center}
	}
\end{table}

\newpage
\clearpage
\begin{table}[h!]
	{\tiny 
		\caption{Simulation results of Study 4 ($M=15$; $n=1,500$);
			$\bb=(1.2,1,1,1)^{\tT}$; $\bga=(0.7,-0.2,0.1,-1.2)^{\tT}$; three sets of $\ba$ are $(2.4,0.6,0.6)^{\tT}$, $(1.6,0.6,0.6)^{\tT}$, and $(1,0.6,0.6)^{\tT}$ to create the observed selection probabilities: 
			$(0.70, 0.12,0.12,0.06)$, $(0.50,0.19,0.19,0.12)$, and $(0.36,0.24,0.24,0.16)$, respectively}
		\vspace{-0.8cm}
		\setlength\tabcolsep{10.0 pt} 
		\begin{center}
			\begin{tabular}{llrrrrrrrrr}
				\hline \\ [-0.7em]
				&&$\hbb_F$&$\hbb_C$&$\hbb_{W}$&$\hbb_{M1}$&$\hbb_{M2}$&$\hbb_{M1n}$&$\hbb_{M2n}$&$\hbb_{RF}$&$\hbb_{EM}$\\
				\hline \\ [-0.7em]
				\multicolumn{11}{c}{Observed selection probabilities: $(0.70, 0.12, 0.12, 0.06)$; $\bm\alpha=(2.4,0.6,0.6)^{\tT}$}\\
				$\beta_0$				
		
				&Bias& 0.0104 & 0.0508 & 0.0144 & 0.0130 & 0.0164 & 0.0130 & 0.0164 & 0.1364 & 0.0130 \\ 
				&SD  & 0.1532 & 0.1839 & 0.1720 & 0.1658 & 0.1619 & 0.1658 & 0.1619 & 0.1532 & 0.1612 \\ 
				&ASE & 0.1521 & 0.1833 & 0.1668 & 0.1583 & 0.1585 & 0.1604 & 0.1631 & 0.1645 & 0.1613 \\ 
				&MSE & 0.0236 & 0.0364 & 0.0298 & 0.0277 & 0.0265 & 0.0277 & 0.0265 & 0.0421 & 0.0261 \\ 
				&CP  & 0.9520 & 0.9430 & 0.9440 & 0.9370 & 0.9490 & 0.9420 & 0.9540 & 0.9080 & 0.9550 \\ 
				
				\cline{2-11}  \\ [-0.7em]
				$\beta_1$		

				&Bias& 0.0197 & 0.0239 & 0.0231 & 0.0243 & 0.0196 & 0.0243 & 0.0196 & -0.1695 & 0.0251 \\ 
				&SD  & 0.2187 & 0.2622 & 0.2677 & 0.2567 & 0.2526 & 0.2567 & 0.2526 & 0.1995 & 0.2502 \\ 
				&ASE & 0.2108 & 0.2579 & 0.2561 & 0.2298 & 0.2311 & 0.2364 & 0.2473 & 0.2302 & 0.2382 \\ 
				&MSE & 0.0482 & 0.0693 & 0.0722 & 0.0665 & 0.0642 & 0.0665 & 0.0642 & 0.0685 & 0.0632 \\ 
				&CP  & 0.9490 & 0.9530 & 0.9450 & 0.9310 & 0.9360 & 0.9420 & 0.9530 & 0.9050 & 0.9480 \\  
				
				\cline{2-11}  \\ [-0.7em]
				$\beta_2$		

				&Bias& -0.0041 & -0.0044 & -0.0072 & -0.0076 & -0.0108 & -0.0076 & -0.0108 & -0.1948 & -0.0026 \\ 
				&SD  & 0.2012 & 0.2518 & 0.2596 & 0.2393 & 0.2353 & 0.2393 & 0.2353 & 0.1824 & 0.2324 \\ 
				&ASE & 0.2099 & 0.2567 & 0.2547 & 0.2287 & 0.2302 & 0.2353 & 0.2459 & 0.2293 & 0.2373 \\ 
				&MSE & 0.0405 & 0.0634 & 0.0674 & 0.0573 & 0.0555 & 0.0573 & 0.0555 & 0.0712 & 0.0540 \\ 
				&CP  & 0.9560 & 0.9680 & 0.9510 & 0.9340 & 0.9450 & 0.9470 & 0.9610 & 0.9000 & 0.9560 \\  
				
				\cline{2-11}  \\ [-0.7em]
				$\beta_3$				
	
				&Bias& 0.0072 & 0.0839 & 0.0108 & 0.0071 & 0.0062 & 0.0071 & 0.0062 & -0.0054 & 0.0102 \\ 
				&SD  & 0.2085 & 0.2675 & 0.2145 & 0.2096 & 0.2090 & 0.2096 & 0.2090 & 0.2078 & 0.2095 \\ 
				&ASE & 0.2104 & 0.2656 & 0.2122 & 0.2111 & 0.2110 & 0.2099 & 0.2114 & 0.2098 & 0.2118 \\ 
				&MSE & 0.0435 & 0.0786 & 0.0461 & 0.0440 & 0.0437 & 0.0440 & 0.0437 & 0.0432 & 0.0440 \\ 
				&CP  & 0.9600 & 0.9520 & 0.9560 & 0.9580 & 0.9610 & 0.9580 & 0.9610 & 0.9540 & 0.9600 \\  
				
				\hline \\ [-0.7em]
				\multicolumn{11}{c}{Observed selection probabilities: $(0.50, 0.19, 0.19, 0.12)$; $\bm\alpha=(1.6,0.6,0.6)^{\tT}$}\\
				$\beta_0$		
	
				&Bias& 0.0104 & 0.0760 & 0.0201 & 0.0188 & 0.0236 & 0.0188 & 0.0236 & 0.2109 & 0.0142 \\ 
				&SD  & 0.1532 & 0.2131 & 0.1880 & 0.1777 & 0.1713 & 0.1777 & 0.1713 & 0.1562 & 0.1697 \\ 
				&ASE & 0.1521 & 0.2151 & 0.1813 & 0.1611 & 0.1618 & 0.1673 & 0.1698 & 0.1725 & 0.1688 \\ 
				&MSE & 0.0236 & 0.0512 & 0.0357 & 0.0319 & 0.0299 & 0.0319 & 0.0299 & 0.0689 & 0.0290 \\ 
				&CP  & 0.9520 & 0.9490 & 0.9460 & 0.9290 & 0.9350 & 0.9340 & 0.9480 & 0.8100 & 0.9530 \\

				\cline{2-11}  \\ [-0.7em]
				$\beta_1$		
	
				&Bias& 0.0197 & 0.0240 & 0.0192 & 0.0174 & 0.0104 & 0.0174 & 0.0104 & -0.2777 & 0.0295 \\ 
				&SD  & 0.2187 & 0.2991 & 0.3098 & 0.2775 & 0.2716 & 0.2775 & 0.2716 & 0.1820 & 0.2677 \\ 
				&ASE & 0.2108 & 0.3053 & 0.2979 & 0.2365 & 0.2402 & 0.2551 & 0.2677 & 0.2399 & 0.2600 \\ 
				&MSE & 0.0482 & 0.0900 & 0.0964 & 0.0773 & 0.0739 & 0.0773 & 0.0739 & 0.1102 & 0.0725 \\ 
				&CP  & 0.9490 & 0.9630 & 0.9440 & 0.9150 & 0.9220 & 0.9400 & 0.9520 & 0.8230 & 0.9570 \\ 
				
				\cline{2-11}  \\ [-0.7em]
				$\beta_2$				
	
				&Bias& -0.0041 & 0.0033 & -0.0005 & -0.0106 & -0.0152 & -0.0106 & -0.0152 & -0.2948 & -0.0003 \\ 
				&SD  & 0.2012 & 0.3009 & 0.3152 & 0.2723 & 0.2615 & 0.2723 & 0.2615 & 0.1786 & 0.2598 \\ 
				&ASE & 0.2099 & 0.3044 & 0.2965 & 0.2355 & 0.2388 & 0.2531 & 0.2658 & 0.2398 & 0.2586 \\ 
				&MSE & 0.0405 & 0.0905 & 0.0994 & 0.0742 & 0.0686 & 0.0742 & 0.0686 & 0.1188 & 0.0675 \\ 
				&CP  & 0.9560 & 0.9630 & 0.9360 & 0.9090 & 0.9230 & 0.9310 & 0.9540 & 0.8060 & 0.9570 \\ 
				
				\cline{2-11}  \\ [-0.7em]
				$\beta_3$			
	
				&Bias& 0.0072 & 0.1173 & 0.0302 & 0.0074 & 0.0057 & 0.0074 & 0.0057 & -0.0103 & 0.0128 \\ 
				&SD  & 0.2085 & 0.3233 & 0.2323 & 0.2092 & 0.2084 & 0.2092 & 0.2084 & 0.2067 & 0.2094 \\ 
				&ASE & 0.2104 & 0.3207 & 0.2157 & 0.2116 & 0.2115 & 0.2077 & 0.2117 & 0.2095 & 0.2127 \\ 
				&MSE & 0.0435 & 0.1183 & 0.0549 & 0.0438 & 0.0435 & 0.0438 & 0.0435 & 0.0428 & 0.0440 \\ 
				&CP  & 0.9600 & 0.9580 & 0.9370 & 0.9570 & 0.9580 & 0.9550 & 0.9570 & 0.9520 & 0.9590 \\ 
				
				\hline \\ [-0.7em]
				\multicolumn{11}{c}{Observed selection probabilities: $(0.36,0.24,0.24,0.16)$; 
					$\bm\alpha=(1,0.6,0.6)^{\tT}$}\\
				$\beta_0$	
	
				&Bias& 0.0104 & 0.0987 & 0.0106 & 0.0204 & 0.0263 & 0.0204 & 0.0263 & 0.2643 & 0.0087 \\ 
				&SD  & 0.1532 & 0.2598 & 0.2208 & 0.1927 & 0.1811 & 0.1927 & 0.1811 & 0.1564 & 0.1796 \\ 
				&ASE & 0.1521 & 0.2567 & 0.1985 & 0.1622 & 0.1637 & 0.1733 & 0.1728 & 0.1788 & 0.1764 \\ 
				&MSE & 0.0236 & 0.0772 & 0.0489 & 0.0376 & 0.0335 & 0.0376 & 0.0335 & 0.0943 & 0.0323 \\ 
				&CP  & 0.9520 & 0.9480 & 0.9310 & 0.8970 & 0.9210 & 0.9220 & 0.9320 & 0.7270 & 0.9490 \\ 
				
				\cline{2-11}  \\ [-0.7em]
				$\beta_1$			
		
				&Bias& 0.0197 & 0.0468 & 0.0494 & 0.0174 & 0.0081 & 0.0174 & 0.0081 & -0.3518 & 0.0374 \\ 
				&SD  & 0.2187 & 0.3867 & 0.4124 & 0.3330 & 0.3099 & 0.3330 & 0.3099 & 0.1799 & 0.3008 \\ 
				&ASE & 0.2108 & 0.3701 & 0.3479 & 0.2385 & 0.2455 & 0.2696 & 0.2772 & 0.2487 & 0.2821 \\ 
				&MSE & 0.0482 & 0.1517 & 0.1725 & 0.1112 & 0.0961 & 0.1112 & 0.0961 & 0.1561 & 0.0919 \\ 
				&CP  & 0.9490 & 0.9460 & 0.9110 & 0.8510 & 0.8860 & 0.8870 & 0.9170 & 0.7550 & 0.9480 \\ 
				
				\cline{2-11}  \\ [-0.7em]
				$\beta_2$			
		
				&Bias& -0.0041 & 0.0258 & 0.0319 & 0.0030 & -0.0077 & 0.0030 & -0.0077 & -0.3637 & 0.0190 \\ 
				&SD  & 0.2012 & 0.3776 & 0.4078 & 0.3288 & 0.3016 & 0.3288 & 0.3016 & 0.1751 & 0.2888 \\ 
				&ASE & 0.2099 & 0.3685 & 0.3458 & 0.2392 & 0.2450 & 0.2684 & 0.2755 & 0.2480 & 0.2808 \\ 
				&MSE & 0.0405 & 0.1432 & 0.1673 & 0.1081 & 0.0910 & 0.1081 & 0.0910 & 0.1629 & 0.0838 \\ 
				&CP  & 0.9560 & 0.9530 & 0.9250 & 0.8580 & 0.8950 & 0.9100 & 0.9320 & 0.7240 & 0.9580 \\  
				
				\cline{2-11}  \\ [-0.7em]
				$\beta_3$
		
				&Bias& 0.0072 & 0.1678 & 0.0933 & 0.0081 & 0.0063 & 0.0081 & 0.0063 & -0.0128 & 0.0161 \\ 
				&SD  & 0.2085 & 0.4101 & 0.2899 & 0.2121 & 0.2109 & 0.2121 & 0.2109 & 0.2076 & 0.2124 \\ 
				&ASE & 0.2104 & 0.3955 & 0.2237 & 0.2121 & 0.2118 & 0.2029 & 0.2112 & 0.2093 & 0.2137 \\ 
				&MSE & 0.0435 & 0.1963 & 0.0928 & 0.0450 & 0.0445 & 0.0450 & 0.0445 & 0.0433 & 0.0454 \\ 
				&CP  & 0.9600 & 0.9500 & 0.8760 & 0.9570 & 0.9570 & 0.9420 & 0.9560 & 0.9540 & 0.9640 \\ 
				
				\hline \\ [-0.7em]
				\multicolumn{11}{l}{$\hbb_{M1}$: MI1 estimator whose value is the solution of $\BU_{M1}(\bb)=\b0$ in Equation~\eqref{eq: score mi1} with Rubin's type}\\
				\multicolumn{11}{l}{estimated variance in Equation~\eqref{MI1: asymptotic var}} \\
				\multicolumn{11}{l}{$\hbb_{M2}$: MI2 estimator whose value is the solution of $\BU_{M2}(\bb)=\b0$ in Equation~\eqref{eq: score mi2} with Rubin's type}\\
				\multicolumn{11}{l}{estimated variance in Equation~\eqref{MI2: asymptotic var}} \\
				\multicolumn{11}{l}{$\hbb_{M1n}$: MI1 estimator whose value is the solution of $\BU_{M1}(\bb)=\b0$ in Equation~\eqref{eq: score mi1} with proposed}\\
				\multicolumn{11}{l}{estimated variance in Equation~\eqref{MI1n: asymptotic var}} \\
				\multicolumn{11}{l}{$\hbb_{M2n}$: MI2 estimator whose value is the solution of $\BU_{M2}(\bb)=\b0$ in Equation~\eqref{eq: score mi2} with proposed} \\
				\multicolumn{11}{l}{estimated variance in Equation~\eqref{MI2n: asymptotic var}}
			\end{tabular}
		\end{center}
		\label{tab: simulation study 4}
	}
\end{table}

\newpage
\clearpage
\begin{table}[h!] 
	{\footnotesize
		\caption{Relative efficiencies in Study 4 ($M=15$; $n= 1,500$); $\bb=(1.2,1,1,1)^{\tT}$; $\bga=(0.7,-0.2,0.1,-1.2)^{\tT}$}
		\vspace{-0.8cm}
		\setlength\tabcolsep{5.5 pt} 
		\begin{center}
			\begin{tabular}{cccccccccccccc}
				\hline  \\[-0.9em]
				\multicolumn{14}{c}{Relative efficiency}\\
				\hline \\[-0.9em]
				
				$\bb$&$C1$&$W1$&$M11$ & $M21$ & $RF1$&$EM1$& $C2$&$W2$&$M12$ & $M22$ & $RF2$&$EM2$&$M12n$\\
				\hline  \\[-0.9em]
				\multicolumn{14}{c}{Observed selection probabilities: $(0.70,0.12,0.12,0.06)$; $\ba=(2.4,0.6,0.6)^{\tT}$}\\				
		
				$\beta_0$ & 1.143 & 1.040 & 0.987 & 0.988 & 1.025 & 1.005 & 1.124 & 1.022 & 0.970 & 0.972 & 1.008 & 0.989 & 0.983 \\ 
				$\beta_1$ & 1.091 & 1.084 & 0.972 & 0.978 & 0.974 & 1.008 & 1.043 & 1.036 & 0.929 & 0.935 & 0.931 & 0.963 & 0.956 \\ 
				$\beta_2$ & 1.091 & 1.082 & 0.972 & 0.978 & 0.974 & 1.008 & 1.044 & 1.036 & 0.930 & 0.936 & 0.933 & 0.965 & 0.957 \\ 
				$\beta_3$ & 1.265 & 1.011 & 1.006 & 1.005 & 1.000 & 1.009 & 1.257 & 1.004 & 0.999 & 0.998 & 0.993 & 1.002 & 0.993 \\ 
				
				\hline \\[-0.9em]
				\multicolumn{14}{c}{Observed selection probabilities: $(0.50,0.19,0.19,0.12)$; $\ba=(1.6,0.6,0.6)^{\tT}$}\\					
	
				$\beta_0$ & 1.286 & 1.084 & 0.963 & 0.967 & 1.031 & 1.009 & 1.267 & 1.068 & 0.949 & 0.952 & 1.016 & 0.994 & 0.985 \\ 
				$\beta_1$ & 1.197 & 1.168 & 0.927 & 0.942 & 0.941 & 1.019 & 1.141 & 1.113 & 0.883 & 0.897 & 0.896 & 0.971 & 0.953 \\ 
				$\beta_2$ & 1.203 & 1.172 & 0.931 & 0.944 & 0.948 & 1.022 & 1.145 & 1.116 & 0.886 & 0.899 & 0.902 & 0.973 & 0.952 \\ 
				$\beta_3$ & 1.544 & 1.038 & 1.019 & 1.018 & 1.009 & 1.024 & 1.515 & 1.019 & 1.000 & 0.999 & 0.990 & 1.005 & 0.981 \\ 
				
				\hline \\[-0.9em]
				\multicolumn{14}{c}{Observed selection probabilities: $(0.36,0.24,0.24,0.16)$; $\ba=(1.0,0.6,0.6)^{\tT}$}\\

				$\beta_0$ & 1.482 & 1.146 & 0.936 & 0.945 & 1.032 & 1.018 & 1.485 & 1.148 & 0.939 & 0.947 & 1.034 & 1.021 & 1.002 \\ 
				$\beta_1$ & 1.373 & 1.291 & 0.885 & 0.911 & 0.923 & 1.047 & 1.335 & 1.255 & 0.860 & 0.885 & 0.897 & 1.018 & 0.972 \\ 
				$\beta_2$ & 1.373 & 1.288 & 0.891 & 0.913 & 0.924 & 1.046 & 1.338 & 1.255 & 0.868 & 0.889 & 0.900 & 1.020 & 0.974 \\ 
				$\beta_3$ & 1.949 & 1.103 & 1.045 & 1.044 & 1.031 & 1.053 & 1.872 & 1.059 & 1.004 & 1.003 & 0.991 & 1.012 & 0.961 \\ 
				
				\hline \\ [-2mm]
				\multicolumn{14}{l}{$Ar$, $A\in\{C,W,M1,M2,RF,EM\}$, $r=1,2$, are the relative efficiencies of CC, SIPW, MI1, MI2, } \\
			\multicolumn{14}{l}{RFMI, and SAEM estimators to MI$r$n estimators for each $\beta_i$, $i=0,1,2,3$. }\\
				\multicolumn{14}{l}{$Ar$ is the ratio of ASE of $A$ estimator to ASE of MI$r$n estimator for each $\beta_i$.}\\
				\multicolumn{14}{l}{$Ar=ASE_A/ASE_{MIrn}$, $M12n=ASE_{MI1n}/ASE_{MI2n}$} \\
				\multicolumn{14}{l}{MI1: MI1 estimation method  with Rubin's type estimated variance in Equation~\eqref{MI1: asymptotic var}} \\
				\multicolumn{14}{l}{MI2: MI2 estimation method  with Rubin's type estimated variance in Equation~\eqref{MI2: asymptotic var}} \\
				\multicolumn{14}{l}{MI1n: MI1 estimation method with proposed estimated variance in Equation~\eqref{MI1n: asymptotic var}} \\
				\multicolumn{14}{l}{MI2n: MI2 estimation method with proposed estimated variance in Equation~\eqref{MI2n: asymptotic var} }			
			\end{tabular}
			\label{tab: RE_04}
		\end{center}
	}
\end{table}

\newpage
\clearpage
\begin{table}[h!]
	\caption{Results of logistic regression analysis of Feng Chia night market (FCNM) survey data with original missing data, i.e., selection probabilities: $(0.666,0.334,0,0)$}
	\vspace{-0.7cm}
	\begin{center}
		\setlength\tabcolsep{4 pt}
		{\small \begin{tabular}{llcccccccc}   
				\hline \\ [-0.8em]
			Parameter   &&$\hbb_C$ &$\hbb_W$ &$\hbb_{M1}$&$\hbb_{M2}$& $\hbb_{M1n}$&$\hbb_{M2n}$& $\hbb_{RF}$ & $\hbb_{EM}$ \\
				\hline  \\ [-0.9em]
				$\beta_0$ 
	
				&est        & -2.5586 & -2.6680 & -2.7531 & -2.7927 & -2.7531 & -2.7927 & -2.7613 & -2.7427 \\ 
				&ASE        & 0.1628 & 0.1497 & 0.1426 & 0.1470 & 0.1448 & 0.1535 & 0.1468 & 0.1446 \\ 
				&$p$-value  & 0.0000 & 0.0000 & 0.0000 & 0.0000 & 0.0000 & 0.0000 & 0.0000 & 0.0000 \\ 
						
				$\beta_1$     	
		
				&est        & 0.5423 & 0.5463 & 0.5456 & 0.6302 & 0.5456 & 0.6302 & 0.5546 & 0.5091 \\ 
				&ASE        & 0.1674 & 0.1656 & 0.1521 & 0.1664 & 0.1644 & 0.1638 & 0.1744 & 0.1669 \\ 
				&$p$-value  & 0.0012 & 0.0010 & 0.0003 & 0.0002 & 0.0009 & 0.0001 & 0.0015 & 0.0023 \\ 
				
				$\beta_2$      
	
				&est        & 1.7978 & 1.8184 & 1.8649 & 1.8789 & 1.8649 & 1.8789 & 1.8820 & 1.8771 \\ 
				&ASE        & 0.1756 & 0.1713 & 0.1483 & 0.1477 & 0.1472 & 0.1745 & 0.1475 & 0.1487 \\ 
				&$p$-value  & 0.0000 & 0.0000 & 0.0000 & 0.0000 & 0.0000 & 0.0000 & 0.0000 & 0.0000 \\ 
				
				$\beta_3$   
				
				&est        & 0.1201 & 0.2294 & 0.2459 & 0.2397 & 0.2459 & 0.2397 & 0.2446 & 0.2479 \\ 
				&ASE        & 0.1108 & 0.1048 & 0.0996 & 0.1005 & 0.0991 & 0.1046 & 0.0921 & 0.0918 \\ 
				&$p$-value  &\bf 0.2783 & 0.0286 & 0.0135 & 0.0170 & 0.0131 & 0.0231 & 0.0079 & 0.0069 \\

				\hline \\[-0.8em]
				\multicolumn{10}{l}{Estimate (est), asymptotic standard error (ASE), and $p$-value for testing parameter = 0} \\
				\multicolumn{10}{l}{$\texttt{X}_1$ ($1= \{1,2,3\}$ times; $0=$ more than three times): response to the question}\\
				\multicolumn{10}{l}{\hskip 2mm  ``What is the number of times you went shopping in the FCNM in the last half year?''.}\\
				\multicolumn{10}{l}{$\texttt{X}_2$ ($1=$ Neighborhood of Taichung and Others City, $0=$ Taichung City): }\\
				\multicolumn{10}{l}{\hskip 2mm response to the question  ``Which county or city is your current place of residence?''.}\\
				\multicolumn{10}{l}{$Z$ ($Z=\{0.1,0.35,0.75,1.25,1.75,3\}$ after taking the median and being divided by $1,000$): }\\
				\multicolumn{10}{l}{\hskip 2mm response to question ``How much do you spend each time you visit FCNM?”.}\\
				\multicolumn{10}{l}{$\hbb_{M1}$: MI1 estimator whose value is the solution of $\BU_{M1}(\bb)=\b0$ in Equation~\eqref{eq: score mi1} with}\\
				\multicolumn{10}{l}{Rubin's type estimated variance in Equation~\eqref{MI1: asymptotic var}} \\
				\multicolumn{10}{l}{$\hbb_{M2}$: MI2 estimator whose value is the solution of $\BU_{M2}(\bb)=\b0$ in Equation~\eqref{eq: score mi2} with}\\
				\multicolumn{10}{l}{Rubin's type estimated variance in Equation~\eqref{MI2: asymptotic var}} \\
				\multicolumn{10}{l}{$\hbb_{M1n}$: MI1 estimator whose value is the solution of $\BU_{M1}(\bb)=\b0$ in Equation~\eqref{eq: score mi1} with}\\
				\multicolumn{10}{l}{proposed estimated variance in Equation~\eqref{MI1n: asymptotic var}} \\
				\multicolumn{10}{l}{$\hbb_{M2n}$: MI2 estimator whose value is the solution of $\BU_{M2}(\bb)=\b0$ in Equation~\eqref{eq: score mi2} with} \\
				\multicolumn{10}{l}{proposed estimated variance in Equation~\eqref{MI2n: asymptotic var}}
		\end{tabular}}
	\end{center}
	\label{tab: example 1a}
\end{table}

\newpage
\clearpage
\begin{table}[h!]
	\caption{Results of logistic regression analysis of Feng Chia night market (FCNM) survey data with artificial missing data, i.e., selection probabilities: $(0.485, 0.255, 0.181, 0.079)$}
	\vspace{-0.7cm}
	\begin{center}
		\setlength\tabcolsep{4 pt}
		{\small \begin{tabular}{llcccccccc}   
				\hline \\ [-0.8em]
			Parameter   & &$\hbb_C$&$\hbb_W$ &$\hbb_{M1}$&$\hbb_{M2}$& $\hbb_{M1n}$&$\hbb_{M2n}$& $\hbb_{RF}$ & $\hbb_{EM}$ \\
				\hline  \\ [-0.9em]
				$\beta_0$  
	
				&est        & -3.0189 & -2.8525 & -2.7711 & -2.8419 & -2.7711 & -2.8419 & -2.5365 & -2.7117 \\ 
				&ASE        & 0.2221 & 0.1923 & 0.1486 & 0.1562 & 0.1580 & 0.1698 & 0.1601 & 0.1531 \\ 
				&$p$-value  & 0.0000 & 0.0000 & 0.0000 & 0.0000 & 0.0000 & 0.0000 & 0.0000 & 0.0000 \\ 
				
				$\beta_1$      
		
				&est        & 0.6458 & 0.6781 & 0.5828 & 0.7056 & 0.5828 & 0.7056 & 0.6027 & 0.6297 \\ 
				&ASE        & 0.2192 & 0.2073 & 0.1563 & 0.1693 & 0.1640 & 0.1657 & 0.1633 & 0.1697 \\ 
				&$p$-value  & 0.0032 & 0.0011 & 0.0002 & 0.0000 & 0.0004 & 0.0000 & 0.0002 & 0.0002 \\ 
				
				$\beta_2$      
		
				&est        & 1.9736 & 2.0838 & 1.9205 & 1.9396 & 1.9205 & 1.9396 & 1.4151 & 1.7793 \\ 
				&ASE        & 0.2346 & 0.2227 & 0.1634 & 0.1685 & 0.1790 & 0.1784 & 0.2424 & 0.1799 \\ 
				&$p$-value  & 0.0000 & 0.0000 & 0.0000 & 0.0000 & 0.0000 & 0.0000 & 0.0000 & 0.0000 \\ 
				
				$\beta_3$        
		
				&est        & 0.0452 & 0.1255 & 0.2277 & 0.2202 & 0.2277 & 0.2202 & 0.3436 & 0.2212 \\ 
				&ASE        & 0.1478 & 0.1002 & 0.0938 & 0.0958 & 0.0897 & 0.0934 & 0.0952 & 0.0945 \\ 
				&$p$-value  &\bf 0.7599 &\bf 0.2103 & 0.0152 & 0.0215 & 0.0111 & 0.0184 & 0.0003 & 0.0193 \\

				\hline \\[-0.8em]
\multicolumn{10}{l}{Estimate (est), asymptotic standard error (ASE), and $p$-value for testing parameter = 0} \\
\multicolumn{10}{l}{$\texttt{X}_1$ ($1= \{1,2,3\}$ times; $0=$ more than three times): response to the question}\\
\multicolumn{10}{l}{\hskip 2mm  ``What is the number of times you went shopping in the FCNM in the last half year?''.}\\
\multicolumn{10}{l}{$\texttt{X}_2$ ($1=$ Neighborhood of Taichung and Others City, $0=$ Taichung City): }\\
\multicolumn{10}{l}{\hskip 2mm response to the question  ``Which county or city is your current place of residence?''.}\\
\multicolumn{10}{l}{$Z$ ($Z=\{0.1,0.35,0.75,1.25,1.75,3\}$ after taking the median and dividing by $1,000$): }\\
\multicolumn{10}{l}{\hskip 2mm response to question ``How much do you spend each time you visit FCNM?”.}\\
\multicolumn{10}{l}{$\hbb_{M1}$: MI1 estimator whose value is the solution of $\BU_{M1}(\bb)=\b0$ in Equation~\eqref{eq: score mi1} with}\\
\multicolumn{10}{l}{Rubin's type estimated variance in Equation~\eqref{MI1: asymptotic var}} \\
\multicolumn{10}{l}{$\hbb_{M2}$: MI2 estimator whose value is the solution of $\BU_{M2}(\bb)=\b0$ in Equation~\eqref{eq: score mi2} with}\\
\multicolumn{10}{l}{Rubin's type estimated variance in Equation~\eqref{MI2: asymptotic var}} \\
\multicolumn{10}{l}{$\hbb_{M1n}$: MI1 estimator whose value is the solution of $\BU_{M1}(\bb)=\b0$ in Equation~\eqref{eq: score mi1} with}\\
\multicolumn{10}{l}{proposed estimated variance in Equation~\eqref{MI1n: asymptotic var}} \\
\multicolumn{10}{l}{$\hbb_{M2n}$: MI2 estimator whose value is the solution of $\BU_{M2}(\bb)=\b0$ in Equation~\eqref{eq: score mi2} with} \\
\multicolumn{10}{l}{proposed estimated variance in Equation~\eqref{MI2n: asymptotic var}}
		\end{tabular}}
	\end{center}
	\label{tab: example 1b}
\end{table}


\begin{thebibliography}{999}
	\providecommand{\natexlab}[1]{#1}
	
	\bibitem[Austin and van Buuren(2022)]{austin2022effect}
	Austin, P. C., and van Buuren, S. (2022). 
	\newblock The effect of high prevalence of missing data on estimation of the coefficients of a logistic regression model when using multiple imputation. 
	\newblock {\em BMC Medical Research Methodology}, 22, 196.
	
	
	
	\bibitem[Buuren and Groothuis-Oudshoorn(2011)]{buuren2011mice}
	Buuren, S. V., and Groothuis-Oudshoorn, K. (2011).
	\newblock mice: Multivariate imputation by chained equations in R.
	\newblock {\em Journal of Statistical Software}, 45(3), 1--67.
	
	
	
	\bibitem[Fay(1996)]{fay1996alternative}
	Fay, R. E. (1996).
	\newblock Alternative paradigms for the analysis of imputed survey data.
	\newblock {\em Journal of the American Statistical Association}, 91, 490--498.
	
	\bibitem[Foutz(1997)]{foutz1997unique}
	 Foutz, R. V. (1977). 
	\newblock On the unique consistent solution to the likelihood equations. 
	\newblock {\em Journal of the American Statistical Association}, 72, 147--48.
	
	\bibitem[Horvitz and Thompson(1952)]{horvitz1952generalization}								
	Horvitz, D. G., and Thompson, D. J. (1952).
	\newblock A generalization of sampling without replacement from a finite universe.
	\newblock {\em Journal of the American Statistical Association}, 47, 663--685.
	
	
	
	\bibitem[Hsieh et al.(2010)]{hsieh2010logistic}
	Hsieh, S. H., Lee, S. M., and Shen, P. S. (2010).
	Logistic regression analysis of randomized response data with missing covariates.
	\newblock {\em Journal of Statistical Planning and Inference} 140, 927--940.			
	
	\bibitem[Hsieh et al.(2013)]{hsieh2013logistic}
	Hsieh, S. H., Li, C. S., and Lee,  S. M.  (2013).
	\newblock Logistic regression with outcome and covariates missing separately or simultaneously.
	\newblock {\em Computational Statistics and Data Analysis}, 66, 32--54.
	
	\bibitem[Jiang et al.(2020)]{jiang2020logistic}
	Jiang, W., Josse, J., Lavielle, M., and  Group, T. (2020).
	\newblock Logistic regression with missing covariates -- Parameter estimation, model selection and prediction within a joint-modeling framework.
	\newblock {\em Computational Statistics and Data Analysis}, 145, 106907.
	
	\bibitem[Josse and Husson(2016)]{josse2016missmda}
	Josse, J., and Husson, F. (2016). 
	\newblock missMDA: a package for handling missing values in multivariate data analysis. 
	\newblock {\em Journal of Statistical Software}, 70, 1-31.
	
	\bibitem[Lee et al.(2011)]{lee2011semiparametric}
	Lee, S. M., Gee, M. J., and Hsieh, S. H. (2011).
	\newblock Semiparametric methods in the proportional odds model for ordinal response data with missing covariates.
	\newblock {\em Biometrics}, 67, 788--798.					
	
	\bibitem[Lee et al.(2016)]{lee2016estimation}
	Lee, S. M., Hwang, W. H., and de Dieu Tapsoba, J. (2016).
	\newblock Estimation in closed capture–recapture models when covariates are missing at random.
	\newblock {\em Biometrics}, 72, 1294--1304.
	
	\bibitem[Lee et al.(2023)]{lee2023estimation}	
	Lee, S. M., Le, T. N., Tran, P. L., and Li, C. S. (2023). 
	\newblock Estimation of logistic regression with covariates missing separately or simultaneously via multiple imputation methods. 
	\newblock {\em Computational Statistics}, 38, 899–934. 
	
	\bibitem[Lee et al.(2012)]{lee2012semiparametric}
	Lee, S. M., Li, C. S., Hsieh, S. H., and Huang, L. H. (2012).
	\newblock Semiparametric estimation of logistic regression model with missing covariates and outcome.
	\newblock {\em Metrika}, 75, 621--653.
	
	\bibitem[Lee et al.(2020)]{lee2020estimation}	
	Lee, S. M., Lukusa, T. M., and Li, C. S. (2020). 
	\newblock Estimation of a zero-inflated Poisson regression model with missing covariates via nonparametric multiple imputation methods. 
	\newblock {\em Computational Statistics}, 35, 725--754.
	
	\bibitem[Lee et al.(2022)]{lee2022goodness}	
	Lee, S. M., Tran, P. L., and Li, C. S. (2022). 
	\newblock Goodness-of-fit tests for a logistic regression model with missing covariates. 
	\newblock {\em Statistical  Methods in Medical Research}, 31, 1031-1050.
	

	
	
	
	\bibitem[Lukusa et al.(2016)]{lukusa2016semiparametric}
	Lukusa, T. M., Lee, S. M., and Li, C. S. (2016).
	\newblock Semiparametric estimation of a zero-inflated Poisson regression model with missing covariates.
	\newblock {\em Metrika}, 79, 457--483.
	
	
	
	\bibitem[Righi et al.(2014)]{righi2014methods}
	Righi, P., Falorsi, S., and Fasulo, A. (2014). 
	\newblock Methods for variance estimation under random hot deck imputation in business surveys. 
	\newblock {\em Rivista Di Statistica Ufficiale}, 16, 45--64.
	
	\bibitem[Rubin(1976)]{rubin1976inference}
	Rubin, D. B. (1976).
	\newblock Inference and missing data.
	\newblock {\em Biometrika}, 63, 581--592.
	
	\bibitem[Rubin(1978)]{rubin1978multiple}
	Rubin, D. B. (1978). 
	\newblock Multiple imputations in sample surveys-a phenomenological Bayesian approach to nonresponse. 
	\newblock In {\em Proceedings of the Survey Research Methods Section of the American Statistical Association}, 1, 20-34.
	
	\bibitem[Rubin(1987)]{rubin1987statistical}
	Rubin, D. B. (1987).
	\newblock Statistical analysis with missing data.
	\newblock {\em John Wiley} \& {\em Sons, New York}.
	
	\bibitem[Rubin(1996)]{rubin1996multiple}
	Rubin, D. B. (1996).
	\newblock Multiple imputation after 18+ years.
	\newblock {\em Journal of the American Statistical Association}, 91, 473--489.
	
	\bibitem[Rubin and Schenker(1986)]{rubin1986multiple}
	Rubin, D. B., and Schenker, N. (1986).
	\newblock Multiple imputation for interval estimation from simple random samples with ignorable nonresponse.
	\newblock {\em Journal of the American Statistical Association}, 81, 366--374.
	
	
	\bibitem[Tran et al.(2023)]{tran2023}
	Tran, P. L., Le, T. N., Lee, S. M., and Li, C. S. (2023).
	\newblock Estimation of parameters of logistic regression with covariates missing separately or simultaneously.
	\newblock {\em Communications in Statistics – Theory and Methods}, 52, 1981-2009.  												
	
	\bibitem[Wang et al.(2002)]{wang2002joint}
	Wang, C. Y., Chen, J. C., Lee, S. M., and Ou, S. T. (2002).
	\newblock Joint conditional likelihood estimator in logistic regression with missing covariate data.
	\newblock {\em Statistica Sinica}, 12, 555--574.
	
	
	\bibitem[Wang et al.(1997)]{wang1997weighted}
	Wang, C. Y., Wang, S., Zhao, L. P., and Ou, S. T. (1997).
	\newblock Weighted semiparametric estimation in regression analysis with missing covariate data.
	\newblock {\em Journal of the American Statistical Association}, 92, 512--525.
	
	\bibitem[Wang and Chen(2009)]{wang2009empirical}											
	Wang, D., and Chen, S. X. (2009).
	\newblock Empirical likelihood for estimating equations with missing values.
	\newblock {\em The Annals of Statistics}, 37, 490--517.
	
	\bibitem[Wang and Wang(2001)]{wang2001note}
	Wang, S., and Wang, C. Y. (2001).
	\newblock A note on kernel assisted estimators in missing covariate regression.
	\newblock {\em Statistics and Probability Letters}, 55, 439--449.
	
	
	
	\bibitem[Zhao and Lipsitz(1992)]{zhao1992designs}
	Zhao, L. P., and Lipsitz, S. (1992).
	\newblock Designs and analysis of two-stage studies.
	\newblock {\em Statistics in Medicine}, 11, 769--782.
	
\end{thebibliography}
\end{document}